\documentclass[12pt]{article}
\usepackage{graphicx}
\usepackage{latexsym,amsmath,amsfonts,amssymb}
\newcommand{\be}{\begin{equation}}
\newcommand{\ee}{\end{equation}}
\newcommand{\beq}{\begin{equation}}
\newcommand{\eeq}{\end{equation}}
\newcommand{\bea}{\begin{eqnarray}}
\newcommand{\eea}{\end{eqnarray}}

\newcommand{\ba}{\begin{eqnarray}}
\newcommand{\ea}{\end{eqnarray}}
\begin{document}
\baselineskip=15.5pt
\pagestyle{plain}
\setcounter{page}{1}


\def\del{{\partial}}
\def\vev#1{\left\langle #1 \right\rangle}
\def\cn{{\cal N}}
\def\co{{\cal O}}
\def\IC{{\mathbb C}}
\def\IR{{\mathbb R}}
\def\IZ{{\mathbb Z}}
\def\RP{{\bf RP}}
\def\CP{{\bf CP}}
\def\Poincare{{Poincar\'e }}
\def\tr{{\rm tr}}
\def\tp{{\tilde \Phi}}

\def\TL{\hfil$\displaystyle{##}$}
\def\TR{$\displaystyle{{}##}$\hfil}
\def\TC{\hfil$\displaystyle{##}$\hfil}
\def\TT{\hbox{##}}
\def\HLINE{\noalign{\vskip1\jot}\hline\noalign{\vskip1\jot}}
\def\seqalign#1#2{\vcenter{\openup1\jot
  \halign{\strut #1\cr #2 \cr}}}
\def\lbldef#1#2{\expandafter\gdef\csname #1\endcsname {#2}}
\def\eqn#1#2{\lbldef{#1}{(\ref{#1})}%
\begin{equation} #2 \label{#1} \end{equation}}
\def\eqalign#1{\vcenter{\openup1\jot
    \halign{\strut\span\TL & \span\TR\cr #1 \cr
   }}}
\def\eno#1{(\ref{#1})}
\def\href#1#2{#2}
\def\half{{1 \over 2}}

\def\ads{{\it AdS}}
\def\adsp{{\it AdS}$_{p+2}$}
\def\cft{{\it CFT}}

\newcommand{\ber}{\begin{eqnarray}}
\newcommand{\eer}{\end{eqnarray}}

\newcommand{\beqar}{\begin{eqnarray}}
\newcommand{\cN}{{\cal N}}
\newcommand{\cO}{{\cal O}}
\newcommand{\cA}{{\cal A}}
\newcommand{\cT}{{\cal T}}
\newcommand{\cF}{{\cal F}}
\newcommand{\cC}{{\cal C}}
\newcommand{\cR}{{\cal R}}
\newcommand{\cW}{{\cal W}}
\newcommand{\eeqar}{\end{eqnarray}}
\newcommand{\tht}{\thteta}
\newcommand{\lm}{\lambda}\newcommand{\Lm}{\Lambda}
\newcommand{\eps}{\epsilon}


\newcommand{\nonu}{\nonumber}
\newcommand{\oh}{\displaystyle{\frac{1}{2}}}
\newcommand{\dsl}
  {\kern.06em\hbox{\raise.15ex\hbox{$/$}\kern-.56em\hbox{$\partial$}}}
\newcommand{\id}{i\!\!\not\!\partial}
\newcommand{\as}{\not\!\! A}
\newcommand{\ps}{\not\! p}
\newcommand{\ks}{\not\! k}
\newcommand{\D}{{\cal{D}}}
\newcommand{\dv}{d^2x}
\newcommand{\Z}{{\cal Z}}
\newcommand{\N}{{\cal N}}
\newcommand{\Dsl}{\not\!\! D}
\newcommand{\Bsl}{\not\!\! B}
\newcommand{\Psl}{\not\!\! P}
\newcommand{\eeqarr}{\end{eqnarray}}
\newcommand{\ZZ}{{\rm \kern 0.275em Z \kern -0.92em Z}\;}

                                                                                                    
\def\del{{\delta^{\hbox{\sevenrm B}}}} \def\ex{{\hbox{\rm e}}}
\def\azb{A_{\bar z}} \def\az{A_z} \def\bzb{B_{\bar z}} \def\bz{B_z}
\def\czb{C_{\bar z}} \def\cz{C_z} \def\dzb{D_{\bar z}} \def\dz{D_z}
\def\im{{\hbox{\rm Im}}} \def\mod{{\hbox{\rm mod}}} \def\tr{{\hbox{\rm Tr}}}
\def\ch{{\hbox{\rm ch}}} \def\imp{{\hbox{\sevenrm Im}}}
\def\trp{{\hbox{\sevenrm Tr}}} \def\vol{{\hbox{\rm Vol}}}
\def\rl{\Lambda_{\hbox{\sevenrm R}}} \def\wl{\Lambda_{\hbox{\sevenrm W}}}
\def\fc{{\cal F}_{k+\cox}} \def\vev{vacuum expectation value}
\def\nodiv{\mid{\hbox{\hskip-7.8pt/}}}
\def\ie{{\em i.e.}}
\def\ie{\hbox{\it i.e.}}

\def\CC{{\mathchoice
{\rm C\mkern-8mu\vrule height1.45ex depth-.05ex
width.05em\mkern9mu\kern-.05em}
{\rm C\mkern-8mu\vrule height1.45ex depth-.05ex
width.05em\mkern9mu\kern-.05em}
{\rm C\mkern-8mu\vrule height1ex depth-.07ex
width.035em\mkern9mu\kern-.035em}
{\rm C\mkern-8mu\vrule height.65ex depth-.1ex
width.025em\mkern8mu\kern-.025em}}}
                                                                                                    
\def\RR{{\rm I\kern-1.6pt {\rm R}}}
\def\NN{{\rm I\!N}}
\def\ZZ{{\rm Z}\kern-3.8pt {\rm Z} \kern2pt}
\def\IB{\relax{\rm I\kern-.18em B}}
\def\ID{\relax{\rm I\kern-.18em D}}
\def\II{\relax{\rm I\kern-.18em I}}
\def\IP{\relax{\rm I\kern-.18em P}}
\newcommand{\CS}{{\scriptstyle {\rm CS}}}
\newcommand{\CSs}{{\scriptscriptstyle {\rm CS}}}
\newcommand{\rc}{\nonumber\\}
\newcommand{\bear}{\begin{eqnarray}}
\newcommand{\eear}{\end{eqnarray}}
\newcommand{\W}{{\cal W}}
\newcommand{\F}{{\cal F}}
\newcommand{\x}{{\cal O}}
\newcommand{\LL}{{\cal L}}
                                                                                                    
\def\mani{{\cal M}}
\def\calo{{\cal O}}
\def\calb{{\cal B}}
\def\calw{{\cal W}}
\def\calz{{\cal Z}}
\def\cald{{\cal D}}
\def\calc{{\cal C}}
\def\to{\rightarrow}
\def\ele{{\hbox{\sevenrm L}}}
\def\ere{{\hbox{\sevenrm R}}}
\def\zb{{\bar z}}
\def\wb{{\bar w}}
\def\nodiv{\mid{\hbox{\hskip-7.8pt/}}}
\def\menos{\hbox{\hskip-2.9pt}}
\def\dr{\dot R_}
\def\drr{\dot r_}
\def\ds{\dot s_}
\def\da{\dot A_}
\def\dga{\dot \gamma_}
\def\ga{\gamma_}
\def\dal{\dot\alpha_}
\def\al{\alpha_}
\def\cl{{closed}}
\def\cls{{closing}}
\def\vev{vacuum expectation value}
\def\tr{{\rm Tr}}
\def\to{\rightarrow}
\def\too{\longrightarrow}


\def\a{\alpha}
\def\b{\beta}
\def\c{\gamma}
\def\d{\delta}
\def\e{\epsilon}           
\def\f{\phi}               
\def\vf{\varphi}  \def\tvf{\tilde{\varphi}}
\def\vp{\varphi}
\def\g{\gamma}
\def\h{\eta}
\def\i{\iota}
\def\j{\psi}
\def\k{\kappa}                    
\def\l{\lambda}
\def\m{\mu}
\def\n{\nu}
\def\o{\omega}  \def\w{\omega}
\def\q{\theta}  \def\th{\theta}                  
\def\r{\rho}                                     
\def\s{\sigma}                                   
\def\t{\tau}
\def\u{\upsilon}
\def\x{\xi}
\def\z{\zeta}
\def\pt{\tilde{\varphi}}
\def\tt{\tilde{\theta}}
\def\lab{\label}  
\def\6{\partial}
\def\wg{\wedge}
\def\atanh{{\rm arctanh}}
\def\bpsi{\bar{\psi}}
\def\bt{\bar{\theta}}
\def\bvf{\bar{\varphi}}

%
                                                                                                    
\newfont{\namefont}{cmr10}
\newfont{\addfont}{cmti7 scaled 1440}
\newfont{\boldmathfont}{cmbx10}
\newfont{\headfontb}{cmbx10 scaled 1728}
\renewcommand{\theequation}{{\rm\thesection.\arabic{equation}}}
\begin{titlepage}

\begin{center} \Large \bf Towards the String Dual of ${\cal 
N}=1$ SQCD-like Theories

\end{center}

\vskip 0.3truein
\begin{center}
Roberto Casero${}^{*}$\footnote{roberto.casero@cpht.polytechnique.fr}, 
Carlos 
N\'u\~nez${}^{\dagger}$\footnote{c.nunez@swansea.ac.uk} and Angel 
Paredes${}^{*}$\footnote{angel.paredes@cpht.polytechnique.fr}
\vspace{0.7in}\\
${}^{*}$ \it{Centre de Physique Th\'eorique\\ \'Ecole Polytechnique\\
and UMR du CNRS 7644\\
91128 Palaiseau, France}
\vspace{0.3in}
\vskip 0.2truein
${}^{\dagger}$ \it{Department of Physics\\ University of Swansea, Singleton 
Park\\
Swansea SA2 8PP\\ United Kingdom.}
\vspace{0.3in}
\end{center}
\vskip 1cm
\centerline{{\bf Abstract}}
We construct supergravity plus branes solutions, which we argue to be 
related to 4d 
$\mathcal{N}=1$ SQCD with a quartic superpotential. 
The geometries depend on the ratio 
$N_f / N_c$ which can be kept of order one, 
present a good singularity at the origin and are weakly curved 
elsewhere. We support our field theory interpretation by studying 
a variety of features 
like $R$-symmetry breaking, instantons, Seiberg duality, Wilson loops and 
pair creation, 
running of couplings and domain walls. \\
In a second part of this paper, we address a different 
problem: the analysis of the interesting physics of 
different members of a family of 
supergravity solutions dual to (unflavored) $\mathcal{N}=1$ SYM plus some 
UV completion.

\vskip1truecm
\vspace{0.1in}
\leftline{CPHT-RR 010.0106}
\leftline{SWAT/06/454}
\leftline{hep-th/0602027 }
\smallskip
\end{titlepage}
\setcounter{footnote}{0}
\tableofcontents

\section{Introduction}

The AdS/CFT conjecture originally proposed by Maldacena \cite{Maldacena:1997re}
and refined in \cite{Gubser:1998bc, Witten:1998qj} is one of the most 
powerful analytic tools for studying strong coupling effects in gauge 
theories. There are many examples that go beyond the initially 
conjectured duality and first steps in generalizing it 
to non-conformal models 
were taken in \cite{Itzhaki:1998dd}. 
Later, very interesting developments led to the construction of the gauge-string 
duality in phenomenologically more relevant 
theories \ie\, minimally or non-supersymmetric gauge theories 
\cite{Girardello:1999bd}. 

Conceptually, the clearer setup for duals to less symmetric 
theories is obtained by breaking  conformality and (partially) supersymmetry 
by deforming ${\cal N}=4$ SYM with relevant operators or VEV's.  
The models put forward in \cite{Girardello:1999bd} 
and (even though there are some 
important technical differences) the  
Klebanov-Tseytlin and
Klebanov-Strassler model(s) 
\cite{Klebanov:2000hb} 
belong 
to this class. 

On the other hand, a different set of models, that are less 
conventional regarding the UV completion of the field theory have been 
developed. The idea here is to start from a set of Dp-branes 
(usually with $p>3$), that 
wrap a $q$-dimensional compact manifold in a 
way such that two conditions are 
satisfied: first, one imposes that the low energy description of the 
system is ($p-q$)-dimensional, that is, the size of 
the $q$-manifold is small and is 
not observable at low energies. Secondly, one also requires that a minimal amount of SUSY 
is preserved, in order to have technical control over the theory 
  (for example, the resolution of the 
 Einstein eqs. is eased).  
According to  intuition, in this second class of phenomenologically 
interesting dualities, the UV completion of the 
(usually four-dimensional) field 
theory of interest is a higher dimensional field (or string) theory.
There are several 
models that belong to this latter class. In this paper we 
will concentrate on the model dual to $\mathcal{N}=1$ SYM 
\cite{Maldacena:2000yy}, that builds on a 
geometry originally found in 4-d gauged supergravity in 
\cite{Chamseddine:1997nm}.
It must be noted that all of the models in this category (and also those 
in the class 
described in the previous paragraph \cite{Girardello:1999bd, Klebanov:2000hb}) are 
afflicted  
by the fact that they are not dual to the ``pure'' field theory of 
interest, but instead, the field theory degrees of 
freedom are entangled with the KK modes (on the 
$q$-manifold  or with the modes in the latest steps in the cascade) in a 
way that depends on the energy scale of the field theory.  
The KK modes enter the theory at an energy scale which is inversely 
proportional to the size of the $q$-manifold and the main problem 
is that this size is comparable to the 
scale at which  one wants to study non-perturbative phenomena such as 
confinement, spontaneous breaking of chiral symmetry, etc. 
Nevertheless, this limitation can be seen as an artifact of the supergravity 
approximation and will hopefully be avoided once the formulation 
of the string sigma model on these RR backgrounds becomes available.
Many articles have studied different 
aspects of these models.  Instead of revisiting the main results here, we 
refer the interested reader to the review articles~\cite{Bertolini:2003iv}. 

An important characteristic of the different models mentioned above, is 
that these 
supergravity backgrounds are conjectured to be dual to field theories 
with adjoint 
matter (also some quiver gauge theories are in this class).
A problem that remains basically unsolved is how to deal with field theories
containing matter fields transforming in the fundamental representation. 
It is a very natural problem to tackle if one is interested in making contact with 
phenomenological theories, like QCD. 

There were some attempts to study this problem. To begin with, the ``quenched" 
approximation (when the number of flavors is negligible compared to the number of 
colors, hence using a probe brane approximation in the dual picture) was studied in 
detail and in the different models, leading to 
a set of interesting results, regarding breaking of symmetries, meson spectrum and 
interactions. This line was developed in detail; for a list of references, 
see the citations to the paper that 
initiated this approach \cite{Karch:2002sh}.

The need to go beyond this ``quenched" approximation 
motivated some authors to search for more complicated backgrounds including the 
effects of a large number of flavors 
\cite{Cherkis:2002ir}. These new backgrounds, that typically depend on two 
coordinates, are afflicted (in the case of four-dimensional field theories), by 
the usual problems caused by the addition of D7-branes: 
a codimension-two
object 
with an associated conical singularity, 
a maximum number of D7-branes (or $O_7$ planes) that can be 
added in a compact space, etc.
There is another line of research that used non-critical 
string theory \cite{KM, othernoncrit} to 
construct duals to interesting field theories. 
The use of non-critical strings cures the problem of 
the existence of extra massive states (the KK modes pointed out above), 
but it typically relies on  backgrounds which have 
large curvature, with the consequent problem that solutions need to be 
stringy-corrected 
(see \cite{Babington:2005dr} for some work in this direction). 
The results obtained  
are nevertheless indicating that this approach has some potential.

Perhaps being conservative, one could say that the results 
regarding the duality 
with flavored field theories using the backgrounds 
of \cite{Cherkis:2002ir, KM}, are not 
as spectacular and clear as the ones 
obtained with duals to field theories with 
adjoint fields only.
All this indicates that a different approach is necessary, which might try to  
combine the 
successes of the previous ones 
and to avoid their inconveniences as much as possible.

In this paper we propose a way of finding a dual to 
$\mathcal{N}=1$ SQCD-like theories using 
critical strings, focusing our attention here on type IIB string theory. 
The basic idea 
will be to add flavors to the background in 
\cite{Maldacena:2000yy}, by using different technical points developed in 
\cite{Nunez:2003cf} and in non-critical string approaches \cite{KM, Bigazzi:2005md}. Below, we 
describe our approach and give a guide to read this paper.

A second part of the paper, namely section \ref{unflavoredcase},
 deals with  backgrounds dual to different UV 
completions of minimally SUSY Super-Yang-Mills.
For these geometries, unlike in the model \cite{Maldacena:2000yy}, the
dilaton does not diverge in the UV.
 We study relevant aspects 
of the dual gauge theory.
\subsection{General Idea}
Let us describe here the procedure we will adopt to add a large number of 
flavors $N_f$ to a given  supergravity background, more particularly to 
the one constructed with wrapped D5-branes and conjectured to 
be dual to $\mathcal{N}=1$ SYM 
plus massive adjoint matter \cite{Maldacena:2000yy}. Even 
though we will concentrate on this particular case, the 
procedure described below could be used to add many flavors in different 
backgrounds dual to $\mathcal{N}=1$, $\mathcal{N}=2$ SYM in $2+1$ and $ 3+1$
dimensions, etc.

We want to preserve SUSY in order to solve BPS eqs rather than Einstein 
eqs; 
since the original background preserves four supercharges, we 
cannot break more SUSY in duals 
to four-dimensional field theories. It is then  
necessary to find a non-compact and holomorphic two-cycle ($\Sigma_2$), 
where we can place $N_f$ D5-branes that share the $3+1$ gauge theory 
dimensions with those $N_c$ D5-branes that generated the 
original background \cite{Maldacena:2000yy}. These surfaces must be 
holomorphic to preserve SUSY, and non-compact so that the symmetry added 
by the flavor branes is global (in other words, the coupling of the 
effective four-dimensional field theory on the flavor branes $g_4^2= 
g_6^2/Vol(\Sigma_2)$, vanishes) \cite{Cachazo:2001jy}.

The problem of finding non-compact SUSY preserving two-cycles in the 
geometry of \cite{Maldacena:2000yy} was solved in \cite{Nunez:2003cf}. We 
will use a particular 
solution found there in order to place {\it many} D5-branes. 
This new stack of $N_f$ branes is heavy, so it will backreact 
and the original background  will be modified. We 
can think of an action describing the dynamics of the backreacted system 
that reads \footnote{It is known that it is not possible to write a 
polynomial action for IIB supergravity due to the self-duality 
condition, we nevertheless will deal with solutions that have $F_5=0$, 
so, what we have in mind in this case is an action of the form $S_{IIB}= \int 
d^{10}x \sqrt{g} \Big(R-\frac{1}{2}(\partial\phi)^2 
-e^{-\phi}\frac{F_3^2}{12}   \Big)$.} 
\beq 
S= S_{IIB} + S_{flavor} 
\label{actionschemat} 
\eeq
Basically, we have added the open string sector in $S_{flavor}$. The 
procedure might be thought of as consisting of two steps; first we take the 
background 
\cite{Maldacena:2000yy}, where the $N_c$ color branes have been 
replaced by a flux, that we represent by $S_{IIB} $ in 
(\ref{actionschemat}). On top of this, we add a  bunch of $N_f$ flavor 
branes and 
find a new background solving new BPS (and Einstein) eqs. that encode the 
presence of the  flavor branes. The  
part of the action (\ref{actionschemat}) corresponding to the flavors 
will be the Dirac+WZ action of 
the $N_f $ five-branes. In Einstein frame it reads
\beq
S_{flavor}=T_5 \sum^{N_f} \left(
-\int_{{\cal M}_6} d^6x e^{\frac{\phi}{2}}\sqrt {-\hat g_{(6)}}
+ \int_{{\cal M}_6} P[C_{6}] \right) \; ,
\label{accionflavor}
\eeq
where the integrals are taken over the six-dimensional
worldvolume of the flavor
branes ${\cal M}_6$
and $\hat g_{(6)}$ stands for the determinant of the pull-back
of the metric on such worldvolume. We will take these D5's
extended along six coordinates that we call  $x_0, x_1, x_2,x_3, r, \psi$ 
at constant $\theta, \varphi, \tilde \theta, \tilde \varphi$. 
These supersymmetric embeddings were called {\it cylinder solutions} in
\cite{Nunez:2003cf}.

The action $S_{IIB}$ is ten-dimensional in contrast with $S_{flavor}$, that is six-dimensional since 
the ``flavor branes'' are localized in the directions $\theta, 
\tilde{\theta}, \varphi, \tilde{\varphi}$. So, trying to find
solutions to the equations of motion derived from (\ref{actionschemat}), will 
involve writing Einstein eqs with Dirac delta functions on the 
coordinates where the branes are localized. Thus, the solution will 
depend on ($\theta,
\tilde{\theta}, \varphi, \tilde{\varphi}$) and in this respect, this 
solution would be similar to those found in \cite{Cherkis:2002ir}. This is 
a very hard problem in principle and it is nice to notice that the 
(six-dimensional) non-critical string approach of \cite{KM} to ${\cal N}=1$
 flavored solutions is not afflicted by this technical difficulty, 
since $S_{flavor}$ is basically a cosmological term \footnote{But as 
we mentioned, the non-critical string approach is seriously afflicted by 
string corrections to the gravity approximation; nevertheless, the $AdS$ 
solutions in \cite {KM} are likely to persist after these corrections.}.  
We will circumvent this difficult 
technical obstacle 
by following a procedure first proposed in 
\cite{Bigazzi:2005md} in the context of flavor branes. In 
this paper, an eight-dimensional non-critical string action was 
used and an  ingenious trick developed. Indeed, it was found in 
\cite{Bigazzi:2005md} that one could 
solve Einstein eqs without Dirac delta functions if one
{\it smears} the very many $N_f$ branes in the extra directions (in our 
case $\theta,\tilde{\theta}, \varphi, \tilde{\varphi}$).

Since $N_f$ and $N_c$ are very large numbers of the same order, 
interchanging the sum in (\ref{accionflavor}) by an integral over the 
extra directions will produce a fully ten-dimensional action, erasing the 
explicit dependence on the extra coordinates. This gives rise to  some 
new global symmetries. We will comment on the interpretation 
of this procedure from the dual field theory perspective in section 
\ref{gen asp gauge}.

We proceed in the rest of the paper by first describing in detail the 
background 
in \cite{Maldacena:2000yy} and a singular 
version of it (that is easier to deal with and illustrates the technical 
details). We 
will then propose a new deformed 
background, where the fingerprint of the flavor branes will be the 
explicit violation of the Bianchi identity (a sort of Dirac-string like singularity). We will find 
BPS eqs describing this situation and in appendix \ref{appendixsup} will 
develop a superpotential approach and compare with a purely IIB
SUSY approach to this problem. The outcome being that the BPS eqs 
obtained from the superpotential 
coincide with those coming from imposing vanishing of the variations of 
the gravitino and dilatino, $\delta \psi_\mu, 
\delta \lambda$, in ten dimensions.

Then, we apply a similar approach to the non-singular case, obtain a 
set of BPS eqs, find some interesting asymptotic  solutions with 
numerical interpolation and study their (strongly-coupled) dual gauge theory 
predictions.

The content of the paper is the following: in section 2, 
we present the dual to $\mathcal{N}=1$ SYM plus adjoint matter that will be the 
arena on which we will construct flavored solutions. The presentation is 
detailed enough to make the paper self-contained and the reader familiar 
with these results may skip it. In 
section~\ref{singularsolution}, we will provide a derivation of the 
flavored-BPS 
equations for the singular background, 
that solve the Einstein eqs derived from (\ref{actionschemat}). This 
derivation is 
complemented in appendix \ref{appendixsup}, using a superpotential 
approach. The presentation 
in this section will be very detailed, and it was written in order to describe and explain in a simpler context the procedure we will follow in the physically interesting non-abelian case.

In section \ref{flavnonsing}
we will write BPS equations for the flavoring of the non-singular 
background and, as an 
interesting particular case, we will also deal with the case 
$N_f=2 N_c$ that presents unique features. Then, we study in detail 
the asymptotics of the 
solutions and provide a careful numerical treatment to the eqs derived in 
this section. In 
section \ref{gaugethnonsing} we first provide arguments to describe the 
field theory dual to our backgrounds, that indicate that (at low energies) 
we are dealing with $\mathcal{N}=1$ SQCD plus a quartic superpotential 
for the quark superfield. 
Then, we will 
initiate the study of the field theory properties of these flavored solutions, 
most notably, we will analyze Wilson loops and SQCD-string breaking, 
instanton action, theta angle, beta function, $U(1)_R$ symmetry breaking, 
domain walls and  Seiberg duality.

In section \ref{nf2ncint} we will come back to the interesting case $N_f= 
2N_c$. Here, we will analyze distinctive gauge theory features; most 
notably (non)-confinement (screening of quarks), $U(1)_R$ symmetry 
preservation and Seiberg duality, that becomes particularly interesting in 
this case. We will also comment on finite temperature aspects of this 
field theory.

In section \ref{sec: fluxes}, a different approach to flavored 
backgrounds will be introduced. We argue that it might be possible 
to account for the violation of the Bianchi identity induced by the 
addition of the flavor branes, by turning on non-trivial fluxes which 
are not present in the unflavored solution of \cite{Maldacena:2000yy}.  

In the second part of the paper (as a byproduct of our results above), in 
section \ref{unflavoredcase}  we will consider the particular case 
$N_f=0$ (which in the following  we will call 
``unflavored case'')
and find a one parameter family  of deformed (non-singular) solutions that 
correspond to different UV completions of $\mathcal{N}=1$ SYM. We present 
a gauge theory interpretation of this  family of solutions and study 
many field theory aspects as 
seen from it; most notably, confinement, k-string 
tensions, PP-waves, rotating strings, and beta function, pointing out in each case 
the similarities and differences between different members of the family 
and the solution in \cite{Maldacena:2000yy}.

Section \ref{concl} is left for conclusions and
 possible future directions. With the aim of making 
 the main text more readable,
we wrote many appendixes, where we have relegated lots of technical 
details. 

\paragraph{Reader's guide}

Given that this is a considerably long paper, but that different
parts may be read almost independently, we feel it is useful to write a
guide in order to help the reader find the results of her/his interest.
Readers interested in the construction of the flavored backreacted
solutions can concentrate on sections \ref{singularsolution} and
\ref{flavnonsing} supplemented with appendices
\ref{appendixsup} and \ref{appendixa} and also read section 
\ref{sec: fluxes}. Readers interested in the field theory features
reproduced by the flavored solutions can go directly to sections
\ref{gaugethnonsing} and \ref{nf2ncint} and just look at the 
explicit expressions for the backgrounds 
(sections \ref{flavnonab}-\ref{nf2nc}) when needed.
Finally, who is interested in the deformed unflavored solutions
can read, in a self-contained way, sections \ref{sect: unflavnonab}
and \ref{unflavoredcase}, supplemented with appendices
\ref{appendixa} and \ref{appd}.

\section{The Dual to $\mathcal{N}=1$ SYM}
\label{dualn=1sym}
\setcounter{equation}{0}

We work with the model 
presented in \cite{Maldacena:2000yy}
(the solution was first found in a 
4d context in \cite{Chamseddine:1997nm}). 
Let us briefly describe the 
main points of this supergravity dual to $\mathcal{N}=1$ SYM and its UV completion.
We start with $N_c$ D5-branes. 
The field theory that lives on them is 
6D SYM with 16 supercharges. Then, 
suppose that we wrap two directions of the 
D5-branes on a curved two-manifold 
that can be chosen to be a sphere.
In order to preserve some fraction of SUSY, a twisting procedure 
has to be implemented \cite{Witten:1994ev}
and actually, there are two ways of doing it. 
The one we will be interested in in this 
paper deals with a twisting that preserves 
four supercharges. In this case 
the two-cycle mentioned above lives inside a Calabi-Yau 3-fold. 
The corresponding supergravity solution can be argued to be dual to a 
four-dimensional field theory only for low energies (small values of the 
radial coordinate). 
Indeed, at high energies, the modes of the gauge theory 
explore also the two-cycle and as the energy is
increased further, the theory first becomes  six-
dimensional and then, blowing-up of the dilaton forces one to
S-dualize. Therefore the UV completion of the model is given by 
a little string 
theory. 

The supergravity solution that interests us, 
preserves four supercharges and has the topology 
$\mathbb{R}^{1,3}\times \mathbb{R}\times S^2\times S^3$. There is a fibration 
of the two spheres in such a way that  ${\cal N}=1$ supersymmetry
is preserved. By going near $r=0$ it can be seen that the 
topology is $\mathbb{R}^{1,6} \times S^3$. The full solution and Killing 
spinors are written in detail in \cite{Nunez:2003cf}. Let us write some 
aspects for future reference and to set conventions.
The metric in the Einstein frame reads,
\beq
ds^2_{10}\,=\,\alpha' g_s N_c e^{{\phi\over 2}}\,\,\Big[\,
\frac{1}{\alpha' g_s N_c }dx^2_{1,3}\,+\,
dr^2+\,e^{2h}\,\big(\,d\theta^2+\sin^2\theta 
d\varphi^2\,\big)\,\,+\,{1\over 4}\,(\tilde{\omega}_i-A^i)^2\,\Big]\,\,,
\label{metricaa}
\eeq
where $\phi$ is the dilaton. The angles
$\theta\in [0,\pi]$ and
$\varphi\in [0,2\pi)$ parametrize a two-sphere. 
This sphere is fibered in the ten-dimensional metric by the one-forms
$A^i$ $(i=1,2,3)$.
They are given in terms of a 
function
$a(r)$ and the angles $(\theta,\varphi)$ as follows:
\beq
A^1\,=\,-a(r) d\theta\,,
\,\,\,\,\,\,\,\,\,
A^2\,=\,a(r) \sin\theta d\varphi\,,
\,\,\,\,\,\,\,\,\,
A^3\,=\,- \cos\theta d\varphi\,.
\label{oneform}
\eeq
The $\tilde{\omega}_i$ one-forms are defined as
\bea\lab{su2}
\tilde{\omega}_1&=& \cos\psi d\tilde\theta\,+\,\sin\psi\sin\tilde\theta
d\tilde\varphi\,\,,\rc
\tilde{\omega}_2&=&-\sin\psi d\tilde\theta\,+\,\cos\psi\sin\tilde\theta
d\tilde\varphi\,\,,\rc
\tilde{\omega}_3&=&d\psi\,+\,\cos\tilde\theta d\tilde\varphi\,\,.
\eea
The geometry in (\ref{metricaa}) preserves four supercharges and is 
non-singular  
when the functions $a(r)$, $h(r)$ and the dilaton $\phi(r)$ are:
\bea
a(r)&=&{2r\over \sinh 2r}\,\,,\rc
e^{2h}&=&r\coth 2r\,-\,{r^2\over \sinh^2 2r}\,-\,
{1\over 4}\,\,,\rc
e^{-2\phi}&=&e^{-2\phi_0}{2e^h\over \sinh 2r}\,\,,
\label{MNsol}
\eea
where $\phi_0$ is the value of the dilaton at $r=0$. 
The solution of type IIB supergravity includes a
RR three-form $F_{(3)}$ that is given by
\beq
\frac{1}{g_s \alpha' N_c} F_{(3)}\,=\,-{1\over 
4}\,\big(\,\tilde{\omega}_1-A^1\,\big)\wedge
\big(\,\tilde{\omega}_2-A^2\,\big)\wedge \big(\,\tilde{\omega}_3-A^3\,\big)\,+\,{1\over 4}\,\,
\sum_a\,F^a\wedge \big(\,\tilde{\omega}_a-A^a\,\big)\,\,,
\label{RRthreeform}
\eeq
where $F^a$ is the field strength of the $SU(2)$ gauge field $A^a$, defined as:
\beq
F^a\,=\,dA^a\,+\,{1\over 2}\epsilon_{abc}\,A^b\wedge A^c\,\,.
\label{fieldstrenght}
\eeq
The different components of $F^a$ read                                                                               
\beq
F^1\,=\,-a'\,dr\wedge d\theta\,\,,
\,\,\,\,\,\,\,\,\,\,
F^2\,=\,a'\sin\theta dr\wedge d\varphi\,\,,
\,\,\,\,\,\,\,\,\,\,
F^3\,=\,(\,1-a^2\,)\,\sin\theta d\theta\wedge d\varphi\,\,,
\eeq
where the prime denotes derivative with respect to $r$.
Since $dF_{(3)}=0$, one can 
represent $F_{(3)}$ in terms of a two-form potential
$C_{(2)}$ as $F_{(3)}\,=\,dC_{(2)}$. 
Actually, it is not difficult to verify that
$C_{(2)}$ can be taken as:
\bea
\frac{C_{(2)}}{g_s \a'N_c}&=&{1\over 4}\,\Big[\,\psi\,(\,\sin\theta 
d\theta\wedge d\varphi\,-\,
\sin\tilde\theta d\tilde\theta\wedge d\tilde\varphi\,)
\,-\,\cos\theta\cos\tilde\theta d\varphi\wedge d\tilde\varphi\,-\rc
&&-a\,(\,d\theta\wedge \tilde{\omega}_1\,-\,\sin\theta d\varphi\wedge \tilde{\omega}_2\,)\,\Big]\,\,.
\label{RR}
\eea
The equation of motion of $F_{(3)}$ in the Einstein frame is
$d\Big(\,e^{\phi}\,{}^*F_{(3)}\,\Big)=0$, where $*$ denotes Hodge duality. 
Let us  stress that the configuration presented above is non-singular.

For future reference, let us quote here the asymptotic  expansions of the 
functions $a(r)$, $e^{2h(r)}$, $e^{2 \phi(r)}$ near $r=0$,
\beq
a(r)\sim 1 - \frac{2}{3}r^2 +..., \;\; \qquad  e^{2h}\sim r^2 +... , \;\; \qquad
e^{2\phi}\sim e^{2\phi_0}(1 + ...)
\label{expansionnearr=0}
\eeq
and for large values of the radial coordinate,
\beq
a(r)\sim 4 r e^{-2r}+..., \;\; \qquad e^{2h}\sim r +... , \;\;\qquad
e^{2\phi}\sim e^{2\phi_0}\frac{e^{2r}}{4\sqrt{r}}
\label{expansionnearr=infinity}
\eeq
The BPS eqs that the configuration in 
(\ref{metricaa})-(\ref{RR}) solves  (for a complete derivation
from the spinor transformations of IIB sugra, see 
appendix A in \cite{Nunez:2003cf}) read 
\bear
\phi'&=&{1\over Q}\,
\Big[\,e^{2h}\,-\,{e^{- 2h}\over 16}\,(a^2-1)^2\,\Big]\,\,,\rc
h'&=&{1\over 2Q}\,
\Big[\,a^2+1+{e^{- 2h}\over 4}\,(a^2-1)^2\,\Big]\,\,,\rc
a'&=&-{2a\over Q}\,
\Big[\,e^{ 2h}+{1\over 4}\,(a^2-1)\,\Big]\,\,,
\label{differentialeqs}
\eear
with 
\beq
Q\,\equiv\,\sqrt{e^{4h}\,+\,{1\over 2}\,e^{2h}\,(a^2+1)\,+\,{1\over 16}\,
(a^2-1)^2}\,\,.
\label{defQ}
\eeq
The Einstein eqs read 
\beq
R_{\mu\nu}-\frac12 g_{\mu\nu} R= \frac12
\left(\partial_\mu \phi \partial_\nu \phi -\frac12 g_{\mu\nu} 
\partial_\lambda \phi \partial^\lambda \phi \right) 
+\frac{1}{12}e^\phi
\left(3 F_{\mu\lambda_2\lambda_3}F_\nu^{\,\lambda_2\lambda_3}
-\frac12 g_{\mu\nu} F_{(3)}^2\right) 
\label{Einseqmn}
\eeq
The Maxwell equation as quoted 
above reads $d\Big(\,e^{\phi}\,{}^*F_{(3)}\,\Big)=0$ and the Bianchi 
identity is $d F_3 =0$. Both of them are solved by 
(\ref{metricaa})-(\ref{RR}).

Finally, in the next section, as a warm up example, we will add many 
flavor branes to a particular singular solution of the system 
(\ref{differentialeqs}), which is characterized by $a(r)=0,\;\; e^{2h}=r,\;\; 
e^{2\phi-2\phi_0}=\frac{e^{2r}}{4\sqrt{r}} $,
\beq
ds^2_{10}\,=\,\alpha' g_s N_c e^{{\phi\over 2}}\,\,\Big[\,
\frac{1}{\alpha' g_s N_c }dx^2_{1,3}\,+\,
dr^2\,+\,e^{2h}\,\big(\,d\theta^2+\sin^2\theta 
d\varphi^2\,\big)\,+\, \frac{1}{4} (\tilde{\omega}_1^2 +\tilde{\omega}_2^2)+{1\over 
4}\,(\tilde{\omega}^3 + \cos\theta d\varphi)^2\,\Big]\,\,,
\label{metricaasing}
\eeq
and a RR potential
\bea
\frac{C_{(2)}}{g_s \a'N_c}&=&{1\over 4}\,\Big[\,\psi\,(\,\sin\theta 
d\theta\wedge d\varphi\,-\,
\sin\tilde\theta d\tilde\theta\wedge d\tilde\varphi\,)
\,-\,\cos\theta\cos\tilde\theta d\varphi\wedge d\tilde\varphi\Big]\,\,.
\label{RRsing}
\eea
Notice that at $r=0$ this background presents a (bad) singularity that is 
solved by the turning-on of the function $a(r)$ in the full solution  
(\ref{metricaa})-(\ref{RR}). On the field theory side, this way of 
resolving the singularity boils into the phenomena of confinement 
and breaking of the R-symmetry. Since they will be useful in the remainder 
of this paper, let us summarize some aspects of the dual field theory.
\subsection{Dual Field Theory}\label{sec dft}
In \cite{Maldacena:2000yy}
the solution presented in (\ref{metricaa})-(\ref{RR}) was argued to be 
dual 
to ${\cal N}=1$ SYM plus some KK massive
adjoint matter.
The 4D field theory is obtained by reduction of $N_c$ D5-branes on
$S^2$ with a twist that we explain below. Therefore, as the energy
scale of the 4D field theory becomes comparable to the inverse volume
of $S^2$, the KK modes begin to enter the spectrum. 

To analyze the spectrum in more detail, we briefly review the twisting
procedure. In order to have a supersymmetric theory on a curved
manifold like the $S^2$ here, one needs globally defined spinors.
In our case the argument goes as follows. As D5-branes wrap the two-sphere, the Lorentz symmetry along the branes decomposes as $SO(1,3)\times 
SO(2)$. 
There is also an $SU(2)_L\times SU(2)_R$ symmetry  
that rotates the transverse coordinates and  
corresponds to the R-symmetry of the supercharges of the field theory
on the D5-branes. One can properly define ${\cal N}=1$ supersymmetry
on the curved space that is obtained by wrapping the D5-branes on the
two-cycle, by identifying a $U(1)$ subgroup of either
$SU(2)_L$ or $SU(2)_R$ R-symmetry with the $SO(2)$ of the
two-sphere.
To fix notations, let us choose the $U(1)$ in $SU(2)_L$. Having 
done the
identification with $SO(2)$ of the sphere, we denote this twisted
$U(1)$ as $U(1)_T$. 
 
After this twisting procedure is performed, the fields in the theory
are labeled by the quantum numbers of $SO(1,3)\times U(1)_T\times
SU(2)_R$. The bosonic fields are (the $a$ indicates an adjoint index)
\beq
A^{a}_{\mu}= (4, 0, 1), \;\; \Phi^a=(1, \pm, 1), \;\; 
\xi^{a}=  (1, \pm, 2). \;\;
\label{twistboson}
\eeq
Respectively they are the gluon, two 
massive scalars that are coming from the reduction of the original 
6D gauge field on $S^2$ 
(explicitly from the $A_{\vf}$ and $A_{\q}$ components)   
and finally four other massive scalars 
(that originally represented the positions of the D5-branes in the
transverse $\mathbb{R}^4$). As a general rule, all the fields that transform
under $U(1)_T$, the second entry in the above charge 
designation, are massive. For the fermions one has, 
\beq
\lambda^a=(2,0,1),\,(\bar{2},0,1),\;\; 
\Psi^a=(2,++,1),\,(\bar{2},--,1),\;\; 
\psi^a= (2,+,2), \, (\bar{2},-,2).
\label{twistfermion}
\eeq
These fields are the gluino plus some massive fermions 
whose $U(1)_T$ quantum number is 
non-zero. The KK modes in the 4D theory are obtained by the harmonic
decomposition of the massive modes, $\Phi,\xi,\Psi$ and $\psi$ that
are shown above. Their mass is of the order of 
$M_{KK}^2= (Vol_{S^2})^{-1}\propto \frac{1}{g_s \alpha' N_c} $. A very 
important point to notice here is that these KK modes are 
charged under $U(1)_T \times U(1)_R$ where the second 
$U(1)$ is a subgroup of the $SU(2)_R$ that is left 
untouched in the twisting procedure. On the other hand, the gluonic gauge
field and the gluino are not charged under 
either of the
$U(1)$'s.     

The dynamics of these KK modes mixes with the dynamics of 
confinement in this model because the strong coupling scale of 
the theory is of the order of the KK mass. One way to evade the mixing 
problem would be to work instead with the full string solution, namely
the world-sheet sigma model on this background (or in the 
S-dual NS5 background) which would give us control over the duality 
to all orders in $\alpha'$, hence we would be able to decouple the
dynamics of KK-modes from the gauge dynamics. This direction is
unfortunately not (yet) available. Meanwhile, in \cite{Gursoy:2005cn} a 
procedure was 
developed to determine when a field theory observable computed from the 
supergravity solution is affected by the presence of the massive KK modes  
or is purely an effect of the massless fields $A_\mu, \lambda$.

In order to get a better intuition of the dynamics, one can schematically 
write a 
lagrangian for these fields as follows:
\beq
L= -Tr[\frac{1}{4} F_{\mu\nu}^2 + i\lambda D\lambda- (D_\mu\Phi_i)^2 
- (D_\mu\xi_k)^2 + \Psi 
(i D - M_{KK})\Psi + 
M_{KK}^2 (\xi_k^2 + \Phi_i^2) + V[\xi,\Phi, \Psi]]
\label{lagrangiantwisted}
\eeq
The potential typically contains the scalar potential for the bosons,
Yukawa type interactions and more. 
This expression is schematic because of (at least) two reasons. First
of all, the potential presumably contains very
complicated interactions involving the KK and massless fields. 
Secondly, there is
mixing between the infinite tower of spherical harmonics that are obtained
by reduction on $S^2$ and $S^3$, see 
\cite{Andrews:2005cv, Andrews:2006aw} for a careful treatment.      

\section{Adding Flavor Branes to the Singular Solution}
\label{singularsolution}
\setcounter{equation}{0}
In this section we will add flavors 
to the particular solution
(\ref{metricaasing})-(\ref{RRsing}). Even though there is little physical 
significance for this solution, the point of this section is to illustrate 
in detail the type of formalism we use and its remarkable 
consistency. Readers more interested in  physically more relevant 
aspects 
of this paper should perhaps skip this section in a first reading, but 
since the formalism and technical subtleties are quite involved, we 
decided to spell them out explicitly here in a simpler context. 
\subsection{Deforming the Space}
The first step will be to deform the background 
(\ref{metricaasing})-(\ref{RRsing}). First, we will study the deformation 
of the backgrounds without the addition of flavors and then we will treat 
a  deformation due to the 
presence of the $N_f$ flavor branes. For this we propose
a set of vielbeins given by,
\ba
& & e^{xi}= e^f dx_i,\;\; e^r= e^f dr, \;\; e^\theta = e^{f+h} d\theta,\;\; e^\varphi= e^{f+h }\sin\theta 
d\varphi,\;\;\nonumber\\
& &  e^1= \frac{e^{f+ g} \tilde{\omega}_1}{2},\;\; e^2= \frac{e^{f+ g} \tilde{\omega}_2}{2}, \;\; e^3=  
\frac{e^{f+ k} 
(\tilde{\omega}_3 + \cos\theta d\varphi)}{2} .  
\label{vielbein}
\ea
that leads to the metric (Einstein frame):
\beq
ds^2 =  e^{2 f(r)} [dx_{1,3}^2 + dr^2 + e^{2 h(r)} 
(d\theta^2 + \sin^2\theta d\varphi^2) 
+\frac{e^{2 g(r)}}{4} (\tilde{\omega}_1^2 + \tilde{\omega}_2^2) + \frac{e^{2 k(r)}}{4} 
(\tilde{\omega}_3 + \cos\theta d\varphi)^2]  \,\,.
\label{metric}
\eeq
Notice that compared to (\ref{metricaasing}), we took $\alpha' g_s =1$, 
while $N_c$ has been absorbed in $e^{2h}$, $e^{2g}$, $e^{2k}$.
In our configuration, we will also have a dilaton and a RR three-form
\beq
\phi(r), \;\;\; F_{(3)}= - 
2 N_c e^{-3 f - 2g -k} e^1 \wedge e^2\wedge e^3 
+\frac{N_c}{2} e^{-3f -2h -k} 
e^\theta \wedge e^\varphi \wedge e^3, 
\label{3form}
\eeq
where $N_c$, the number of color D5-branes,
 comes from the quantization condition:
\beq
\frac{1}{2\kappa_{(10)}}\int_{S^3} F_{(3)} = N_c T_5\,\,.
\label{quantcond}
\eeq
The $S^3$ on which we integrate is parameterized by $\tilde\theta,
\tilde\varphi,\psi$, so only the first term in (\ref{3form}) contributes.

We plug this configuration  into the IIB SUSY transformations
(see eq (\ref{IIBproja})).
The projections for the Killing  spinor are:
\beq
\Gamma_{r123}\epsilon= \epsilon, \;\;\qquad \Gamma_{r\theta\varphi 
3}\epsilon= \epsilon , \;\;\qquad\epsilon= 
i\epsilon^*\,\,.
\eeq
(The sign of the two first expressions can be changed, but we fix
it in order to match the full non-abelian case).
After some algebra, the following eqs. arise,
\ba
4f&=&\phi \\
h'&=& \frac{1}{4} N_c e^{-2 h -k} +\frac14 e^{-2 h +k}
\label{bps1} \\
g' &=& -N_c e^{-2g-k} + e^{-2g+k}
\label{bps2} \\
k' &=&\frac{1}{4} N_c e^{-2h -k} 
-N_c e^{-2g -k} 
-\frac14 e^{-2h +k} - e^{-2g +k} +2 e^{-k}
\label{bps3}  \\
\phi'&=&-\frac{1}{4} N_c  e^{-2 h-k} + N_c  e^{-2 g-k} 
\label{bps4}
\ea

\subsection{Flavoring the Space}
\label{secabflavor}
We want to backreact the space with D5-flavor branes. Let us try
to do this following the procedure introduced in \cite{KM}, 
{\it i.e} adding an open string sector to the 
gravity action. It is important to remark that the construction of
\cite{KM} involved non-critical strings and thus curvatures of
order of the string scale. This fact hindered the possibility of
finding reliable quantitative results from a supergravity approximation.
On the other hand, we are using ten-dimensional string theory and
weak curvature can be obtained by taking $g_s N_c \gg 1$
as usual. The 
action of the system is:
\beq
S=S_{grav} + S_{flavor}
\label{stotal}
\eeq
where $S_{grav}$ (Einstein frame) is given by:
\beq
S_{grav}=\frac{1}{2\kappa_{(10)}^2}
\int d^{10}x \sqrt{-g} \left[R-\frac12 (\partial_\mu \phi)
(\partial^\mu \phi)-\frac{1}{12}e^{\phi}F_{(3)}^2\right]
\label{gravaction}
\eeq
whereas $S_{flavor}$
is the Dirac+WZ action for the $N_f$ D5-flavor 
branes (we take the worldvolume gauge field to be zero). In Einstein frame:
\beq
S_{flavor}=T_5 \sum^{N_f} \left(
-\int_{{\cal M}_6} d^6x e^{\frac{\phi}{2}}\sqrt {-\hat g_{(6)}}
+ \int_{{\cal M}_6} P[C_{6}] \right)
\label{sflavor}
\eeq
where the integrals are taken over the six-dimensional
worldvolume of the flavor
branes ${\cal M}_6$
and $\hat g_{(6)}$ stands for the determinant of the pull-back
of the metric in such worldvolume. We will take these D5's
extended along $x_0,\dots x_3, r, \psi$ at constant
$\theta, \varphi, \tilde \theta, \tilde \varphi$. Notice that
this configuration preserves the $U(1)_R$ symmetry associated to shifts
in $\psi$.
This is one of the 
embeddings which were called {\it cylinder solutions} in
\cite{Nunez:2003cf}. 
Moreover, notice that this brane configuration makes clear
 the need of the deformed ansatz  (\ref{metric}) where 
  $k(r)\neq g(r)$
(compare to (\ref{metricaa})).

Following a procedure analogous to
\cite{Bigazzi:2005md} we think of the $N_f \to \infty$ branes as being
homogeneously
smeared along the two transverse $S^2$'s parameterized by
$\theta, \varphi$ and $\tilde \theta, \tilde \varphi$. 
We will elaborate on how the smearing may affect the dual field theory
in section \ref{sect: gauge}.
The smearing erases
the dependence on the angular coordinates and makes it 
possible to consider an ansatz with functions only depending on $r$,
enormously simplifying  computations.
We use the same ansatz (\ref{metric}) for the metric and
also fix $\phi = 4f$.
We have:
\ba
-T_5 \sum^{N_f} 
\int_{{\cal M}_6} d^6x e^{\frac{\phi}{2}} \sqrt {-\hat g_{(6)}} &\to&
-\frac{T_5 N_f}{(4\pi)^2} \int d^{10}x
\sin\theta \sin \tilde \theta e^{\frac{\phi}{2}} \sqrt {-\hat g_{(6)}}
\label{BIterm}\\
T_5 \sum^{N_f} \int_{{\cal M}_6} P[C_{6}] &\to& \frac{T_5 N_f}{(4\pi)^2}
\int Vol({\cal Y}_4) \wedge C_{(6)}
\label{WZterm}
\ea
where we have defined $Vol({\cal Y}_4)=\sin \theta \sin \tilde \theta
d\theta \wedge d\varphi \wedge d\tilde\theta \wedge d\tilde\varphi$
and the new integrals span the full space-time.
We will need the expressions (we choose $\alpha'=g_s=1$).
\beq
T_5=\frac{1}{(2\pi)^5}\,\, ,\qquad\qquad
\frac{1}{2\kappa_{(10)}^2} =\frac{1}{ (2\pi)^7 }
\label{T5value}
\eeq
Let us turn our attention to the effect of the WZ term of 
the flavor brane
action (\ref{WZterm}). Since it does not depend on the metric nor on
the dilaton, it does not enter the Einstein equations. However, it
 alters the equation of motion for the 6-form $C_{(6)}$, which
before  was $d*F_{(7)}\equiv dF_{(3)} =0 $. Now we have a source
term so\footnote{In general, if for a form $F_{(n)}=dA_{(n-1)}$
there is an action $-\frac{1}{2n!}\int \sqrt{|g|} F^2
+\int G \wedge A$, the equation of motion for the form reads
$d*F = sign(g)(-1)^{D-n+1} G$. In this case, the relevant part of
the action (go to string frame for this computation) reads:
$-\frac{1}{2\kappa_{(10)}^2}\frac{1}{2\cdot 7!}\int \sqrt{|g|} F_{(7)}^2
+\frac{T_5 N_f}{(4\pi)^2} \int Vol({\cal Y}_4)\wedge C_{(6)}$
so the equation of motion is 
$\frac{1}{2\kappa_{(10)}^2} d*F_{(7)}=-\frac{T_5 N_f}{(4\pi)^2}
Vol({\cal Y}_4)$. Taking into account $F_{(3)}=- *F_{(7)}$
and  (\ref{T5value}) we arrive at (\ref{newdF}).}:
\beq
dF_{(3)} = \frac14 N_f Vol({\cal Y}_4) =
\frac14 N_f \sin \theta \sin \tilde \theta
d\theta \wedge d\varphi \wedge d\tilde\theta \wedge d\tilde\varphi
\label{newdF}
\eeq
Notice that this particular form for the violation of the Bianchi identity is an effect of the
smearing: we replace the sum of delta functions on the position of each
of the $N_f$ branes by a constant density,
so there is a continuous distribution of charge which acts as a source
for the RR form.
We can slightly modify (\ref{3form}) 
to solve (\ref{newdF})\footnote{
More generally, we could consider:
\begin{displaymath}
F_{(3)}= -\frac{N_c + N_c'}{4} \sin\tilde\theta d\tilde\theta \wedge 
d\tilde\varphi \wedge (d\psi + \cos \theta d\varphi)
-\frac{N_f - N_c - N_c'}{4} 
\sin\theta d\theta \wedge d\varphi \wedge ( d\psi + \cos 
\tilde \theta d\tilde\varphi)\,\,
\end{displaymath}
but $N_c'$ is just a shift in $N_c$.
As explained below (\ref{quantcond}), it is the coefficient of the
first term the one identified with the number of colors.
}:
\beq
F_{(3)}= -\frac{N_c}{4} \sin\tilde\theta d\tilde\theta \wedge 
d\tilde\varphi \wedge (d\psi + \cos \theta d\varphi)
-\frac{ N_f - N_c}{4} 
\sin\theta d\theta \wedge d\varphi \wedge ( d\psi + \cos 
\tilde \theta d\tilde\varphi)
\label{new3form}
\eeq
On the other hand, we still have $dF_{(7)}=0$.
In order to find the new system of BPS equations, we
recompute (\ref{IIBproja}) using the {\it modified} RR three-form  
field strength (\ref{new3form}). It is quite straightforward in this way to get: 
\ba
h'&=& \frac{1}{4} (N_c-N_f) e^{-2 h -k} +\frac14 e^{-2 h +k}
\label{newbps1} \\
g' &=& -N_c e^{-2g-k} + e^{-2g+k}
\label{newbps2} \\
k' &=&\frac{1}{4} (N_c-N_f) e^{-2h -k} 
-N_c e^{-2g -k} 
-\frac14 e^{-2h +k} - e^{-2g +k} +2 e^{-k}
\label{newbps3}  \\
\phi'&=&-\frac{1}{4} (N_c-N_f)  e^{-2 h-k} + N_c  e^{-2 g-k} 
\label{newbps4}
\ea
Nevertheless, it is important to point out that, apart
from the IIB sugra action, we also have the flavor branes action, so,
in order to preserve supersymmetry, it
is necessary that their action preserves $\kappa$-symmetry. As stated
above, this is indeed the case in this construction.
In appendix \ref{appendixsup}, we reobtain this system using the
so-called ``superpotential" approach and in section 
\ref{sect: eins} we check that this first order system automatically implies
that the set of second order equations of motion are satisfied.

The system (\ref{newbps1})-(\ref{newbps4}) has a very simple
solution when $N_f = 2N_c$ that we present in section \ref{nf2nc}
and discuss in section \ref{nf2ncint}. In this
work, we will not undertake the study of the system for 
$N_f \neq 2 N_c$.

 Using this toy-example, we  have clarified our 
proposal. In order to add backreacting flavors, we will 
consider a deformed background, solution of the eqs of motion derived 
from the sugra action plus
an action like (\ref{BIterm})-(\ref{WZterm}). These last terms
come from the action of the (susy preserving) flavor branes.
We modify the 
RR form so that the failure of the  Bianchi identity indicates the 
presence of the smeared flavor 
branes. After that, we impose vanishing of the IIB SUSY 
variations. This will 
produce a set of BPS eqs that satisfy the Einstein 
and Maxwell eqs and ensure the susy of the whole construction.
\subsection{The Einstein Equations}
\label{sect: eins}
In order to check for good the consistency of the approach, in this
section we will check that the solutions found above indeed satisfy the
full set of equations of motion.

First of all, we have already guaranteed the equation for the RR-form
(see (\ref{newdF}) and (\ref{new3form})). The equations for the 
worldvolume fields on the flavor branes are also satisfied since
 $\kappa$-symmetry is preserved.
 We still have to
check the equations for the dilaton and the metric. The relevant action is
given by the sum of (\ref{gravaction}) and (\ref{BIterm}), since
(\ref{WZterm}) does not depend on the dilaton nor on the metric.
The equation for the dilaton is:
\beq
\frac{1}{\sqrt{-g_{(10)}}}\partial_\mu
\left( g^{\mu\nu} \sqrt{-g_{(10)}} \partial_\nu \phi \right)
-\frac{1}{12} e^{\phi} F_{(3)}^2 -
\frac{N_f}{8} e^{\frac{\phi}{2}}\frac{{\sqrt{-\hat 
g_{(6)}}}}{\sqrt{-g_{(10)}}} \sin\theta \sin \tilde \theta = 0
\label{dileq}
\eeq
whereas the equations coming from variations of the metric are (compare 
with (\ref{Einseqmn})):
\beq
R_{\mu\nu}-\frac12 g_{\mu\nu} R= \frac12
\left(\partial_\mu \phi \partial_\nu \phi -\frac12 g_{\mu\nu} 
\partial_\lambda \phi \partial^\lambda \phi \right) 
+\frac{1}{12}e^\phi
\left(3 F_{\mu\lambda_2\lambda_3}F_\nu^{\,\lambda_2\lambda_3}
-\frac12 g_{\mu\nu} F_{(3)}^2\right) + T_{\mu\nu}^{flavor}
\label{Einseq}
\eeq
where the energy-momentum tensor generated by the flavor branes comes
from the variation of the lagrangian of the flavor branes (\ref{BIterm}):
\beq
{\cal L}_{flavor}=-\frac{T_5 N_f}{(4\pi)^2} 
\sin\theta \sin \tilde \theta e^{\frac{\phi}{2}} \sqrt {-\hat g_{(6)}}
\eeq
and therefore is:
\beq
T^{\mu\nu}_{flavor}=\frac{2\kappa_{(10)}}{\sqrt{-g_{(10)}}}
\frac{\delta {\cal L}_{flavor}}{\delta g_{\mu\nu}}=
-\frac{N_f}{4}\sin\theta \sin \tilde \theta \frac12
e^{\frac{\phi}{2}}\, \delta_\alpha^\mu\, \delta_\beta^\nu\, 
\hat g_{(6)}^{\alpha\beta}\frac{\sqrt {-\hat g_{(6)}}}{\sqrt {- g_{(10)}}}
\eeq
where $\alpha,\beta$ span the brane worldvolume.
In components, after lowering the indices:
\ba
&&T^{flavor}_{x_i x_j}=-\frac{N_f}{2}\eta_{ij}e^{-2h-2g}\,\,,
\qquad\qquad\
T^{flavor}_{rr}=-\frac{N_f}{2}e^{-2h-2g}\,\,,\rc
&&T^{flavor}_{\psi\psi}=-\frac{N_f}{8}e^{2k-2h-2g}\,\,,\qquad\qquad\ \
T^{flavor}_{\varphi\psi}=-\frac{N_f}{8}e^{2k-2h-2g}\cos\theta\,\,,\rc
&&T^{flavor}_{\varphi\varphi}=-\frac{N_f}{8}e^{2k-2h-2g} \cos^2\theta \,\,,\qquad
T^{flavor}_{\tilde\varphi\psi}=-\frac{N_f}{8}e^{2k-2h-2g} \cos\tilde\theta\,\,,\rc
&&T^{flavor}_{\tilde\varphi\tilde\varphi}=-\frac{N_f}{8}e^{2k-2h-2g}
\cos^2 \tilde \theta \,\,,\qquad
T^{flavor}_{\varphi\tilde\varphi}=-\frac{N_f}{8}e^{2k-2h-2g}\cos\theta
\cos\tilde\theta  \,\,.
\label{Tflavor}
\ea
Now it is straightforward to check (using {\it Mathematica}) that the equations 
(\ref{newbps1})-(\ref{newbps4}) imply that (\ref{dileq}) and (\ref{Einseq}) are satisfied.

To summarize, we have explicitly showed in this section that a consistent 
procedure 
to add flavor branes to a given background is to consider the BPS eqs 
obtained by imposing the vanishing of  SUSY 
transformations for the gravitino and dilatino in a system of a ``deformed 
spacetime" (like (\ref{metric}) with respect  to (\ref{metricaasing})) and 
modified
RR forms so that the failure of the Bianchi identity is indicating 
the presence of the smeared
 flavor branes. In the next section, we will apply this 
proposal to add flavor branes to the background dual to $\mathcal{N}=1$ SYM 
(\ref{metricaa})-(\ref{RR}).

\section{Adding Flavor Branes to the Non-singular Solution }
\label{flavnonsing}
\setcounter{equation}{0}
In this section we 
will solve the important problem of adding $N_f$ flavor branes to
the background dual to $\mathcal{N}=1$ SYM (\ref{metricaa})-(\ref{RR}) 
\footnote{In the probe approximation also the paper \cite{Wang:2003yc} 
discussed the problem.}. As pointed 
out in the introduction, 
the resolution of this problem is instrumental
in the duality between string theory and $\mathcal{N}=1$ SQCD-like 
theories. We will be more sketchy here than in section 
\ref{singularsolution} and just present the ansatze and corresponding
BPS equations for different cases.
Many technical details are left for 
appendix~\ref{appendixa}.
 Our procedure (as explained in detail in section 
\ref{singularsolution}), will be first to propose a deformation of the 
background dual to $\mathcal{N}=1$ SYM 
(\ref{metricaa})-(\ref{RR}). Then, we will obtain the BPS eqs that 
describe the deformation due to the presence of flavor branes, that will 
be extended along the directions $x_0,x_1,x_2,x_3, r, \psi$ and smeared 
over the directions ($\theta,\tt, \varphi,\tilde{\varphi}$). As before, 
these flavor branes will be sources for RR forms that will flux the 
deformed background and the mark of their existence as singular sources 
will be the violation of the Bianchi identity. 

\subsection{Deforming the Space}
\label{sect: unflavnonab}
We start by proposing a deformation to the background 
(\ref{metricaa})-(\ref{RR}).
The ansatz we use below
is a subcase (with no fractional branes $H_{(3)}=F_{(5)}=0$)
of that first proposed in \cite{Papadopoulos:2000gj}
and further analyzed in \cite{Butti:2004pk}.
We borrow some results and notation from \cite{Butti:2004pk}.
Although it is not the main motivation of this work, this 
unflavored setup encodes some interesting physics which we will 
analyze in section \ref{unflavoredcase}.
We consider the Einstein frame 
metric:
\ba
ds^2 &=& e^{2 f(r)} \Big[dx_{1,3}^2 + dr^2 + e^{2 h(r)} 
(d\theta^2 + \sin^2\theta d\varphi^2) +\rc
&+&\frac{e^{2 g(r)}}{4} 
\left((\tilde{\omega}_1+a(r)d\theta)^2 
+ (\tilde{\omega}_2-a(r)\sin\theta d\varphi)^2\right)
 + \frac{e^{2 k(r)}}{4} 
(\tilde{\omega}_3 + \cos\theta d\varphi)^2\Big]  \,\,.
\label{nonabmetric}
\ea
The vielbein we consider for this metric is the straightforward
generalization of (\ref{vielbein}) by the inclusion of the $a(r)$
dependence in $e^1$ and $e^2$.
Apart from the dilaton
\beq
\phi = 4f,
\label{phif4}
\eeq
we also excite the RR 3-form field strength which
we take to be of the same form as (\ref{RRthreeform}):
\ba
F_{(3)}=\frac{N_c}{4}\Bigg[-(\tilde{\omega}_1+b(r) d\theta)\wedge
(\tilde{\omega}_2-b(r) \sin\theta d\varphi)\wedge
(\tilde{\omega}_3 + \cos\theta d\varphi)+\rc
b'dr \wedge (-d\theta \wedge \tilde{\omega}_1  + \sin\theta d\varphi \wedge 
\tilde{\omega}_2) + (1-b(r)^2) \sin\theta d\theta\wedge d\varphi \wedge
\tilde{\omega}_3\Bigg]\,\,.
\label{F3unflav}
\ea
The condition $dF_{(3)}=0$ is
automatically ensured by this ansatz. It is useful to
rewrite this expression in terms of the vielbein forms:
\ba
F_{(3)}=-2 N_c e^{-3f-2g-k} e^1\wedge e^2 \wedge e^3 +
\frac{N_c}{2} b' e^{-3f-g-h} e^r\wedge (-e^\theta \wedge e^1
+e^\varphi \wedge e^2)+\qquad\rc
\frac{N_c}{2}e^{-3f-2h-k}(a^2 -2ab +1) e^\theta \wedge e^\varphi
\wedge e^3 +
N_c e^{-3f-h-g-k}(b-a)\left(-e^\theta \wedge e^2 +
e^1 \wedge e^\varphi \right) \wedge e^3
\label{F3beinunflav}
\ea
We now analyze the dilatino and gravitino transformations 
(\ref{IIBproja}) (in this case, the superpotential approach is
far more complicated because, as we will see, there are algebraic
constraints).
After very lengthy algebra (that is explicitly reported in 
appendix \ref{appendixa}), we 
obtain  a system of BPS eqs and constraints, some of which can be 
solved, leaving us with two differential eqs for $a$ and $k$. We define: 
\beq
e^{-k(r)} dr \equiv d\rho\ .
\label{rrhochange}
\eeq
The differential equations read:
\ba
\partial_\rho a&=&\frac{-2}{-1+2\rho \coth 2\rho}
\left[\frac{e^{2k}}{N_c}\, \frac{(a \cosh 2 \rho -1)^2}{\sinh 2\rho}
+a (2\rho - a \sinh 2\rho)\right]\,\,,\rc
\partial_\rho k&=&\frac{2(1+a^2-2a \cosh 2\rho)^{-1}}{(-1+2\rho \coth 2\rho)}
\left[\frac{e^{2k}}{N_c}\,a \sinh 2\rho (a \cosh 2 \rho -1)
+(2\rho -4a \rho \cosh 2\rho +\frac{a^2}{2} \sinh 4\rho)\right]\rc
\label{akeqs}
\ea
The equation for $f= \frac{\phi}{4} $ reads
\beq
\partial_\rho f=-\frac{(-1+a \cosh 2 \rho)^2 
\sinh^{-2} (2 \rho)  (-4 \rho +\sinh 4 \rho )}{4 (1+a^2-2 a \cosh 2 \rho)
 (-1+2 \rho  \coth 2 \rho )},
 \label{eqforf1}
\eeq
and can be integrated once $a(\rho)$ is known. The functions $b(\rho)$, $ 
g(\rho)$, $h(\rho)$ can be solved to be:
\bea
& & b(\rho)= \frac{2\rho}{\sinh(2\rho)}, \;\;\; e^{2g}=N_c \frac{b 
\cosh(2\rho) -1}{a \cosh(2\rho) -1}\nonumber\\
& &  
e^{2h}=\frac{e^{2g}}{4} (2 a \cosh(2\rho) -1 - a^2).
\label{solnonab}
\eea

\subsection{Flavoring the Space}
\label{flavnonab}

We consider the same ansatz for the metric:
\ba
ds^2 &=& e^{2 f(r)} \Big[dx_{1,3}^2 + dr^2 + e^{2 h(r)} 
(d\theta^2 + \sin^2\theta d\varphi^2) +\rc
&+&\frac{e^{2 g(r)}}{4} 
\left((\tilde{\omega}_1+a(r)d\theta)^2 
+ (\tilde{\omega}_2-a(r)\sin\theta d\varphi)^2\right)
 + \frac{e^{2 k(r)}}{4} 
(\tilde{\omega}_3 + \cos\theta d\varphi)^2\Big]  \,\,.
\label{nonabmetric42}
\ea
and the dilaton is still
(\ref{phif4}).

One should now think of introducing backreacting flavor branes along the
lines of section~\ref{singularsolution}.
We incorporate smeared flavored D5-branes on the non-singular
solution and the analysis of the equations is quite similar.
We again
consider D5's extended along $x_0,\dots,x_3,r,\psi$, each at constant
$\theta,\varphi,\tilde\theta,\tilde\varphi$. The analysis in \cite{Nunez:2003cf}
shows that these branes preserve the same supersymmetry as the background 
for any
value of the angles $\theta,\varphi,\tilde\theta,\tilde\varphi$. One can
therefore think of smearing along this space\footnote{ 
Notice that in the unflavored construction the spheres parameterized
by $\theta,\varphi,\tilde\theta,\tilde\varphi$ do not play an important
role from the point of view of the IR ${\cal N}=1$ SYM theory. Since we want
to add flavor to this theory, we can expect that the effect of the
smearing will not alter dramatically the resulting IR
${\cal N}=1$ SQCD theory. We will comment more about this in
section \ref{gen asp gauge}.}.
As in section \ref{secabflavor},
the Bianchi identity gets modified to (\ref{newdF}):
\beq
dF_{(3)} = \frac14 N_f Vol({\cal Y}_4) =
\frac14 N_f \sin \theta \sin \tilde \theta
d\theta \wedge d\varphi \wedge d\tilde\theta \wedge d\tilde\varphi
\label{newdF2}
\eeq
We solve this by adding to the 3-form written in (\ref{F3unflav}) 
the same as in (\ref{new3form}):
\beq
F_{(3)}^{flavor} = 
-\frac{N_f}{4} 
\sin\theta d\theta \wedge d\varphi \wedge ( d\psi + \cos 
\tilde \theta d\tilde\varphi)
\label{flav3form}
\eeq
Written in flat indices, the 3-form now reads:
\ba
F_{(3)}&=&-2 N_c \,e^{-3f-2g-k} e^1\wedge e^2 \wedge e^3 +
\frac{N_c}{2}\, b' e^{-3f-g-h} e^r\wedge (-e^\theta \wedge e^1
+e^\varphi \wedge e^2)+\rc
&+&\frac{N_c}{2}\,e^{-3f-2h-k}\left(a^2 -2ab +1-x 
 \right)e^\theta \wedge e^\varphi
\wedge e^3 + \rc
&+& N_c\, e^{-3f-h-g-k} (b-a)
\left(-e^\theta \wedge e^2 +
e^1 \wedge e^\varphi \right) \wedge e^3
\label{F3beinflav}
\ea
where we have defined
\be
x\equiv \frac{N_f}{N_c} \, .
\ee
We now insert the expression (\ref{F3beinflav}) for $F_{(3)}$ in the transformation of the spinors equations
(\ref{IIBproja}).
We find a set of BPS eqs: seven
first order equations and two algebraic constraints (for 
details, the reader is referred to appendix \ref{appendixa}).
This system\footnote{These expressions are clearly not
valid when $N_f = 2N_c$. We will deal with this special case in
section~\ref{nf2nc} and appendix \ref{appendixnf2nc}.} 
can be partially solved, leaving us with
(here we use  the definition (\ref{rrhochange})):
\bea
& &
b=\frac{\left(2-x\right) \rho }
{\sinh (2\rho)}, \;\quad e^{2g}= \frac{N_c}{2}\, \frac{2b 
\cosh 2\rho -2+x}{a \cosh2\rho -1}\nonumber\\
& & e^{2h}=\frac{e^{2g}}{4} (2 a \cosh(2\rho) -1 - a^2)
\label{bghexpression}
\eea
plus
two coupled first order eqs for $a(\rho), k(\rho)$ and  a differential 
equation 
for $f =\frac{\phi}{4}$ that can be integrated in terms of $a(\rho)$. 
These 
equations read ($\tilde{k}\equiv k-\frac{1}{2}\log N_c$):
\bea
\partial_\rho a&=&\frac{2}{2\rho \coth 2\rho -1}
\left(-\frac{2e^{2\tilde{k}}-x}{2-x} \frac{(a\cosh 2\rho -1)^2}
{\sinh 2\rho}+a^2 \sinh 2\rho-2a\rho\right)\label{eq1}\\
\partial_\rho \tilde{k}&=& \frac{2}{(2\rho \coth 2\rho -1)
(1-2a\cosh 2\rho+a^2)}\left(  \frac{2e^{2\tilde{k}}+x}{2-x} a\sinh 
2\rho (a\cosh 2\rho -1)  +\right. \nonumber \\
&&\left.+2\rho(a^2\frac
{\sinh 2\rho}{2\rho}\cosh 2\rho-2a\cosh 2\rho +1)  \right)
\label{eq2}
\eea
and
\ba
\partial_\rho f &=&\frac{(-1+a \cosh 2 \rho)
\sinh^{-2} (2 \rho)  }{4 (1+a^2-2 a \cosh 2 \rho)
 (-1+2 \rho  \coth 2 \rho )}\Big[
  -4 \rho +\sinh 4 \rho + \rc
 && + 4a \rho \cosh 2\rho
 -2a \sinh 2\rho - \frac{4}{(2-x)}
 a (\sinh 2\rho)^3 \Big]
\label{feqflav}
\ea
The equations (\ref{bghexpression})-(\ref{feqflav}) constitute an 
important result of this paper.
In order to summarize the flavored setup,
 the metric is given by (\ref{nonabmetric42}), the dilaton by (\ref{phif4})
 and the RR three-form is the sum of (\ref{F3unflav}) and (\ref{flav3form})
 (or, in flat indices, it is (\ref{F3beinflav})), while all the functions
 in the ansatz are determined by (\ref{bghexpression})-(\ref{feqflav}).
One can check (using {\it Mathematica})
that these conditions solve the full system of second
order equations given by (\ref{dileq}), (\ref{Einseq}) and
$d*F_{(3)}=0$. Notice that the expressions 
(\ref{Tflavor}) are still valid in
this case.
So, if we can find solutions to (\ref{eq1})-(\ref{feqflav}), we will 
have the string dual to 
a family of SQCD-like theories with $N_c$ colors and 
$N_f$ flavors; we now turn into this.

\subsection{Asymptotic Solutions ($N_f\neq 2N_c$)}
\label{asymptoticsols}
Unfortunately, it was not possible for us to find exact explicit 
solutions to the system (\ref{eq1})-(\ref{feqflav}). 
We study the problem in the following way, we first find asymptotic 
solutions near $\rho\to \infty$ (the UV of the dual theory) and near 
$\rho=0$ (the IR of the dual gauge theory). Then, we study a numerical 
interpolation showing that both expansions can be joined smoothly; for 
many purposes, this is 
as good as finding an exact solution.

\subsubsection{Expansions for $\rho \to \infty$}\label{sub:infty}
Let us disregard exponentially suppressed terms, and assume that
$a=e^{-2\rho} j$ where $j$ is some function that can be Laurent
expanded in $\rho^{-1}$. Then, when $x<2$,
the large $\rho$ expansion reads:
\be
a=e^{-2\rho}\left( (4-2x)\rho + \frac{x}{2} + \frac{x(4-x)}{8(x-2)}
\rho^{-1} + {\cal O}(\rho^{-2}) \right) + \ldots 
\ee
When $x=0$ we have $a=4\rho e^{-2\rho} + \ldots $
which agrees with the usual solution (\ref{metricaa})-(\ref{RR}), see 
(\ref{expansionnearr=infinity}) for comparisons. The 
other functions read:
\bea
e^{2k}&=& N_c \left(1 - \frac{x}{8(2-x)}\rho^{-2}+ \dots \right)\rc
e^{2g}&=& N_c \left(1 + \frac{x}{4(2-x)}\rho^{-1}+ \dots \right)\rc
e^{2h}&=& \frac{N_c}{2} \left((2 - x) \rho + \frac{x -1}{2} + \dots\right)\rc
\partial_\rho f &=& \frac14 - \frac{1}{16}\rho^{-1}+ \dots
\label{expansionnearinfinity}\eea
Equation (\ref{expansionnearinfinity}) is clearly not valid
for $x>2$. In this case we find a different expansion:
\bea
a&=&e^{-2\rho}\left( 1 + \frac{x-1}{2(x-2)}
\rho^{-1} + \frac{x-1}{8(x-2)}
\rho^{-2} \right) +\dots\rc
e^{2k}&=& N_c \left(x-1 - \frac{x(x-1)}{8(x-2)}\rho^{-2}+ \dots \right)\rc
e^{2g}&=& N_c \left(2(x-2) \rho + 1 + 
\frac{x(x-1)}{4(x-2)}\rho^{-1}+ \dots \right)\rc
e^{2h}&=& N_c \left(\frac{x-1}{4} + \frac{x(x-1)}{16(x-2)} \rho^{-1} 
+ \dots \right)\rc
\partial_\rho f &=& \frac14 - \frac{1}{16}\rho^{-1}+ \dots
\label{expansionnearinfinityx>2}
\eea
The other possible behavior at $\rho\to\infty$ is a geometry 
asymptoting to Minkowski times the deformed conifold.
Since we will not use such solutions in our physical interpretation,
we relegate their description to appendix \ref{appendixb}.

\subsubsection{Expanding around $\rho \to 0$}
\label{rho0expand}

A series expansion can be found for the functions $a(\rho), 
k(\rho)$ that solve the BPS eqs near $\rho=0$. This is actually a 
two-parameter family of solutions labelled by two free numbers which we denote 
$c_1 $ and $ c_2$.
The solution for the functions reads in this case
\bea
& & a= 1- 2 \rho^2 + c_1 \rho^3 +\frac{(80-40x +9 x 
c_1^2)}{12(2-x)}\rho^4+...,\rc
&& e^{2k}= N_c \left(c_2 \rho^2 + \frac{3 c_1c_2 
x}{2(x-2)}\rho^3+
\frac{c_2 (256-256x+x^2(64+27c_1^2))}{48(2-x)^2}\rho^4
+... \right) \nonumber\\
& & e^{2g}= N_c \left(\frac{2 (2-x)}{3c_1 }\frac{1}{\rho} -\frac{x}{2}
+\frac{8(2-x)}{9c_1}\rho+... \right) ,\rc
&& e^{2h}= N_c \left(\frac{2(2-x)}{3c_1}\rho -\frac{x}{2}\rho^2
-\frac{8(2-x)}{9c_1} \rho^3+... \right) , \nonumber\\
& & e^{2\phi-2\phi_0}= 1+ \frac{3c_1 x}{2(2-x)}\rho + 
\frac{27x^2 c_1^2}{16(2-x)^2}\rho^2+...
\label{solutionnerar=0b}
\eea
Some comments about this expansion are in order.
First, we should notice that (\ref{solutionnerar=0b}) above does not 
reduce to (\ref{expansionnearr=0})
when $x=0$. This is not very surprising, since the behavior of gauge 
theories  in the IR changes radically in the presence of massless flavors. 
Second and perhaps more important,  we notice that this solution is {\it 
singular}. Indeed, when computing
the Ricci scalar when $\rho \to 0$,
 one gets $R\sim \rho^{-2}$ and there does not seem to 
exist a choice of the constants ($c_1, c_2$) that avoids this.

The presence of a singularity might be source of concern and cast doubts on the
reliability of the solutions. An important point that we would like to
emphasize, though, is that there is an important and significant difference
between the singularity in (\ref{solutionnerar=0b}) and others present in the
literature, as the one in (\ref{metricaasing}) for example. There is one criterium developed 
in \cite{Maldacena:2000mw} (see also \cite{Gubser:2000nd} for other 
criteria)
that suggests a way of deciding when an (IR) singularity in a supergravity 
solution should be accepted or rejected as unphysical. This criterium is 
quite easy to apply and coincides with other different criteria developed 
in previous literature. It consists in analyzing the 
(Einstein-frame) component of the metric $g_{tt}$. If this is bounded, the 
singularity should be accepted as a good background.
 The idea in this criteria is that an excitation 
travelling towards the origin (that is a low energy object on the gauge 
theory dual) will have 
an energy as measured by an inertial gauge theory observer given  by 
$E=\sqrt{|g_{tt}|} E_0$ (with $E_0$ the proper energy of the object). 
So, if $g_{tt}$ 
diverges (the singularity is repulsive), the object 
gains unbounded energy from the field theory inertial observer point of view and this 
is to be considered as unphysical. 

According to this criterium, the singularity encoded in the solution 
(\ref{solutionnerar=0b}) above should be accepted as being physically 
relevant; indeed $g_{tt}= e^{\phi/2}= e^{\phi_0/2}(1+ ...)$, see 
(\ref{solutionnerar=0b}). 
This means 
that even though the (good) singularity we find might be a signal of some
IR dynamics that we are overlooking, we can still perform computations to try to
answer non-perturbative field theory questions. As we will see in section
 \ref{gaugethnonsing}, the background matches several gauge theory expectations
 even when we are performing computations near $\rho=0$.

To complete the study of these solutions, it is worthwhile to 
do a detailed numerical analysis to which we turn now.

\subsection{Numerical Study ($N_f < 2N_c$)}
\label{numericstudy}
\begin{figure}[!htb] 
   \includegraphics[width=0.45\textwidth]{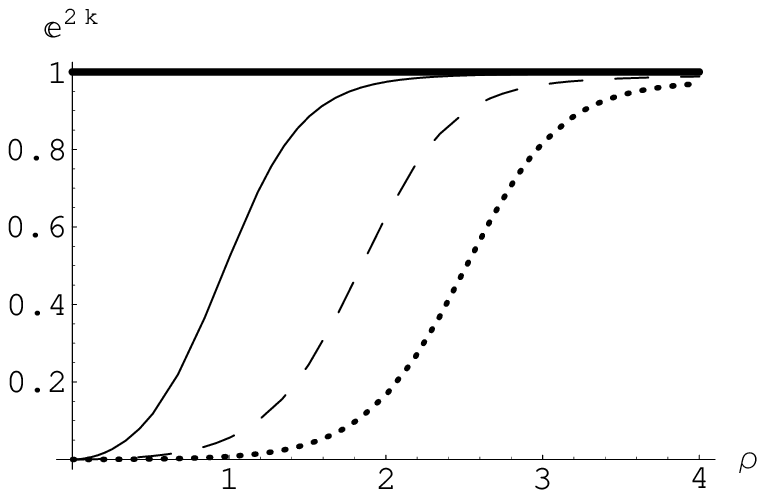} 
   \includegraphics[width=0.45\textwidth]{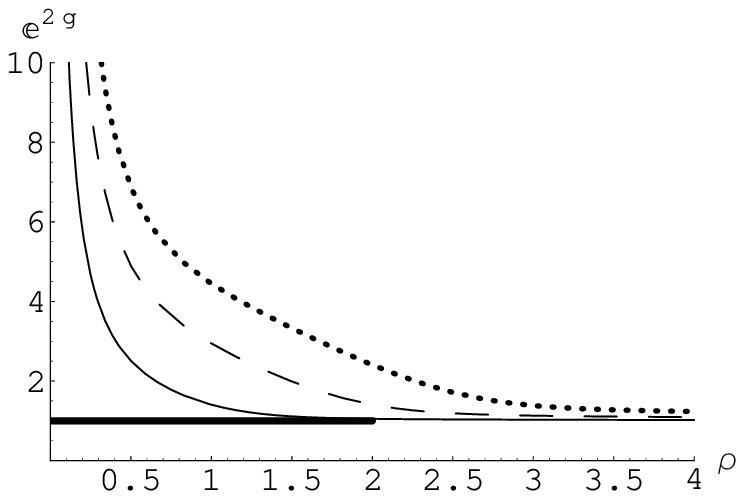} \\
   \includegraphics[width=0.45\textwidth]{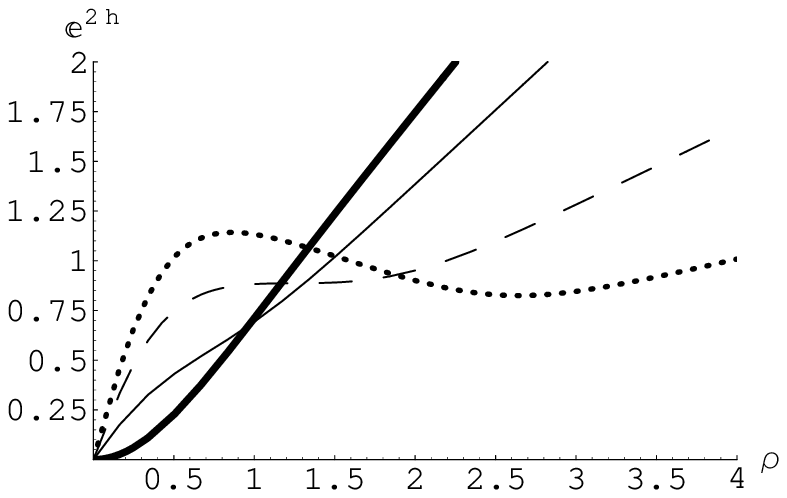} 
   \includegraphics[width=0.45\textwidth]{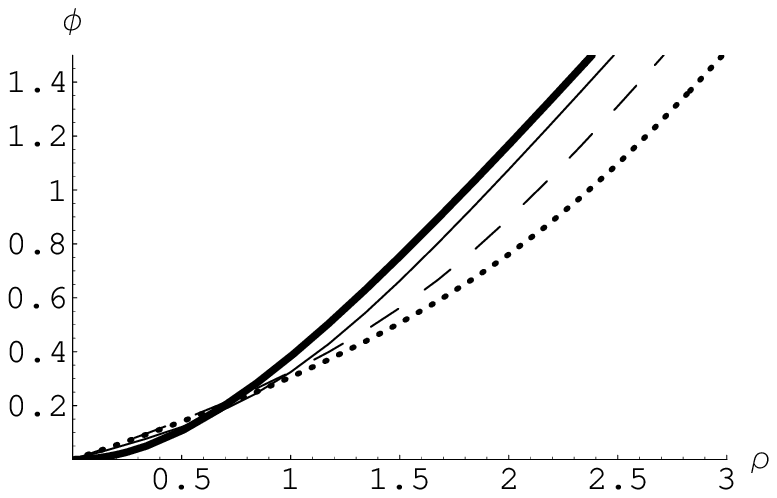} \\
   \includegraphics[width=0.45\textwidth]{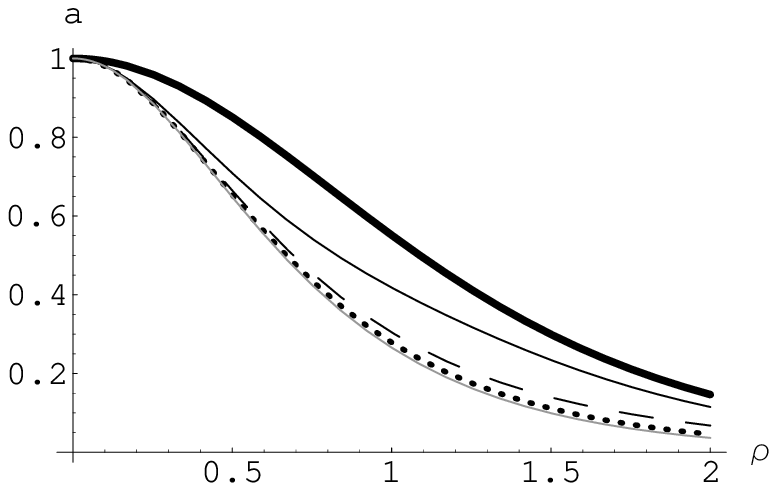} 
   \includegraphics[width=0.45\textwidth]{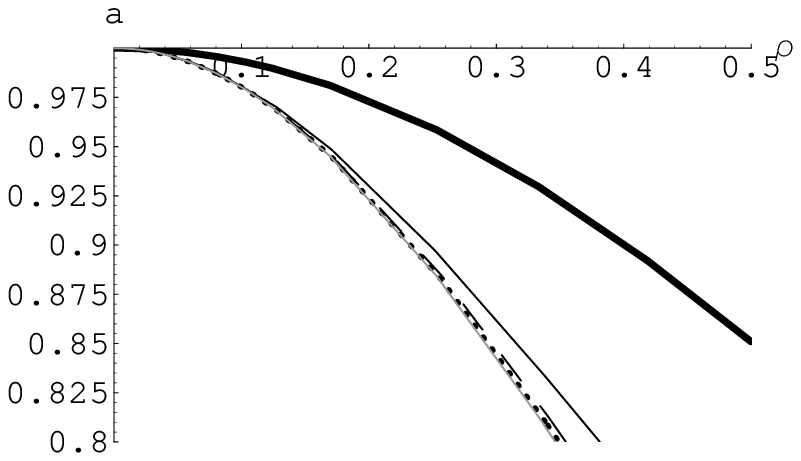} 
   \caption{Some functions of the flavored solutions for
   $N_f = 0.5 N_c$ (thin solid lines), $N_f=1.2 N_c$ (dashed lines),
   $N_f = 1.6 N_c$ (dotted lines). For comparison, we also plot
   (thick solid lines) the usual unflavored solution. The graphs
   are $e^{2k}$, $e^{2g}$, $e^{2h}$, $\phi$, $a$ and a zoom
   of  the plot for $a$ near $\rho=0$. In the figures for $a$ we have
   also plotted with a light solid line $a=\frac{1}{\cosh (2\rho)}.$}
   \label{flavoredgraphs}
\end{figure}

In order to get a qualitative idea of the behavior of the solutions, 
we have studied numerically the equations 
(\ref{bghexpression})-(\ref{feqflav}) with the initial
conditions (\ref{solutionnerar=0b}) near $\rho=0$.
We also impose the
large $\rho$ asymptotic behavior (\ref{expansionnearinfinity}), what
gives a relation between the parameters $c_1$ and $c_2$. We also 
fix $c_2 = 9c_1^2 / 16$.
This amounts to requiring that the domain wall tension scales with
$N_c - \frac{N_f}{2}$ (see section \ref{sect: dw}).
However, the qualitative behavior of the plots does not depend on this
constraint. With these conditions, all the functions are uniquely
determined for a given value of $x \equiv \frac{N_f}{N_c}$, up to the
constant $\phi_0$ which we will set to zero.
Some results are reported in figure \ref{flavoredgraphs}.

Notice that the $x \to 0$ limit is not continuous (at least in the
small $\rho$ region) as it is apparent from the expansion
(\ref{solutionnerar=0b}). In order to show it graphically, we have included
in figure \ref{flavoredgraphs} a plot of the the function $a$ near
$\rho=0$.

\subsection{The Simple $N_f= 2N_c$ Solution}
\label{nf2nc}

From the expressions (\ref{bghexpression})-(\ref{feqflav}), one can readily see
that $N_f = 2N_c$ is a special point and, in fact, the manipulations leading to
some of those expressions are ill-defined in this case. Nevertheless,
from the first equation in (\ref{bghexpression}) we expect to 
have
$b=0$. Then,
considering the system (\ref{fequationflava})-(\ref{alg3flava}), it is easy to
see that it can be solved taking $a=b=0$
and ${\cal A}=1$, ${\cal B}=0$ (however, this
is not the only possibility, see appendix \ref{appendixnf2nc}).
In fact, $a=b=0$ leads us back to the
construction of section \ref{secabflavor}. For the particular
$N_f = 2 N_c$ case, there is an extremely simple solution, in which
 the other functions satisfy:
\beq
e^{2h}=\frac{N_c}{\xi}\,,\qquad
e^{2g}=\frac{4N_c}{4-\xi}\,,\qquad
e^{2k}=N_c\,,\qquad 
\phi = 4f = \phi_0 + r
\label{nf2ncvalues}
\eeq
where we have introduced the constant $0<\xi<4$.
The resulting  Einstein
frame metric, dilaton and RR three-form read:
\bear
ds^2&=&e^\frac{\phi}{2}\Big[dx_{1,3}^2+ N_c \Big(
dr^2 +\frac{1}{\xi} ( d\th^2+\sin^2\th d\varphi^2)+\rc
&&+\frac{1}{4-\xi}(d\tilde\th^2+\sin^2\tilde\th d\tilde\varphi^2)
+\frac14 (d\psi + \cos \theta d\varphi + \cos \tilde \th d\tilde\varphi)^2
\Big)\Big]\rc
e^\phi &=& e^{\phi_0} e^r\rc
F_{(3)}&=&-\frac{N_c}{2}\left(\sin\tilde\theta 
d\tilde\th\wedge d\tilde\varphi +
\sin\theta d\th\wedge d\varphi \right)\wedge (d\psi 
+ \cos \theta d\varphi + \cos \tilde \th d\tilde\varphi)
\label{nf2ncsol}
\eear
 We will discuss the
gauge theory associated to this background in section 
\ref{nf2ncint}.

\subsubsection{A Flavored Black Hole}

In section \ref{rho0expand}, we discussed the fact that the 
$0<N_f < 2N_c$ solutions present a curvature singularity, which is
good according to the criterium of having bounded $g_{tt}$. 
The $N_f = 2N_c$ case presents the same kind of singularity
at $r \to -\infty$. However, the simplicity of this metric allows
us to check that this singularity also satisfies an alternative criterium, 
that basically states that a gravity singularity is good if
it can be smoothed out by turning on some temperature in the dual field 
theory and thus finding
a black hole solution \cite{Gubser:2000nd}. 
So, in order
to construct the black hole, let us first make a change of variable,
\beq
r=2 \log z   \,\,,
\eeq
and define:
\beq
{\cal F}= 1 - \left(\frac{z_0}{z}\right)^4
\eeq
Keeping the same dilaton and three-from 
but deforming the metric to:
\bear
ds^2&=&e^\frac{\phi_0}{2}\,z
\Big[-{\cal F}dt^2 + dx_1^2+ dx_2^2+ dx_3^2+ N_c \Big(
\frac{4}{z^2}{\cal F}^{-1}
dz^2 +\frac{1}{\xi} ( d\th^2+\sin^2\th d\varphi^2)+\rc
&&+\frac{1}{4-\xi}(d\tilde\th^2+\sin^2\tilde\th d\tilde\varphi^2)
+\frac14 (d\psi + \cos \theta d\varphi + \cos \tilde \th d\tilde\varphi)^2
\Big)\Big]
\label{bhflav}
\eear
provides a regular solution of the Einstein equations (\ref{dileq}),
(\ref{Einseq}). As a technical comment, notice that, due to the change
in the metric, the expressions for $T_{tt}^{flavor}$ 
and $T_{rr}^{flavor}$ get multiplied,
with respect to 
those in (\ref{Tflavor}) by ${\cal F}$ and ${\cal F}^{-1}$ respectively.
The rest of (\ref{Tflavor}) remains unchanged.

Let us stress that this solution has its own interest since it is, to our 
knowledge, the first solution dual to a field theory in four dimensions
with adjoints and 
fundamentals at finite 
temperature\footnote{In an interesting work
\cite{Gomez-Reino:2004pw}, the addition of finite temperature to 
a three dimensional gauge theory with adjoint and fundamental fields
was studied.}. We will explore some of the gauge 
theory aspects of this solution (\ref{bhflav}) in section 
\ref{nf2ncint}.

Finally, to close this section, we mention a numerical solution described 
in detail in 
appendix 
\ref{appendixnf2nc}. This solution for the case 
$N_f= 2N_c$, appears after a careful analysis of the BPS eqs and 
has remarkable features. We comment more on it at 
the end of section \ref{nf2ncint}.

\section{Gauge Theory Aspects and Predictions of the Solution
($N_f \neq 2N_c$)} 
\label{gaugethnonsing}
\setcounter{equation}{0}

In this section we will study the family of gauge 
theories  dual to  the solution(s) 
we have presented in the previous section. 
We will first try to give arguments towards
a definite lagrangian dual to the background 
(\ref{nonabmetric42})-(\ref{feqflav}) and then 
we will study 
different strongly-coupled aspects of these field theories to 
support our proposal. These aspects include 
$R$-symmetry breaking, pair creation and Wilson loops, instantons, Seiberg 
duality, domain walls and asymptotic beta functions.

\subsection{General Aspects of the Gauge Theory}
\label{gen asp gauge}
\label{sect: gauge}

Below we will present some arguments to motivate a lagrangian description 
for the field theory dual to our backgrounds 
(\ref{nonabmetric42})-(\ref{feqflav}). 
We will argue that the dual field theory is related in the IR to $\mathcal{N}=1$ SQCD plus a 
quartic superpotential for the quark superfields. Some aspects related to the smearing of the flavor branes and to the six-dimensional little string UV completion of the theory are not completely under control, and in particular we could expect that some operator we are overlooking might still be present and slightly deform  the IR dynamics of the theory. Nonetheless we are confident that 
the result to which we arrive is robust and we present many 
tests in the following, which support our interpretation.

Our argument starts by reminding the discussion in section \ref{sec dft},  that 
the 
field theory dual to the 
unflavored solution of \cite{Maldacena:2000yy} is six-dimensional 
$\mathcal{N}=1$ $SU(N_c)$ SYM 
compactified on a twisted two-sphere in such a way that the 
low-energy theory is four-dimensional and $\mathcal{N}=1$ supersymmetric.
The spectrum of the corresponding $U(1)$
 theory is related to that of the four-dimensional 
$U(M)$ $\mathcal{N}=1^*$ theory, that is $\mathcal{N}=4$ SYM with equal 
masses for the adjoint chiral multiplets, in its Higgs vacuum 
\cite{Andrews:2005cv}.  Let us briefly review the 
main features of this four-dimensional theory 
before arguing how to extend this remarkable result to the 
large $N_c$ case. The superpotential  of the $\mathcal{N}=1^*$ 
theory can be written (in $\mathcal{N}=1$ notations) as
\be
W=\mathrm{Tr}_{U(M)}\left( i\Phi_1[\Phi_2,\Phi_3]+\frac{\mu}{2}\sum_i \Phi_i^2\right)
\ee
and therefore the F-term equations of motion read
\be\label{F term su2}
[\Phi_i,\Phi_j]=i\mu \epsilon_{ijk} \Phi_k
\ee
It is immediate to recognize  an $su(2)$ algebra in this equation. This 
allows for non-trivial solutions to  (\ref{F term su2}) where the adjoint
 fields $\Phi_i$ are taken to be proportional to the generators 
 $J_i^{(n)}$ of any irreducible representation of $su(2)$. For 
 any dimension $n$ there is only one such representation and 
 therefore the classical vacua of this theory correspond to 
 the partition of $M$ into positive integer numbers \cite{Vafa:1994tf}
\be\label{class vac}
\sum_{n=1}^M n\,k_n=M
\ee 
where $k_n$ is the number of times the $n$-dimensional irreducible 
representation appears in the expectation value of the adjoint fields. 
Any specific such vacuum has residual gauge group $\prod_n U(k_n)$. 
To obtain a vacuum with gauge group $U(1)$, one has to take $k_M=1$ and 
all other $k_n$'s to zero, whereas to  obtain a vacuum 
which has exactly $SU(N_c)$ residual gauge group, one needs to start from 
$SU(M)$  $\mathcal{N}=1^*$ with $M$ an integer multiple 
of $N_c$, $M=mN_c$, and  to take $k_m=N_c$ in   (\ref{class vac}). 
The expectation values of the adjoint fields in this Higgs vacuum satisfy the 
relation
\be\label{fuzzy Higgs}
\Phi_1^2+\Phi_2^2+\Phi_3^2=\mu^2\,\frac{M^2-1}{4},
\ee
that is they describe a (fuzzy) two-sphere.

In the particular case $U(1)$, it was shown in  
\cite{Andrews:2005cv}  that the matching of the spectrum to that of 
the compactified six-dimensional  theory we started with, 
is exact in the $M\rightarrow \infty$ limit.  
Because of the ways the $U(1)$ and the $SU(N_c)$ vacua are built, 
it is natural to argue that the results of  
\cite{Andrews:2005cv,Andrews:2006aw} extend 
to the more general $SU(N_c)$ case, that is the spectrum and lagrangian of 
the 
$k_m=N_c$ Higgs vacuum of $\mathcal{N}=1^*$ is exactly the same 
as that of the six-dimensional $SU(N_c)$ SYM wrapped on a twisted 
two-sphere when $m\rightarrow \infty$.

Let us consider now the case we are interested in, that is the addition of flavors to the compactified six-dimensional  theory. 
When we add $N_f$ flavor branes into the background, we are effectively 
adding a set of massless chiral multiplets transforming in the fundamental 
and anti-fundamental of the $SU(N_c)$ gauge group. Let us call these 
multiplets $Q=(S, q)$ and $\tilde{Q}=(\tilde{S}, \tilde{q})$. These 
fields are massless because the flavor branes are at zero distance from 
the color branes that generated the background. 

A generic lagrangian for this new system could then be written as
\beq
L= \mathrm{Tr}\Big[\int d^4\theta \left(\Phi_{KK}^\dagger e^V \Phi_{KK}^{\phantom{\dagger}}  + Q^\dagger 
e^V Q +  \tilde{Q}^\dagger e^V \tilde{Q} \right)+ \int d^2\theta \left(W_\alpha^2 + {\cal W}\right)\Big]
\label{lagrangianon=2}
\eeq
where $\mathcal{W}$ is a chiral superpotential describing the mass terms and 
self-interaction of the adjoint Kaluza-Klein states, and their coupling to the 
fundamental fields.  
Because of the argument above, we can think of the 
Kaluza Klein states as the three massive adjoint fields of 
the $\mathcal{N}=1^*$ theory. The most natural way 
to couple the fundamental fields to an adjoint state 
is mediated from the only allowed term in $\mathcal{N}=2$ SQCD, which is the one 
 typically appearing in intersecting branes setups. It reads
\be\label{superpotw}
\mathcal{W}=\kappa\, \tilde{Q}\Phi Q
\ee
with $\kappa$ a coupling constant whose value will not affect our discussion. In a theory like the $\mathcal{N}=1^*$ in a Higgs vacuum described above, if the fundamentals couple to a single adjoint field through (\ref{superpotw}), the $SU(2)$ global symmetry that rotates the three adjoints breaks down to a $U(1)$ subgroup. The term (\ref{superpotw}) breaks explicitly also the $SU(N_f)\times SU(N_f)$ flavor group to its diagonal $SU(N_f)$ subgroup.

The explicit flavor group breaking is exactly matched in our dual string
 picture by the presence of flavor D5-branes alone, rather than of
  flavor D5 and anti-D5 branes. This feature is in contrast with other 
  models of theories with fundamental degrees of freedom where the 
  chiral symmetry breaking is spontaneous and flavors are added to 
  the picture via both D-branes and anti D-branes, which reconnect 
  in the IR to account for the spontaneous breaking \cite{Sakai:2004cn}. 
 Since in our case the breaking of the global symmetry
 is explicit rather than spontaneous, we do not expect to find
Goldstone bosons in the spectrum.

Regarding the breaking of the $SU(2)$ to $U(1)$, a little more care is 
required since  the smearing of the flavor branes
restores the $SU(2)$ symmetry in our model. 
Let us start, therefore, from a different distribution of the flavor 
branes to try to understand what is happening. We could smear the flavor branes on the ($\theta$, $\varphi$) two-sphere, 
while putting all of them on a single point on the ($\tilde{\theta}$, 
$\tilde{\varphi}$) directions, for simplicity let us say at 
$\tilde{\theta}=0$. 
Since  the exact unflavored solution of 
\cite{Maldacena:2000yy} is invariant under an $SU(2)$ 
acting on the $\psi$, $\tilde{\theta}$ and $\tilde{\varphi}$ angles, 
this new configuration breaks this $SU(2)$ background isometry 
to $U(1)$. We claim that this is the configuration dual to the 
fundamental-adjoint coupling (\ref{superpotw}). From the field theory 
point of view this would probably be the most natural brane 
configuration to consider, but as we already mentioned above, 
finding a solution to the Einstein and flux equations for this less 
symmetric configuration would probably turn out to be technically 
impossible. This is the reason why we considered a smeared configuration 
for the flavor branes. Let us try, therefore, to understand 
what this means in field theory terms. 
We know from (\ref{fuzzy Higgs}) that the three massive adjoint fields 
describe a two-sphere (in the limit $M\rightarrow \infty$ it is 
a smooth sphere). Therefore, we can take them to be 
parameterized by two angles ($\tilde{\theta}$ and $\tilde{\varphi}$), by 
the usual projections over the coordinate planes of 
$\mathbb{R}^3$ of a two-sphere of radius $R^2=\mu^2(M^2-1)/4$ centered at 
the origin. We will denote these adjoints parametrizing the 
$S^2(\tilde{\theta},\tilde{\varphi})$ as 
$\Phi_{(\tilde{\theta},\tilde{\varphi})}^{(i)}$. 

Then, we propose that 
the fundamental-adjoint coupling corresponding to our model is,
schematically:
\be\label{su2 W}
\mathcal{W}\sim\kappa\, \int_{S^2} d\tilde{\theta}\,
d\tilde{\varphi}\sin\tilde{\theta}\,
\tilde{Q}\left(\Phi_{(\tilde{\theta},\tilde{\varphi})}^{(1)}
+\Phi_{(\tilde{\theta},\tilde{\varphi})}^{(2)}
+\Phi_{(\tilde{\theta},\tilde{\varphi})}^{(3)}\right) 
Q
\ee
Let us quickly go through now the low-energy effective 
lagrangian corresponding to the superpotential (\ref{su2 W}). 
In general, if the parameter $\mu$ is much bigger than the scale of 
strong coupling $\Lambda_{sqcd}$ we can integrate out the 
KK fields from (\ref{lagrangianon=2}) to 
end up with an $\mathcal{N}=1$ SQCD-like theory that looks
\beq
L= Tr\Big[\int d^4\theta \left(Q^\dagger e^V Q +
\tilde{Q}^\dagger e^V \tilde{Q}\right) + \int d^2\theta\, W_\alpha^2 + {\cal 
W'}\Big]
\label{lagrangianon=1}
\eeq
where the effective superpotential $\mathcal{W}'$ will depend on 
the way the adjoints couple to the fundamentals and to 
themselves in (\ref{lagrangianon=2}). For 
our case (\ref{su2 W}), after integrating out the massive KK fields, 
we might expect to obtain some 
combination of the quartic operators~(\ref{superpotw}):
\beq
{\cal W'}\sim \frac{\kappa}{\mu}( Tr (\tilde{Q}Q \tilde{Q}Q )+ 
(Tr\tilde{Q}Q)^2 ) 
\label{wp}
\eeq
and possibly some more complicated operator due to the cubic coupling of the adjoint KK fields.

Could we take  the KK masses to be very large ($\mu\to \infty$), we would end up
with pure SQCD. Unfortunately, there is no known way to sensibly separate 
$\mu$  from $\Lambda_{sqcd}$ without requiring the full 
string dynamics\footnote{This is the typical problem that afflicts 
all the supergravity 
duals to confining field theories. This will be cleanly solved when a 
worldsheet CFT is found for these models lifting the limitation 
$\alpha'\to 0$. Meanwhile, from a supergravity perspective, the 
KK modes can be disentangled form the gauge theory dynamics following 
\cite{Gursoy:2005cn, Pal:2005nr, Bobev:2005ng}.}. 
In fact, one can expect that taking $\mu \to \infty$ requires
reducing the ``external" space
(that is topologically $S^2\times S^3$) to a stringy scale.
In this context, it is very suggestive that our construction 
has some similarity with the non-critical string model
 of Klebanov and Maldacena \cite{KM}, which, in fact, should be the limit
of our construction when continuously increasing $\mu \to \infty$.
Let us emphasize once again that, in the process, the curvature of 
space-time reaches the string scale and therefore 
one cannot expect to get reliable results from supergravity
(although it has been shown that it is possible to get some
qualitative information from that kind of setups using a gravity 
action as a toy model
\cite{KM,othernoncrit,Bigazzi:2005md}).

We can now complete the comparison of the isometries of the field theory 
and of 
our background. We have already 
analyzed the matching of the flavor symmetry and the $SU(2)$ 
Kaluza-Klein symmetry. The field theory at weak coupling also  
has a $U(1)_B\times U(1)_R$ symmetry, 
where $U(1)_B$ is an exact baryonic symmetry, while $U(1)_R$ 
is the anomalous $R$-symmetry. 
The unbroken baryonic 
symmetry could be associated with rotations along  $\varphi$, whereas the anomalous $R$-symmetry corresponds to shifts of the coordinate $\psi$. As we will analyze in detail in section \ref{u1rbreak}, the supergravity background reproduces faithfully  the breaking pattern $U(1)_R 
\to Z_{2N_c-N_f}\to Z_2$. 

Finally, let us  mention that in the special case $N_f=2 N_c$ we can reasonably claim that all the extra uncontrolled terms appearing along with (\ref{wp}) in the effective superpotential $\mathcal{W}'$ can be turned off, giving on the string side our simple solution (\ref{nf2ncsol}). As we show in section~\ref{nf2ncint}, this background has very special features which we relate to the scale invariance of the dual field theory. On the other hand, for $N_f<2N_c$, we could not find a way on the gravity side to turn off these extra terms, as the comparison between the solutions of section \ref{numericstudy} and of appendix \ref{appendixnf2nc} to the simple solution (\ref{nf2ncsol}) seems to confirm. Nonetheless, as all the rest of this section shows, the string dual  perspective 
of many strong-coupling effects,  remarkably agrees with 
expectations from the field theory described by 
(\ref{lagrangianon=1})-(\ref{wp}), and allows therefore to make  interesting predictions for its non-perturbative dynamics.

As a concluding remark, it would certainly prove very interesting to try to match with our approach, the results of the  careful 
analysis  done in \cite{Argyres:1996eh} on the vacuum structure of $\mathcal{N}=2$ SQCD with supersymmetry broken to $\mathcal{N}=1$ by a mass term for the adjoint superfield, which is closely related to the theory (\ref{lagrangianon=1}), (\ref{wp}).

\subsection{Wilson Loop and Pair Creation}
\label{sect: wilson}
\begin{figure}[htb] 
   \includegraphics[width=0.45\textwidth]{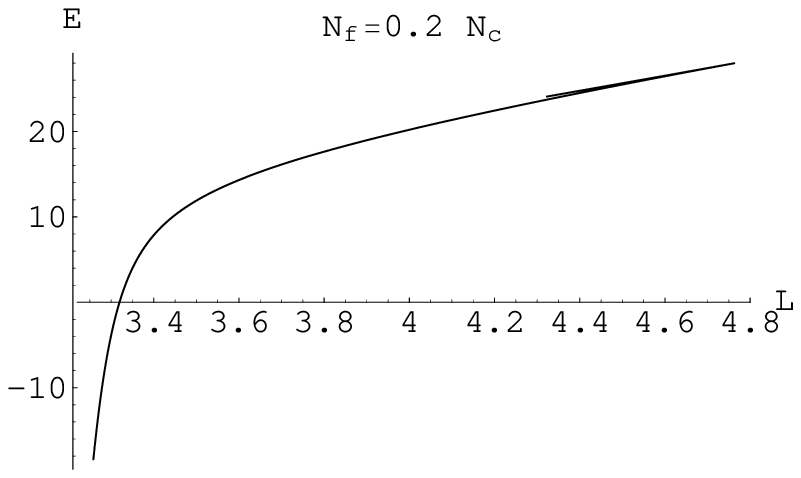} 
   \includegraphics[width=0.45\textwidth]{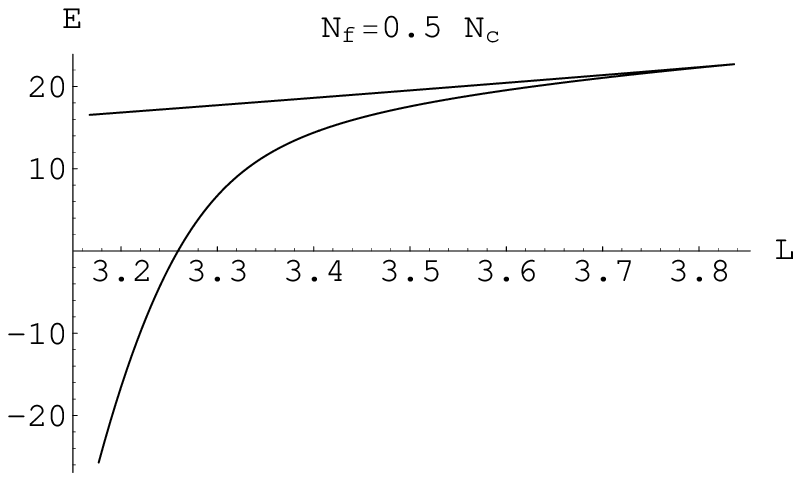} 
   \caption{Energy vs. length of the Wilson loop for 
   $\frac{N_f}{N_c}=0.2$ and $\frac{N_f}{N_c}=0.5$. Energies
    are in units of $\frac{e^{\phi_0}\sqrt{g_s N_c}}{\sqrt\a'}$, 
    whereas lengths are in units of $\sqrt{g_s N_c \a'}$ .There is a value
   of the integration constant $\rho_0$ for which the string reaches
   its maximum $L$ and $E$. Longer strings are not solutions of the Nambu-Goto
   action. We interpret it as string breaking due to quark-antiquark 
   pair creation.}
   \label{figwilson}
\end{figure}

One natural question is what happens to the gauge theory Wilson loop, now 
that we have massless flavors. In principle, as in QCD we should not 
observe an area law, but the SQCD-string should elongate until its tension 
is equal to the mass of the lightest meson, and then break. So, if we find that the 
(very massive) quarks that we introduce in the system feel each other  
up to a 
maximal distance only, this will be indication that we are seeing a 
phenomenon like pair creation\footnote{In 
\cite{Kirsch:2005uy}, screening due to backreacting flavors
was found as a change in the slope of Regge trajectories.}.
In order to study this, we will follow a 
very careful treatment to compute the Wilson loop in gravity 
duals \cite{Maldacena:1998im}  
developed in  \cite{Brandhuber:1998er} (for a good summary of the results, 
see pages 19-25 in \cite{Sonnenschein:1999if}). 
As usual, we propose that the Wilson loop for two 
non-dynamical quarks (strings stretching up to $\rho \to \infty$)
separated a distance $L$ in the gauge theory 
coordinates, should be computed as the action of a fundamental string that 
is parametrized by $t=\tau$, $x_1=\sigma$, $\rho=\rho(\sigma)$ 
\cite{Maldacena:1998im}. By 
 solving the Nambu-Goto action, one obtains a one-parameter family
of solutions depending on an integration constant which we will define
to be $\rho_0$ (the minimal $\rho$ reached by the string).
We convert to string frame the metric (\ref{nonabmetric42}) and
use (\ref{phif4}) and (\ref{rrhochange}). Then,
the length and energy (renormalized by subtracting the infinite masses
of the non-dynamical quarks) read:
\bear
L(\rho_0) &=& 2 \int_{\rho_0}^{\rho_1} 
e^{k(\rho)}
\frac{e^{\phi(\rho_0)}}{\sqrt{e^{2\phi(\rho)} - 
e^{2\phi(\rho_0)}}} d\rho \,,\rc
E(\rho_0) &=& 
\frac{1}{2\pi\a'}\left[
2 \int_{\rho_0}^{\rho_1} \frac{e^{2\phi(\rho)+k(\rho)}}
{\sqrt{e^{2\phi(\rho)} - 
e^{2\phi(\rho_0)}}} d\rho - 2 \int_{0}^{\rho_1} 
e^{\phi(\rho)+k(\rho)}  d\rho \,\right]
\label{length}
\eear
where $\rho_1$ is a cutoff
related to the mass of the heavy quarks that can be taken to
infinity smoothly. 
These integrals can be performed numerically. 
In order to do so, we have to fix the parameters $c_1$ and $c_2$ 
of the expansion (\ref{solutionnerar=0b}) in order to give
initial conditions for the equations (\ref{eq1}), (\ref{eq2}).
Imposing the large $\rho$ asymptotic behavior (\ref{expansionnearinfinity})
gives a relation between them. Moreover, we also 
fix $c_2 = 9c_1^2 / 16$ so that the domain wall tension scales with
$N_c - \frac{N_f}{2}$ (see section \ref{sect: dw}). However, this
condition is not essential for the qualitative behavior. 
The numerical result is reported
in Figure \ref{figwilson}. 
There is a finite value of $\rho_0$,
say $\rho_0^*$,
for which the solution reaches a maximum $L_{max}$ (at this point,
$\frac{dL}{d\rho_0}=\frac{dE}{d\rho_0}=0$). There is also a minimum value
of the length $L_{min}=\pi \sqrt{g_s N_c \alpha'}$. Our interpretation
is that $L_{min}$ is related to the KK scale at which the UV completion
sets in, and, therefore, the background loses predictivity.
For a range of $L_{min}<L<L_{max}$, there are two solutions of the Nambu-Goto
action. The physical one (the one of lower energy), corresponds
to $\rho_0 > \rho_0^*$ which in particular implies that the string
never approaches $\rho = 0$ and therefore it does not approach the curvature
singularity. 

As explained above, when one stretches a flux tube in
a confining theory with dynamical quarks, the tube must break at some
finite length when there is enough energy to create a light meson
which screens the color charge. 
The geometry can reproduce this
behavior (the breaking happens at $L_{max}$)
because the backreaction of the flavors has been taken into
account. The graphs qualitatively match those found on the lattice
in similar (non-supersymmetric) setups \cite{Bali:2005fu}.
Thus, our solution is showing that there is pair creation and giving a 
prediction for the lightest meson in terms of the parameters of the model.
The fact that $L_{min}$ and $L_{max}$ are roughly of the same order reflects
the known fact
that, when constructing the gravity dual of a confining theory,
 the KK scale and $\Lambda_{qcd}$ are of the same order and cannot
be separated within a weakly curved framework.

The discussion above
 should be compared with what happens in unflavored backgrounds, like 
(\ref{metricaa})-(\ref{RR}), the KS background \cite{Klebanov:2000hb} or 
the ones we present in section \ref{unflavoredcase}, where there is no 
possibility of extremizing the function $L(\rho_0)$.
However, notice that the behavior found above
is very similar to the one observed in the paper 
\cite{Evans} for a gravity dual of YM${}^*$ without flavors.
Such a background, like ours, presents a good singularity in the IR and
the authors of \cite{Kirsch:2005uy} considered the 
Wilson loop behavior unphysical
and related to the fact that the string wanted actually to fall in the
singularity. Nevertheless, there is an important difference between the two
models since $g_{tt}$ (and therefore the
QCD string tension) never vanishes in our setup.

\subsection{Instanton Action}
Let us compute the action of an instanton in the field theory. We will 
propose that a gauge theory instanton should be thought of 
as a euclidean D1-brane wrapping some contractible 
two-cycle in the geometry (\ref{nonabmetric}). The action 
of the D1-brane is:
\bea
S=  T_1\int d^2 z e^{-\phi}\sqrt{det g} + T_1\ i \int C_{(2)}=
\frac{8\pi^2}{g_{sqcd}^2} + i\theta_{sqcd}
\label{actiond1a}
\eea
where we have 
compared with the action of the instanton in order to get an expression for the 
four-dimensional gauge coupling and  theta angle.
Since $dF_{(3)} \neq 0$, it is not possible to define a two-form potential
such that $F_{(3)} = dC_{(2)}$. But if we consider the subspace given by $\tilde\theta=\theta, 
\, \tilde{\varphi}=2\pi-\varphi$ (and take
also large constant $\rho$) we can write, from (\ref{F3beinflav}):
\beq
F_{(3)eff}=\frac{N_c}{4} (2-x) \sin\theta d\theta \wedge d\varphi \wedge 
d\psi
\eeq
and therefore:
\beq C_{(2)eff}=(\psi-\psi_0)\,\frac{N_c}{4} (2-x)  \sin\theta d\theta \wedge 
d\varphi
\eeq
We now pick the two-cycle defined by \cite{{DiVecchia:2002ks}} :
\beq
\theta=\tilde{\theta}, \qquad \tilde{\varphi}=2\pi-\varphi, \qquad \psi= (2n+1)\pi
\label{cyclefavorito}
\eeq
So we can identify\footnote{We remind the 
reader that we have used that the tension of the D1-brane  
is $T_1=\frac{1}{2\pi\alpha' g_s}$ and we are using $\alpha'=g_s=1$ along 
our computations.}:
\ba
\theta_{sqcd}&=& \frac12 (2N_c- N_f)(\psi-\psi_0), \;\;,
\label{instcosas}\\
\frac{4\pi^2}{g_{sqcd}^2}&=& e^{2h} + \frac{e^{2g}}{4}(a-1)^2
\label{gtheta}
\ea
It is quite nice to see that the coupling constant satisfies the confining 
behavior we expect for $N_f<2 N_c$. Indeed, plugging in the expansions 
for the functions near the origin (\ref{solutionnerar=0b}), we see that 
even though the solution is 
singular, this singularity is not causing problems and the coupling 
grows at low energies. Conversely, in the UV, using the expansion in 
(\ref{expansionnearinfinity}), we observe that the 
coupling is asymptotically small.
We will further discuss the quantities 
written in (\ref{instcosas}), (\ref{gtheta}) in sections \ref{u1rbreak} and
\ref{betafflavor} respectively.

\subsection{$U(1)_R$ Breaking }
\label{u1rbreak}
In the previous subsection, we have identified a contractible two-cycle
on which we wrapped a euclidean D1-brane that we identified as an 
instanton. Since this cycle contracts to zero near the origin of the 
coordinates,  the partition function for this 
euclidean brane should be equal to one, up to a phase. For the original 
presentation of this argument see \cite{Klebanov:2002gr}.
The partition function of this instantonic string is
\beq
Z= (...) \exp[\frac{i}{2\pi}\int_{\Sigma_2} C_{(2)eff}]
\eeq
where $\Sigma_2$ is the cycle (\ref{cyclefavorito}) and (...) 
represents the Born-Infeld part of the action, see (\ref{actiond1a}). 
Under a 
shift in the $R$-symmetry $\psi \to \psi+2\epsilon$ 
(we write the factor of 2 since $\psi \in [0,4\pi)$ in order
to have $\epsilon \in [0,2\pi)$) we will 
have
\beq
Z\to Z[0] \exp[i \epsilon(2 N_c - N_f ) ]
\label{chisb}
\eeq
so, asking for invariance of $Z$ imposes that (this is equivalent
to requiring that the $\theta_{sqcd}$ defined in (\ref{instcosas})
is unchanged modulo $2\pi$ under shifts of $\psi$): 
\beq
\epsilon= \frac{2 \k \pi}{2N_c - N_f} \;\qquad\k=1,...,2N_c-N_f
\eeq
So we see that  quantum effects select a discrete subset $ Z_{2N_c- N_f}$ out of the possible rotations of the angle $\psi$ associated with the $R$-symmetry, that is consequently 
broken according to $U(1)_R \to Z_{2N_c- N_f}\to Z_2$, with the last step in 
the chain of breakings  due to the fact that the full background is only 
invariant under $\psi\to \psi+ 2\pi$. 
This last step in breaking is understood as the 
formation of a condensate, to which the functions $a(\rho)$, $b(\rho)$ 
should be dual to \cite{Apreda:2001qb}. 

Notice that when $N_f=2 N_c$ the breaking above does not happen and 
the $U(1)_R$ symmetry preservation indicates a theory with higher 
invariance, in this 
case, scale invariance: the $U(1)_R$ anomaly 
is in the same anomaly multiplet as the beta function so, when $N_f=2N_c$ 
the coupling in (\ref{gtheta}) should not run. We study this
 point in more detail in section \ref{nf2ncint}.

\subsection{Beta Function}
\label{betafflavor}
Let us start by summarizing the result of the wilsonian beta function for 
SQCD 
\beq
\beta_g= -\frac{g_{sqcd}^3}{16\pi^2} (3 N_c - N_f (1-\gamma_0)),
\label{betafunctionwilson} 
\eeq
where $\gamma_0$ is the anomalous dimension of the quark (and anti-quark) 
superfields, while for the coupling of the 
quartic superpotential we have
\be
\beta_h \sim (1+ 2 \gamma_0).
\ee

An approach to the calculation of the beta function of SYM from a gravity 
dual is presented in \cite{DiVecchia:2002ks}. Without entering into the many 
possible discussions about the validity of this computation (more on 
this in section \ref{betaunflavoredsection}), let us just 
repeat the steps in \cite{DiVecchia:2002ks} for our case.
The coupling is defined in (\ref{gtheta}) and we define the radius-energy 
relation near the UV of the field theory (at large values of $\rho$) as
\beq
(\frac{\Lambda}{\mu})^3\sim a(\rho)\sim b(\rho) \sim e^{-2\phi}
\label{uvir}
\eeq
where with the symbol $\sim$ above we mean that all these definitions, 
when taken at large values of the radial coordinate, give the same result. 
The idea of the first two definitions is that these functions
should be dual to the gaugino condensate or other condensate as proposed 
in \cite{Apreda:2001qb}. On the other hand, one might consider that the 
exponential of the dilaton is related to the strong coupling scale as 
proposed in 
\cite{Maldacena:2000yy, Hori:2002cd}.
Considering only the large $\rho$ expansion of the coupling defined in 
(\ref{gtheta}) and the functions in 
(\ref{uvir}), we have
\beq
\frac{4\pi^2}{g_{sqcd}^2}\sim N_c (1-\frac{x}{2})\rho+....,\;\; \qquad
\log\frac{\mu}{\Lambda}\sim \frac{2}{3}\rho \;\;
\eeq
So, a straightforward computation leads us to
\beq
\beta=\frac{d g_{sqcd}}{d\log\frac{\mu}{\Lambda}}= -\frac{3 g_{sqcd}^3}
{32 \pi^2}(2 N_c 
- N_f)+...
\label{betafalsa}
\eeq
where the last equality should be understood as valid only in an expansion 
for large values of $\rho$. This shows that in the 
regime described by the large $\rho$ region of the background, 
the anomalous dimension of the quark superfield is 
$\gamma_0=-1/2$ (and therefore $\beta_h = 0$).
Once again, this computation should  be severely  criticized regarding the 
approximations and assumptions that are done (see section 
\ref{unflavoredcase} for a different sort of criticism to this approach), 
but we 
just wanted to spell  it here to show that the coefficient $2N_c - N_f$ 
appears. This also reinforces the fact that 
for the case in which $N_f=2N_c$ 
something special is happening (and we know the solution changes 
qualitatively for that case). It would be interesting to understand
what happens when $N_f > 2N_c$.

\subsection{Domain Walls}
\label{sect: dw}
The domain walls of ${\cal N}=1$ SYM associated to the IR spontaneous
breaking of the R-symmetry $\ZZ_{2N_c}\rightarrow \ZZ_2$ correspond to
D5-branes wrapping the finite $S^3$ at $\rho =0$ \cite{Maldacena:2000yy}.
Their associated worldvolume $\kappa$-symmetry matrix is:
\beq
i\Gamma_{x_0x_1x_2123}\epsilon^* = \epsilon\,
\eeq
As pointed out in \cite{Canoura:2005fc},
this projection commutes with those in (\ref{killingproj})
if and only if
${\cal A}=0$, {\it i.e}, when
the brane is placed at $\rho=0$.
So from the brane point of view, one finds, consistently, that these 
objects are one half BPS. 

In the flavored theory we are dealing with, we have argued that there is
a also a spontaneous IR breaking 
$\ZZ_{2N_c-N_f}\rightarrow \ZZ_2$ and therefore
one would also expect to have domain walls. It is easy to see, using the
expansion (\ref{solutionnerar=0b}) and the expressions in appendix
\ref{appendixa} that it is also true in our flavored case that 
$\lim_{\rho\to 0} {\cal A}=0$, so the same kind of embedding at $\rho =0$
preserves again half of the supersymmetry. Notice also that the action
for the domain wall D5-brane  wrapping the three-cycle parameterized
by $\tilde\theta,\tilde\varphi,\psi$ at constant $\rho$ is:
\beq
E_{D5} \propto T_5\ e^{2\phi+2g+k} \propto 
(g_s N_c)^\frac12  N_c \, e^{2\phi_0}
\frac{2(2-x)\sqrt c_2}{3 c_1} (1 + 2\rho^2 +...)
\label{dwenergy}
\eeq
and it has
indeed a minimum at $\rho=0$ (we have used eq (\ref{solutionnerar=0b})
and restored the dependence on $g_s$). Also, 
despite $e^{2g}$ and $e^{2k}$ tend to infinity and zero respectively,
the relevant combination 
(\ref{dwenergy}) goes to a constant at $\rho=0$, rendering a finite tension 
for the domain wall (its precise value would depend on the integration
constants $c_1,c_2$ and could be used to fix them).
Thus, we would like to stress that, even if at $\rho =0$ we are precisely 
on top of the 
curvature singularity, it is appealing to find sensible results from
these simple computations.
\subsection{Seiberg Duality}
\label{seibergsubsection}
A remarkable feature of many $\mathcal{N}=1$ gauge theories that 
include matter transforming in the fundamental representation of the 
gauge group (or even in the bifundamental representation for a quiver 
gauge 
theory) is Seiberg duality \cite{Seiberg:1994pq}. 
In its simplest form, this can be stated as saying that an 
$\mathcal{N}=1$ $SU(N_c)$ gauge theory with $N_f$ quarks 
and  an $\mathcal{N}=1$ $SU(N_f-N_c)$ gauge 
theory with $N_f$ quarks and a gauge-singlet meson have the same IR physics. When the theory contains a quartic superpotential for the fundamental fields, the duality is a bit more complicated \cite{Strassler:2005qs} but nonetheless it retains the same relations between the number of colors and flavors of the two Seiberg-dual theories as in the simpler case.

In this section we will argue that Seiberg 
duality appears in our construction in a remarkably simple and elegant way: it corresponds to switching between two alternative ways of writing the background\footnote{A first hint of what we will develop for the non-abelian background can be easily gained in the simpler abelian case by looking at the way the value of the integral of $F_{(3)}$ (\ref{new3form}) changes when it is evaluated over the 3-sphere $(\psi, \theta, \varphi)$ or $(\psi, \tilde{\theta}, \tilde{\varphi})$.}.

The central point on which the realization of Seiberg duality in 
our model is based, is that the internal space of 
our background has $S^2\times S^3$ topology, but the choice to use either the 
space spanned by the pair $(\theta, \varphi)$ or 
by  $(\tilde{\theta}, \tilde{\varphi})$ as the $S^2$ base  of the $S^3$ is 
(at least at first sight)  arbitrary.
In fact, let us take the metric and $F_{(3)}$  flux  to be
written as, from 
(\ref{nonabmetric42}),
\be\label{ds2 for Seib}
ds^2=e^\frac{\phi}{2}\left( dx^2_{1,3}+dr^2+e^{2h}(\mathfrak{e}_1^2+\mathfrak{e}_2^2)+\frac{e^{2g}}{4}\left((\tilde{\omega}_1+a(r) \mathfrak{e}_1)^2+(\tilde{\omega}_2+a(r) \mathfrak{e}_2)^2\right) +\frac{e^{2k}}{4}\hat{\omega}_3^2 \right)
\ee
and from (\ref{F3unflav}) and (\ref{flav3form})
\be\label{F3 for Seib}
\begin{split}
F_{(3)}=\frac{N_c}{4}&\left\{-\tilde{\omega}_1\wedge\tilde{\omega}_2\wedge\hat{\omega}_3-b(r)(\mathfrak{e}_1\wedge\tilde{\omega}_2-\mathfrak{e}_2\wedge\tilde{\omega}_1)\wedge\hat{\omega}_3+\right.\\ &\left.-b'(r) dr\wedge(\mathfrak{e}_1\wedge\tilde{\omega}_1+\mathfrak{e}_2\wedge\tilde{\omega}_2)-(1-x)\mathfrak{e}_1\wedge \mathfrak{e}_2\wedge \hat{\omega}_3\right\}
\end{split}
\ee
where the functions $\phi$, $a$, $g$, $h$ and $k$ are numerically evaluated in section  \ref{numericstudy}, $b$ is given by~(\ref{bghexpression})
\be\label{b for Seib}
b(\rho)=(2-x)\frac{\rho}{\sinh 2\rho}
\ee
and for convenience we have defined the following 
1-forms
\begin{align}
&\mathfrak{e}_1=d\theta& &\tilde{\mathfrak{e}}_1=d\tilde{\theta}\nonumber\\
&\mathfrak{e}_2=-\sin\theta \,d\varphi & &\tilde{\mathfrak{e}}_2=-\sin\tilde{\theta} \,d\tilde{\varphi}\nonumber\\
& \omega_1=\cos\psi \,\mathfrak{e}_1-\sin\psi \,\mathfrak{e}_2 & &\tilde{\omega}_1=\cos\psi\, \tilde{\mathfrak{e}}_1-\sin\psi\, \tilde{\mathfrak{e}}_2 \label{new 1-forms}\\
&\omega_2=-\sin\psi\, \mathfrak{e}_1-\cos\psi \,\mathfrak{e}_2 & &\tilde{\omega}_2=-\sin\psi\, \tilde{\mathfrak{e}}_1-\cos\psi \,\tilde{\mathfrak{e}}_2\nonumber\\
&\omega_3=d\psi+\cos\theta \,d\varphi & & \tilde{\omega}_3=d\psi+\cos\tilde{\theta} \,d\tilde{\varphi}\nonumber\\
&\hat{\omega}_3=\omega_3+\cos\tilde{\theta} \,d\tilde{\varphi}=\tilde{\omega}_3+\cos\theta \,d\varphi \nonumber
\end{align}

We know that the number of colors of the dual theory can be evaluated by integrating the three-form flux over the internal $\tilde{S}^3$ spanned by the angles $\tilde{\theta}$, $\tilde{\varphi}$, $\psi$,  as in (\ref{quantcond})
\be
N_c T_5=\frac{1}{2 \kappa_{(10)}}\int_{\tilde{S}^3}F_{(3)}
\ee
whereas the number $N_f$ of flavors in the gauge theory is determined by the failing of the Bianchi identity for $F_{(3)}$, equation (\ref{newdF2}).

It is easy to show that the 1-forms we defined in (\ref{new 1-forms}) satisfy the following identities\footnote{We thank Agostino Butti for bringing these identities to our attention while describing  to us the realization of the $SU(2)\times SU(2)$ isometry of the baryonic branch backgrounds of \cite{Butti:2004pk}.}
\begin{align}
&\omega_1^2+\omega_2^2=\mathfrak{e}_1^2+\mathfrak{e}_2^2& & \tilde{\omega}_1^2+\tilde{\omega}_2^2=\tilde{\mathfrak{e}}_1^2+\tilde{\mathfrak{e}}_2^2\nonumber\\
&\mathfrak{e}_1\, \tilde{\omega}_1+\mathfrak{e}_2\,\tilde{\omega}_2=\omega_1\,\tilde{\mathfrak{e}}_1+\omega_2\,\tilde{\mathfrak{e}}_2\nonumber\\
&\omega_1\wedge\omega_2=- \mathfrak{e}_1\wedge \mathfrak{e}_2 & & \tilde{\omega_1}\wedge\tilde{\omega_2}=- \tilde{\mathfrak{e}_1}\wedge \tilde{\mathfrak{e}_2}\\
&\tilde{\mathfrak{e}}_1\wedge \omega_2 -\tilde{\mathfrak{e}}_2\wedge\omega_1=-(\mathfrak{e}_1\wedge \tilde{\omega}_2-\mathfrak{e}_2\wedge\tilde{\omega}_1)& & \tilde{\mathfrak{e}}_1\wedge \omega_1 +\tilde{\mathfrak{e}}_2\wedge\omega_2=-(\mathfrak{e}_1\wedge \tilde{\omega}_1+\mathfrak{e}_2\wedge\tilde{\omega}_2)\nonumber
\end{align}
We can use these expressions to rewrite the metric and $F_{(3)}$  in a new form. It is very important to notice that even though these two quantities will look quite different from above, what we do is only a rewriting of (\ref{ds2 for Seib}) and (\ref{F3 for Seib}). For the metric we obtain then
\be\label{new ds2 S}
ds^2=e^\frac{\phi}{2}\left( dx^2_{1,3}+dr^2+e^{2\bar{h}}(\tilde{\mathfrak{e}}_1^2+\tilde{\mathfrak{e}}_2^2)+\frac{e^{2\bar{g}}}{4}\left((\omega_1+\bar{a}(r) \tilde{\mathfrak{e}}_1)^2+(\omega_2+\bar{a}(r) \tilde{\mathfrak{e}}_2)^2\right) +\frac{e^{2k}}{4}\hat{\omega}_3^2 \right)
\ee
and for the three-form flux
\be\label{new F3 S}
\begin{split}
F_{(3)}=\frac{\bar{N}_c}{4}&\left\{-\omega_1\wedge\omega_2\wedge\hat{\omega}_3-\bar{b}(r)(\tilde{\mathfrak{e}}_1\wedge\omega_2-\tilde{\mathfrak{e}}_2\wedge\omega_1)\wedge\hat{\omega}_3+\right.\\ &\left.-\bar{b}'(r) dr\wedge(\tilde{\mathfrak{e}}_1\wedge\omega_1+\tilde{\mathfrak{e}}_2\wedge\omega_2)-(1-\bar{x})\tilde{\mathfrak{e}}_1\wedge \tilde{\mathfrak{e}}_2\wedge \hat{\omega}_3\right\}
\end{split}
\ee
where in (\ref{new ds2 S})  we have defined
\be\label{barred}
\begin{split}
&e^{2\bar{h}}=\frac{e^{2g}}{4}\left(1-\frac{a^2e^{2g}}{4 e^{2h}+a^2 e^{2g}}\right)\\
&e^{2\bar{g}}=4 e^{2h}+a^2 e^{2g}\\
&\bar{a}=a\, \frac{e^{2g}}{4 e^{2h}+a^2 e^{2g}}
\end{split}
\ee
and in (\ref{new F3 S}) we have set
\be\label{barred2}
\bar{N}_c=N_f-N_c\qquad\qquad \bar{x}=\frac{x}{x-1}\qquad \qquad\bar{b}(r)=\frac{b(r)}{1-x}
\ee
in order for the new expressions of the metric and the three-form flux to 
have again the same apparent form as in the original versions 
(\ref{ds2 for Seib}) and (\ref{F3 for Seib}). By using the 
definition of $x=N_f/N_c$ and the expression for $\bar{N}_c$, it is 
immediate to show that $\bar{N_f}=N_f$. In the same way, it follows 
from (\ref{barred2}) and (\ref{b for Seib}) that
\be
\bar{b}(\rho)=(2-\bar{x})\frac{\rho}{\sinh 2\rho}
\ee
that is, exactly the same form as in (\ref{b for Seib}) with the new value of the $N_f/N_c$ ratio.

Because of the way the metric looks now (\ref{new ds2 S}), it seems natural to identify  as the three-sphere of the internal manifold, the one spanned by $\theta$, $\varphi$ and $\psi$. Therefore evaluating the integral of $F_{(3)}$ over this $S^3$, we find that the number of colors of the field theory dual to the background (\ref{new ds2 S})-(\ref{new F3 S}) is $\bar{N}_c=N_f-N_c$, whereas the number of flavors is still $N_f$.  This is very interesting because this is exactly the way the rank of the gauge group changes under Seiberg duality.

Let us now compare the geometries in the IR, where Seiberg duality tells us that the two field theories are the same. By using the small $\rho$ expansions of section \ref{asymptoticsols}, we can derive the behavior of $\bar{h}$, $\bar{g}$ and $\bar{a}$ near the origin
\be\label{SeibergIR}
\begin{split}
&e^{2\bar{h}}=e^{2h}-\frac{4}{3}(2-x)\rho^4+\ldots\\
&e^{2\bar{g}}=e^{2g}+\frac{4}{3}(2-x)\rho^2+\ldots\\
&\bar{a}=a-2c_1 \rho^3+\ldots
\end{split}
\ee
and remarkably we see that the newly defined functions correspond to their original homonyms up to the third term in the small $\rho$ series expansions. This is the holographic dual of the statement that the Seiberg pair theories have the same IR behavior.

This result might look a little puzzling at first sight: in (\ref{new ds2 S}) and (\ref{new F3 S}) we have only written in a different way the metric and three-form flux but at the same time claim that the background written in this way corresponds to a different theory from the one that is dual to the (same) original solution (\ref{ds2 for Seib}) and (\ref{F3 for Seib}). 
We believe the right way to look at this puzzle is
in terms of the dictionary we use to translate the geometry data into field theory predictions. So in a sense,  we have two different dictionaries we can use on the same background, corresponding to the two ways of identifying the $S^3$ inside the internal space.

This is a remarkable result: just by rewriting the flavored background in a different form, we can capture the most relevant features of Seiberg duality for theories with flavors. In fact in a simple and elegant way we can reproduce the change in the rank of the color gauge group while leaving the number of flavors untouched, and we can show that the IR dynamics of the two theories is exactly the same thanks to (\ref{SeibergIR}). It would be very interesting to study how information  about the Seiberg-dual field theory is encoded into string theory in this setup, and which predictions this allows us to make.

\section{Features of the $N_f= 2N_c$ Solution}
\label{nf2ncint}
We now discuss the 
gauge theory interpretation of the
particularly simple $N_f= 2N_c$ solution (or $x=2$ in the notation 
introduced 
above), presented in section \ref{nf2nc}, see eq. (\ref{nf2ncsol}). For 
this 
solution (and only for
this one), the effective four-dimensional Yang-Mills coupling,
see (\ref{gtheta}), is constant (notice  that the
coupling is constant regardless of the cycle used to define it)
\beq
\frac{4\pi^2}{g_{sqcd}^2} = e^{2h} + \frac{e^{2g}}{4} (a -1)^2 =
N_c \frac{4}{\xi (4-\xi)}\,\,.
\label{constantcoupling}
\eeq
It is also the only solution for which the $U(1)_R$ symmetry is preserved.
There is not anomalous UV breaking (we can apply the discussion of
section \ref{u1rbreak}, eq (\ref{chisb})), while 
there is not spontaneous IR breaking since $a=b=0$. 
We would therefore not expect to have domain walls and this is in
fact the case since its tension as defined in (\ref{dwenergy}) would vanish
($e^{2\phi} \to 0$ when $r \to -\infty$).
We think that these two features
are signaling an IR conformal fixed point in the effective four-dimensional
theory. 

That this happens only for $N_f = 2N_c$ is in precise agreement
with the field theory interpretation of section \ref{sect: gauge}.
Notice that, unlike in pure ${\cal N}=1$ SQCD, there should
not be a conformal window for $N_f\leq 2N_c$. In fact, having an IR effective quartic
superpotential for the quarks (\ref{wp}) fixes their R-charge to
$\frac12$. The NSVZ beta-function, which can be written in terms of the
(IR) R-charges of the fields can consequently only vanish
when $N_f = 2 N_c$. 
This can also be seen using the wilsonian beta-functions
for the gauge coupling $\beta_g \sim (3N_c - N_f (1-\gamma_0))$
and the quartic coupling $\beta_h \sim 1 + 2\gamma_0$ which
 vanish simultaneously if and only if  $\gamma_0 = -\frac12$ and 
 $N_f = 2N_c$  ($\gamma_0$ is the anomalous dimension of the quark superfield).

Nevertheless, we cannot claim that the solution (\ref{nf2ncsol}) is 
a faithful dual
to a conformal four-dimensional
theory. Since there is no $AdS_5$ space, the conformal
group $SO(2,4)$ does not show up in the gravity side and also there
is no superconformal enhancement of the supersymmetry.
We believe that the explanation resides on the fact that we start with 
a six-dimensional theory living on the D5-branes compactified with a twist 
on an $S^2$ and then add flavor branes. The four-dimensional
theory  never completely decouples from the Kaluza-Klein degrees 
of freedom and this spoils the would-be conformal symmetry. Indeed, even 
though the parameters and couplings might be tuned so that the massless 
modes are in a ``conformal phase", the KK modes spoil this symmetry and the 
background reflects it by not allowing an $AdS_5$ to appear. To add to 
this interpretation we can, for instance, make a dilatation of the four 
dimensional 
spacetime $x_i\to \lambda x_i$, that can be compensated (we take string 
frame)
by a shift  $r \to r -2 \log \lambda$, which should be regarded
as a rescaling of the energy scale (see (\ref{uvir})). This is not
an isometry of the metric since the internal space gets 
rescaled, thus rescaling the KK scale.
However, we find appealing that the geometry is
not deformed in any other way, since the functions $h,g,k,a$ do not
depend on $r$. Notice that this happens only for this simple  $N_f = 2N_c$ 
solution.

In any case, one could not expect to find a weakly curved $AdS_5$
since this would imply that the coefficients of the conformal anomaly are equal $a=c$
at leading order in $N_c$
\cite{Henningson:1998gx}. Since our theory has $N_f \sim N_c$,
one has $a\neq c$ at leading order \cite{Christensen:1978gi}, ruling out
the possibility of having a faithful supergravity dual.
Nevertheless, we think it is very nice that some features 
can be matched in this setup.

Finally, let us conjecture on the gauge theory meaning of the parameter
$\xi$. ${\cal N}=1$ SQCD with $N_f = 2N_c$ and a quartic superpotential
has a one complex-dimensional family of marginal deformations
in the two complex-dimensional parameter space of the gauge coupling
and of the coupling of the quartic term (see \cite{Strassler:2005qs}).
 The parameter $\xi$ changes the volume of the
spheres (\ref{nf2ncvalues}) so it is natural to think that it would change
the masses of the adjoints coming from the 
KK modes and, indeed, modify the
value of the coupling at the ``conformal fixed point" 
(\ref{constantcoupling}). Thus, it may be a parameter 
signaling this conformal line.

Let us see this more precisely. In this geometry, we could define 
couplings (not necessarily the SQCD coupling in (\ref{constantcoupling})), 
to be proportional to the inverse volume of two-cycles
\beq
\frac{1}{{g_4}^2} \sim \frac{Vol \,\Sigma_2}{g_6^2}. 
\label{defrara}
\eeq
Here $g_6^2= \alpha' g_s N$ is the coupling of the little string 
theory. So, different 
definitions would imply the choice of different two-cycles in the 
geometry, like
\beq
\frac{1}{\hat{g}^2}= \frac{Vol\, \hat{S}^2(\theta,\varphi)}{g_6^2},\;\;\; 
\frac{1}{\tilde{g}^2}= \frac{Vol \,
\tilde{S}^2(\tilde{\theta},\tilde{\varphi})}{g_6^2},\;\;\;
\frac{1}{g_{sqcd}^2}= \frac{Vol\, S^2(\theta, 
\tilde{\theta},\varphi, \tilde{\varphi})}{g_6^2}
\label{diffrentcycles}
\eeq
Let us be more precise and define the cycles above. The first two cycles 
are given by 
the effective geometry obtained from (\ref{nf2ncsol}) when imposing 
that all the coordinates 
are constant except ($\theta,\varphi$), or 
$(\tilde{\theta},\tilde{\varphi})$ in the case of 
$\hat{S}^2(\theta,\varphi)$ and 
$\tilde{S}^2(\tilde{\theta},\tilde{\varphi})$ respectively. The last cycle 
is defined like in (\ref{cyclefavorito}) and gives the coupling in 
(\ref{constantcoupling}).

So, computing these ``couplings" one gets
\bea
& & \frac{1}{ \hat{g}^2 N_c}\sim  \frac{1}{\sqrt{\xi}} E[\pi, 
\frac{\xi-4}{\xi}]\nonumber\\
& & \frac{1}{ \tilde{g}^2 N_c}\sim  \frac{1}{\sqrt{4-\xi}} E[\pi, 
\frac{\xi}{\xi-4}]\nonumber\\
& & \frac{1}{ g_{sqcd}^2 N_c}\sim  \frac{4}{\xi(4-\xi)}
\label{couplingsdif}
\eea
Where $E[\pi,k]$ is the second kind Elliptic function. Now, let us observe 
something quite nice. If we interchange the cycles
$\hat{S}^2(\theta,\varphi) \leftrightarrow
\tilde{S}^2(\tilde{\theta},\tilde{\varphi})$, this is 
equivalent to the interchange $\xi \leftrightarrow (4-\xi)$, or 
$\hat{g}\leftrightarrow \tilde{g}$. The 
Seiberg-like duality defined in section \ref{seibergsubsection} does 
precisely this job. So, we 
have a line of fixed points,  parametrized by $\xi$ (solutions where the two-spheres have 
constant radius),  which is reflected under Seiberg duality.  This is what is known to happen in a field theory like 
the one discussed in section \ref{gaugethnonsing}, that is ${\cal N} =1$ 
SQCD plus a quartic potential, when the theory is at its conformal point 
($N_f=2 N_c$, and the anomalous dimension of the quark superfields is 
$\gamma_0=-\frac{1}{2}$). For a review of these results, see Strassler's 
beautiful lecture notes \cite{Strassler:2005qs}.
Indeed, following \cite{Strassler:2005qs}, we can see that the line of 
fixed 
points has a self-dual point (under Seiberg duality), that in our 
case 
is $\xi=2$. Then the duality, in our case the interchange $\xi\leftrightarrow 
(4-\xi)$ or the interchange of two cycles, interchanges points on the 
fixed 
line that lie on different 
sides of the self-dual point. We can see that this is happening in 
eqs. (\ref{couplingsdif}), for 
a holomorphic coupling 
that should be a combination of $g_{sqcd}$ and the quartic coupling.

This suggests that as for the coupling 
(\ref{constantcoupling}), it might be possible to find another two-cycle 
whose volume is inversely proportional to the quartic coupling.
This is a very nice check that points to the consistency of 
our approach and interpretation.

Finally, we would like to stress two very interesting facts.
First, let us compute the Wilson loop (see section \ref{sect: wilson},
eq. (\ref{length}))
for this particular case.
It can be seen that 
for any value of $\rho_0$, one obtains that the quark 
antiquark pair has energy $E_{q\bar{q}}=0$ and
that the separation of the pair is always $L=\pi \sqrt{N_c g_s \a'}$. 
Following the interpretation of
 section \ref{sect: wilson}, below this length
this computation
is unphysical due to the UV
completion.
For $L>\pi \sqrt{N_c g_s \a'}$, the configuration in 
which the ${q\bar q}$ are joined by
a string is never preferred to having two independent 
particles (notice that in this case, the QCD string tension vanishes
$g_{tt} ( r = -\infty) =0$).
We interpret this observation as the existence of total screening, the 
``mesons" have zero energy, in agreement with the fact that this theory does not 
dynamically generate a scale.

The second observation concerns the black hole solution in
(\ref{bhflav}). We can compute the viscosity of the dual quark-gluon 
plasma following the approach developed in \cite{Kovtun:2003wp}. Since our 
background (\ref{bhflav}) satisfies the hypothesis of the theorem in 
\cite{Buchel:2003tz}, we 
can see that the relation between the shear viscosity and the temperature 
is of the predicted form ${\cal D} =\frac{1}{4\pi T}$.
It may be interesting to study other hydrodynamical 
quantities using the dual solution (\ref{bhflav}). This is a very simple 
background dual to a field theory with adjoints, fundamentals and at 
finite temperature.

Finally, to close this section, we would like to call the  
reader's attention to the  solution presented in appendix 
\ref{appendixnf2nc}. There we described a numerical solution for the case 
$N_f= 2N_c$, that appears after a careful analysis of the BPS eqs. 
Interesting features of this solution are that it has a behavior 
similar to the flavored case (with $N_f\neq 2N_c$), asymptotes in 
the UV to 
the simple solution presented in eq.~(\ref{nf2ncsol}) and the fact 
that the solution is Seiberg-duality invariant. We do not know how to 
interpret this new solution from a dual field theory viewpoint. It seems 
to behave as a flow from a fixed point where a relevant operator is being 
inserted, breaks ``conformality" and goes down to an IR theory similar to 
the one described by the solutions with $N_f<2N_c$. 
We leave this interesting problem for future study.

\section{An Alternative Approach: Flavors and Fluxes} \label{sec: fluxes}
\setcounter{equation}{0}

As we have seen in section \ref{flavnonab}, the addition of many flavor 
branes to the MN background is encoded in the non-closure of the three-form RR flux $F_{(3)}$. When the flavor branes are smeared over their transverse directions, the failing of the Bianchi identity is proportional to the volume form of the transverse $S^2\times S^2$
\be
dF_3=\frac{N_f}{4}\sin\theta \sin\tilde{\theta}\;d\theta\wedge d\varphi\wedge d\tilde{\theta}\wedge d\tilde{\varphi}.
\ee
In sections \ref{singularsolution} and \ref{flavnonsing} we shown that 
this effect gives rise to a background which has a uniform distribution of $F_{(3)}$ charge along the two two-spheres. As we have shown in the preceding sections this procedure works very nicely and gives rise to a remarkable matching between  gauge theory expectations and string predictions.

Nonetheless, we think it could also be very interesting to 
approach the same problem from a  complementary approach. 
In this section we sketch the general idea and some basic 
features of this possible new scenario, but setting up and 
solving the whole problem  would require a much deeper  
development and thorough discussion, which we leave as a 
possibility for  the future.

In this new approach, the failing of the Bianchi identity for the three-form $F_{(3)}$ is achieved  through non-trivial fluxes only. Indeed, we know from supergravity that the presence of an NS three-form background deforms the definition of the field strengths of the RR fluxes. Let us suppose to turn on some axion $C_{(0)}$ and NS flux $H=dB_{(2)}$, then the field strength $F_{(3)}$ is defined by
\be
F_3=dC_{(2)}-C_{(0)} H
\ee
and we see that if the axion is non-constant we can have a non-trivial Bianchi identity for $F_{(3)}$
\be
dF_3=-dC_{(0)}\wedge H=-F_{(1)}\wedge H
\ee
Notice that in this case the decomposition of $F_{(3)}$ in an exact and non-exact part is natural,  and therefore the definition of $C_{(2)}$ is unambiguous (up to gauge transformations, obviously).

 In the preceding sections we have always assumed that the susy 
transformations of the flavored background are exactly the same as those 
of the MN solution \cite{Maldacena:2000yy, Nunez:2003cf}. This is due
to the fact that we are adding to the MN solution an (infinite) set of 
BPS branes, which by definition preserve the same supersymmetries as the 
original background. But when we look at the flavoring of the background 
from the flux side, this could fail, and in fact, this is what one usually 
expects when turning on new fluxes in a background. The only condition 
we think it is still sensible to impose, is that the background can 
accommodate supersymmetrically embedded branes.

Regarding the geometrical properties of the 
six-dimensional internal manifold. There are two possibilities to 
build backgrounds with fluxes that preserve four-dimensional $\mathcal{N}=1$ 
supersymmetry: $SU(3)$-structure  manifolds \cite{GMPT} and a particular 
class of $SU(2)$-structure manifolds \cite{SU(2)}. The  unflavored MN 
background the internal manifold has $SU(3)$-structure \cite{GMPT} and 
since we can most likely think of  the flavored background as a 
deformation of the MN solution, it seems reasonable to 
assume that its internal part is an $SU(3)$-structure manifold as well.

Starting from these two assumptions, we can write the spinor ansatz for the supersymmetry transformations. As usual for solutions with four-dimensional Lorentz invariance, we split the IIB ten-dimensional spinors in their four-dimensional and internal six-dimensional parts
\be
\begin{split}
&\epsilon_1=\zeta_+\otimes \eta_+^{1}+\zeta_-\otimes\eta_-^1\\
&\epsilon_2=\zeta_+\otimes \eta_+^{2}+\zeta_-\otimes\eta_-^2
\end{split}
\ee
where $\eta^1$ and $\eta^2$ are spinors of the internal manifold. Notice that $\eta_-^i=(\eta_+^i)^*$. Since $SU(3)$-structures allow only for one globally-defined spinor $\eta$, we need to take $\eta^1$ and $\eta^2$ to be proportional
\be\label{eta}
\eta_+^1=\mathfrak{a}\eta_+\qquad\qquad \eta_+^2=\mathfrak{b}\eta_+
\ee

A necessary condition to have consistent supersymmetric embeddings of branes in an $SU(3)$-structure manifold, is that  the two spinors $\eta^1$ and $\eta^2$  have the same norm, $\|\eta^1\|^2=\|\eta^2\|^2$ \cite{MS}, which requires that the functions $\mathfrak{a}$ and $\mathfrak{b}$ in (\ref{eta}) satisfy
\be\label{embcondbis}
\frac{\mathfrak{a}}{\mathfrak{b}}=e^{i\tilde{\phi}}\qquad \qquad \frac{\mathfrak{a}}{\mathfrak{b}^*}=e^{i\tau}
\ee
with $\phi$ and $\tau$ real functions of the internal coordinates. For the smeared configuration we consider in this paper, symmetry imposes that they only depend on the radius of the internal manifold. Notice that the  spinor  $\eta$ is defined up to a phase, and therefore $\tau$ in (\ref{embcondbis}) can be arbitrarily gauge fixed by this symmetry \cite{GMPT}.  On the contrary $\phi$ does not depend on the gauge choice.

Instead of $\mathfrak{a}$ and $\mathfrak{b}$ we can define $\alpha$ and $\beta$
\be
\alpha=\mathfrak{a}+i\mathfrak{b}\qquad\qquad\beta=\mathfrak{a}-i\mathfrak{b}
\ee
in such a way that when we define $\epsilon=\epsilon_1+i\epsilon_2$, we get
\be\label{eps}
\begin{split}
&\epsilon=\alpha\,\zeta_+\otimes \eta_+ + \bar{\beta}\, \zeta_-\otimes\eta_- \\
&\epsilon^*=\beta\,\zeta_+\otimes \eta_+ + \bar{\alpha}\, \zeta_-\otimes\eta_-
\end{split}
\ee
In terms of $\alpha$ and $\beta$ the first condition in (\ref{embcondbis}) reads
\be\label{embcond}
\frac{\alpha}{\beta}=-i\cot{\frac{\tilde{\phi}}{2}}
\ee

At this point, one should write a smart ansatz for the metric and 
background fluxes, and then the $SU(3)$-structure techniques developed 
in \cite{GMPT, SU(3)} would allow to derive first order differential 
equations describing the solution. This is beyond the scope of the 
present work.


\section{ The Unflavored Case $N_f=0$: Deformed Solutions}
\label{unflavoredcase}
\setcounter{equation}{0}
This section might be read almost independently of the rest of the paper
and addresses the problem of finding gravity duals to different UV 
completions to $\mathcal{N}=1$ SYM from the one described in section 2.
Concretely, we will deal with the setup presented in section
\ref{sect: unflavnonab}.
These solutions  have appeared, in the context of this
work, because in the procedure 
we devised to add flavors, we first deformed the original (unflavored) 
background 
(\ref{metricaa})-(\ref{RR}), see the details in section \ref{flavnonsing}. 
As explained in section \ref{sect: unflavnonab}, these
backgrounds are an interesting subcase
of the ansatz studied in \cite{Butti:2004pk}.

Below, we will study a little 
more the 
interesting solutions of the unflavored system 
(\ref{nonabmetric})-(\ref{solnonab}). In section \ref{sect: descr}, we 
describe
the backgrounds arising from the study of these equations and 
in sections \ref{gauge an unfl}-\ref{betaunflavoredsection}
we discuss their gauge theory interpretation.
\subsection{Description of the Solutions}
\label{sect: descr}
Equations (\ref{akeqs}) have a one-parameter
family of regular solutions.  In fact, for small 
values of the 
radial coordinate, we find a very well controlled 
expansion in terms of a parameter $\mu$ taking values in the interval 
$(-2, -2/3]$:
\beq
a(\rho)= 1 + \mu \rho^2 +..., \;\;\;\; 
e^{2k}= N_c \left(\frac{4}{6 + 3 \mu} -
\frac{20 + 36 \mu + 9 \mu^2}{15(2+\mu)}\rho^2 +....\right)\,\,.
\label{e1}
\eeq
By inserting it in (\ref{eqforf1}), (\ref{solnonab})
 we find, for the other functions:
\beq
 e^{2g}= N_c \frac{4}{6 + 3\mu} +..., \;\;\; 
 e^{2h}= N_c \frac{4 \rho^2}{6 + 3\mu} +..., \;\;
e^{2f-2f_0}= 1+ \frac{(2+\mu)^2}{8}\rho^2+....
\label{e2}
\eeq
 In the case in 
which 
the parameter $\mu=-2/3$ we reobtain the 
known  solution 
(\ref{metricaa})-(\ref{RR}), that is:
\beq
a=\frac{2\rho}{\sinh 2\rho}\,\,,\qquad\quad e^{2k}=N_c
\label{exacta1}
\eeq 
and in the case $\mu=-2$ (and 
$N_c=0$) we 
get  four-dimensional Minkowski times the 
deformed 
conifold (see appendix \ref{appendixb}). 
For $\rho \to \infty$ (except in the particular $\mu=-\frac23$ case),
the solutions asymptote to the deformed conifold 
metric (for details, see appendix \ref{appendixb}), 
{\it i.e} very far away from the branes their
effect becomes negligible and the background asymptotes to a Ricci flat
geometry.

\begin{figure}[!htb] 
   \includegraphics[width=0.45\textwidth]{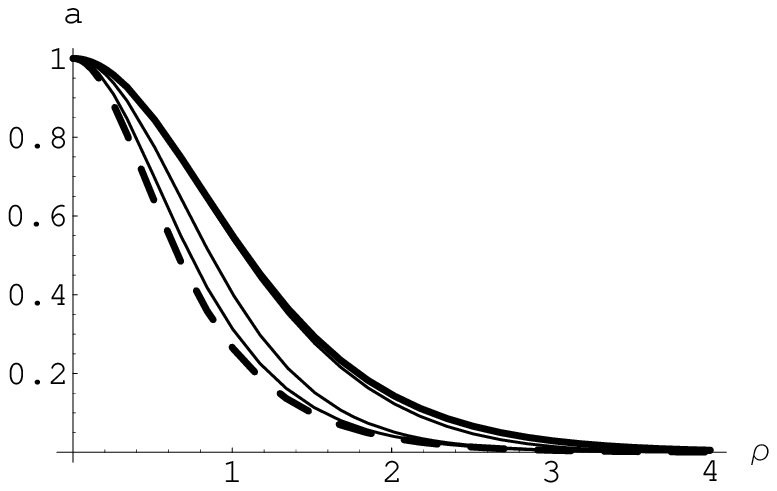} 
   \includegraphics[width=0.45\textwidth]{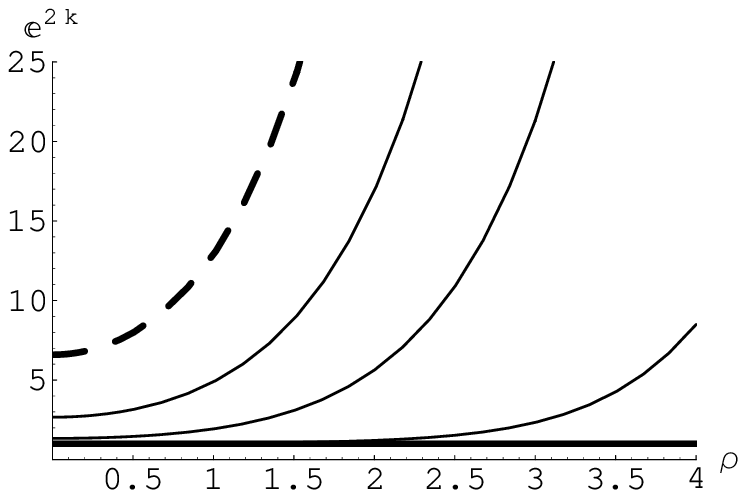} \\
   \includegraphics[width=0.45\textwidth]{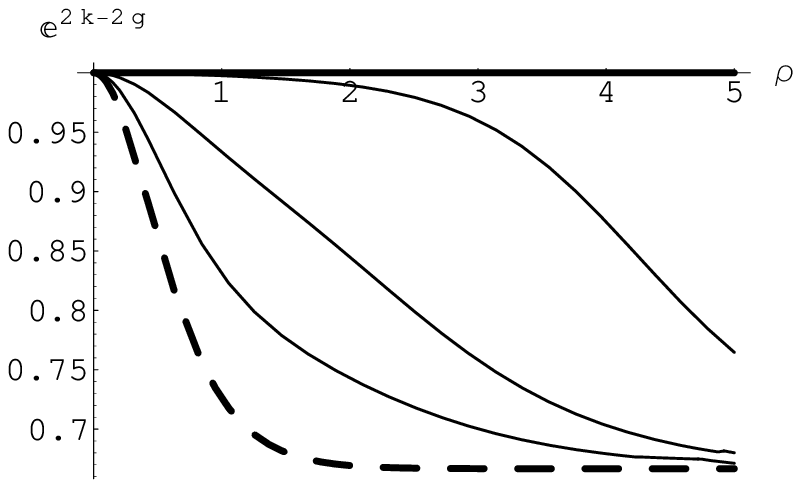} 
   \includegraphics[width=0.45\textwidth]{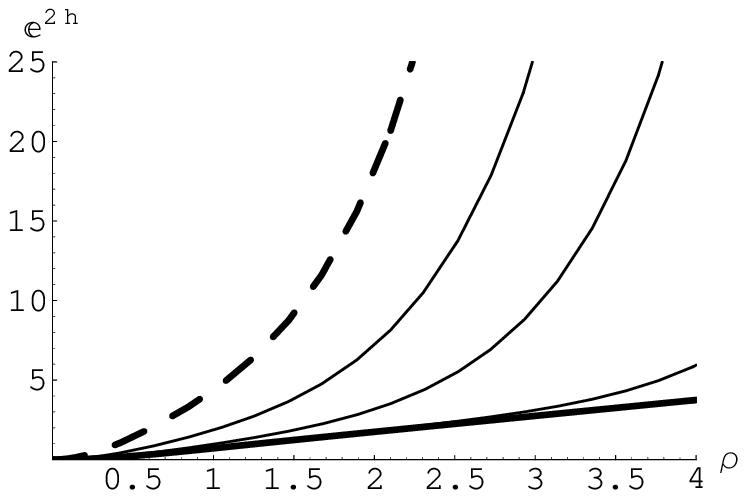} \\
   \centerline{\includegraphics[width=0.45\textwidth]{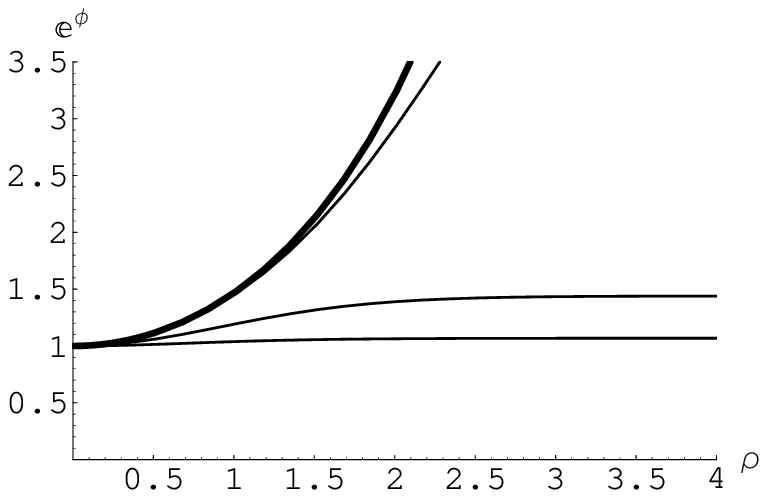} }
   \caption{We plot the different functions for different values
   of the parameter $\mu$. The thick solid line corresponds to
   $\mu=-\frac23$ (the usual MN case), the solid thin lines
   correspond, respectively, to $\mu=-.68,-1,-1.5$ and the dashed line
   is a deformed conifold.}
   \label{fig1}
\end{figure}

In the figure \ref{fig1} we present numerical plots of several of the functions
defining the geometry for different values of $\mu$.
Notice in particular that, unlike the usual  solution (\ref{metricaa})-(\ref{RR}), here the dilaton
does not diverge\footnote{A usual criticism to solutions like 
 (\ref{metricaa})-(\ref{RR}) is that 
since the dilaton diverges for large values of the radial coordinate, one 
must S-dualize and we should be dealing with NS5 
branes, hence, the UV completion is a little string theory, which is an 
unconventional completion for a field theory.}.

\subsection{Gauge Theory Analysis}
\label{gauge an unfl}
All the deformed solutions presented in section 
\ref{sect: descr}
are non-singular duals to 
field theories with field content being a non-abelian gauge field, a 
Majorana fermion and a set of massive adjoint scalar multiplets. 
These field theories in principle confine since the warp factor 
in the metric (\ref{nonabmetric}) goes to a  constant value near $\rho=0$.
This will be analyzed in more detail in section~\ref{unflavwl}.
Regarding the string theory picture, this is a compactification of 
$N_c$ D5-branes on a two-sphere, that preserves 
$\mathcal{N}=1$ SUSY. As studied in many 
places, and recently reviewed in detail in \cite{Andrews:2005cv}, this 
leads to a  field theory whose lagrangian is 
the one of $\mathcal{N}=1$ SYM coupled 
to a set of massive scalar multiplets. We will assume this picture below, 
and will argue that the dynamics of the massive modes is related to the 
parameter that labels the family of solutions.

Contrary to 
what one might think, this is not the ``brane side'' of the solution 
(\ref{metricaa}) (for 
the case of D6-branes this ``brane side'' solution was found in 
\cite{Brandhuber:2001yi}). If these new solutions were the ``brane-side'', 
one should see a resolved conifold and the metric structure that 
characterizes 
the D5-branes, near the origin of the space. One might want to interpret 
this new solution as the 
``non-near-horizon'' version of the solution of D5-branes wrapping a 
two-cycle inside the conifold. This must be at least partially correct,
since at $\rho \to \infty$ one goes to Ricci flat geometry and the effect
of the fluxes disappears (notice also that
a metric like (\ref{nonabmetric}) cannot be obtained by uplifting
from 7d gauged sugra in the same way as (\ref{metricaa})). 
However, it cannot be the whole story since, as
we will see, the IR physics is also modified within the family of solutions.

The proposal we want to put 
forward is that different members of this family (that can 
be thought of as different values of $\mu$ in the window $(-2, -2/3]$)
differ in the superpotential that affects the dynamics and masses of the 
KK modes. At the same time, it is also possible that different members of this family 
differ  in VEV's for some operator. Then, the UV completion of the theory
changes, modifying the large $\rho$ behavior of the background
 (this is in analogy with turning on an irrelevant operator in 
 $AdS_5 \times S^5$). But this change in the UV dynamics also alters the
 infrared (in analogy with what would be a dangerously irrelevant operator).

Indeed, we know that all these 
infinite solutions 
preserve $\mathcal{N}=1$ SUSY by virtue of a 
twisting procedure in the gauge theory.
It should be quite interesting to analyze how the developments in 
\cite{Andrews:2005cv, Andrews:2006aw} apply to this new set of 
configurations.
It is also interesting to notice that any expansion that starts out of the 
window $(-2 ,-2/3]$ for the parameter $\mu$ mentioned above
generates pathological solutions. This suggests that the dynamics of the KK 
modes is such that it shows some instability or similar sick behavior for 
solutions that are outside this window (as the supergravity solution 
becomes singular) \footnote{One might speculate with the existence of a 
potential term in the KK lagrangian of the form 
($\Phi$ denotes a generic KK mode) 
$V= \mu^2/2 \Phi^2 - (\mu+2)(\mu+2/3) \Phi^{2n} +...$ that captures this 
behavior.}. 
Besides, we stress that the values $\mu=-2, \mu=-2/3$ 
correspond to conifolds with maximal and zero deformation parameter. So, 
from the gravity side the existence of this window is clear, see
appendix \ref{appendixb}. 

A first check of our proposal comes from constructing new backgrounds doing 
solution-generating transformations on these 
$\mu$-family of solutions, that have a $U(1)\times U(1)$ isometry. These 
isometries are global symmetries of the KK part of the spectrum, hence, 
constructing a new background by $SL(3,R)$ transformations based on those 
$U(1)$'s (following, for example, the techniques developed in 
\cite{Lunin:2005jy}), 
will only affect the dynamics of the KK modes. This idea was 
proposed in \cite{Gursoy:2005cn, Pal:2005nr} to be the way to distinguish between a pure SYM effect 
and the dynamics of the KK modes influencing some observable.
When we perform the transformation on this family of backgrounds
(we will not report here the formulas, since they are quite lengthy), we 
observe a mixing between the parameter of deformation $\gamma$ and the 
parameter $\mu$ (changing the dynamics of the KK modes).

In the following, we will study a sample of gauge theory observables
that range from the possible existence of massless axionic glueballs, 
k-string tensions, annulons and PP-waves spectrum, Wilson loops and beta 
function. Technical details can be found in appendix \ref{appd}.
These examples will support the validity of the proposal that we 
have made above.

\subsection{Axionic Massless Glueballs}
In the paper \cite{Gubser:2004qj}, the authors studied the possibility of 
finding a massless glueball in the exact solution 
(\ref{metricaa})-(\ref{RR}). 
They found that this glueball was not normalizable, which led them  not 
to consider it to be a physical excitation. This is in contrast to 
what happens for the Klebanov-Strassler solution or a non-supersymmetric
deformation of it in which there are, in fact, physical massless glueballs
\cite{Gubser:2004qj,Schvellinger:2004am}.

In the exact (\ref{metricaa})-(\ref{RR}) solution ($\mu=-\frac23$), the 
dilaton diverges at large values of the radial coordinate and this is the 
root of the non-normalizability of this mode. In this new set of 
solutions, we have that for $\mu\neq -2/3$ the dilaton is 
bounded at infinity, see figure \ref{fig1}. It is natural, therefore, to wonder 
whether this 
family of solutions might have a massless glueball. Let us analyze this in 
some detail.
In principle, we do not expect this massless glueball to exist, since 
there is not in this model any apparent $U(1)_B$ symmetry that is 
spontaneously broken, leading to a massless supermultiplet of composites.

In the paper \cite{Caceres:2005yx} the fluctuated eqs of motion were 
studied in detail.
Let us assume that the background fields vary according to
\beq
g_{\mu\nu}\to g_{\mu\nu}+ \epsilon h_{\mu\nu},\;\; \phi\to 
\phi+\epsilon\delta\phi, \;\;F_{\mu\nu\rho}\to 
F_{\mu\nu\rho}+\epsilon \delta F_{\mu\nu\rho}
\label{fluct2ba}
\eeq
So, keeping only linear order in the parameter $\epsilon$, we get 
eqs for the fluctuated fields that we report in detail in appendix 
\ref{appd}. 
One can easily see that a solution to these eqs is the one found by the 
authors of \cite{Gubser:2004qj}
\beq
\delta F_3= *_4 d a, \;\;\; h_{\mu\nu}= \delta\phi=0; \;\; F_3 .\delta 
F_3=0
\eeq
Once again, the point in  the paper \cite{Gubser:2004qj}, is that 
this 
fluctuation, due to the diverging 
dilaton is not normalizable, so they reject the excitation as a physical 
state. We might wonder what will happen now that we have a new family 
of solutions with a bounded dilaton.
The norm of the fluctuation is
\beq
|\delta F_3|^2 = (*_4 d a )^2\int_0^\infty dr e^{2k + 2g + 2 h + 5\phi}
\label{normnew}
\eeq
In this case the dilaton does not diverge, but the functions $g,h,k$ are 
unbounded making again the mode under consideration not normalizable. Even in this  family of solutions, therefore, we do not seem to have 
this massless glueball. There might be  another combination of fields 
that could play that role, but this does not seem likely, due to the 
fact that this family of theories (if they differ from (\ref{metricaa})-(\ref{RR}) in the masses and 
dynamics for the 
KK-modes and possibly condensates)
should not have a $U(1)$ symmetry that is spontaneously broken leading 
to a massless excitation. 

\subsection{Confining $k$-Strings}\label{sub: k-string}
One very interesting problem in $SU(N)$-QCD and similar theories
is the study of flux tubes induced by color sources in higher 
representations. If the source has $k$ fundamental indexes, the flux tube 
is called the $k$-string. The law for the tensions of these $k$-strings 
has been subject of interest, from the lattice viewpoint and also from 
a more  formal side. 

In the very nice paper \cite{Herzog:2001fq}, Herzog and Klebanov tackled 
this problem using AdS/CFT-like dualities for confining theories. They
considered a confining string (connecting quarks in the antisymmetric 
representation) as a D3-brane wrapping a cycle near the IR of the 
geometry. They found the tensions of these strings. Let us briefly 
summarize their approach and apply it to our cases of interest.
This can be done easily with the expansion written in 
(\ref{e1})-(\ref{e2}), since we have to look at the geometry near 
$\rho=0$.
It is easy to see that our geometry (after the choice of the cycle 
$\theta=\tt, \varphi=2\pi- \tilde{\varphi}, \psi$ and a rescaling $\psi\to 
\psi/2$) turns exactly into the geometry in eq.(12) of 
\cite{Herzog:2001fq}.
So,  near $\rho=0$ the metric induced on these D3-branes that wrap the 
cycle above and  extend in $(t,x)$ forming a confining string, reads
\beq
ds^2=e^{2f_0}\Big[  dx_{1,1}^2 + e^{2k(0)}\big(  
d\psi^2 + 
\cos^2\psi (d\theta^2 + \sin^2\theta d\varphi^2)   \big) \Big].
\eeq
Following the approach of Herzog and Klebanov, we can compute the tension of 
the $q$-confining string as the tension of a D3-brane wrapping the finite 
three-cycle above
\beq
T\sim \Big[ b^2 \sin^4\psi +(\psi -\frac{\sin2\psi}{2} -\frac{\pi q}{N_c})^2  
\Big]^{1/2}.
\label{strtension}
\eeq
The coefficient is $b= \frac{e^{2k(0)}}{N_c} = 
\frac{4}{6+ 3\mu} $. Only for $\mu=-2/3$, that is 
for the exact solution (\ref{exacta1}), we have
$b=1$.
It is precisely this coefficient the one that propagates into formulas
(12)-(18) of \cite{Herzog:2001fq}.
Then, when minimizing the action of this confining string, 
our formulas will be exactly like the ones 
in \cite{Herzog:2001fq}.
The tension of the confining string is minimal when
\beq
\psi=\frac{1-b^2}{2} \sin\psi +\frac{\pi q}{N_c},\;\;\quad
T_q= b\sin\psi\sqrt{1 + (b^2-1)\cos^2\psi}
\eeq
The law that is conjectured to work for $\mathcal{N}=1$ SYM (see for example 
\cite{Hanany:1997hr}) is
\beq
\frac{T_k}{T_k'}= \frac{\sin(\pi k/N_c)}{\sin(\pi k'/N_c)}
\eeq
whereas the string tension in this family of models is not obeying a sine 
law, and only for the exact solution ($\mu=-2/3$) this behavior is reproduced.
It is interesting to notice that observables like this seem to diverge for 
$\mu=-2$, but we know that this point is out of our family.

We mentioned above that we could interpret the parameter $\mu$
as different masses, and most likely also different parameters in the superpotential, for the KK modes.
This 
matches nicely with  the findings of the paper \cite{Edelstein:2000za}. 
The authors there have pointed out that if one considers $\mathcal{N}=1$ SYM as 
broken $\mathcal{N}=2$ SYM, the sine law is not universal. The set of models we have 
resembles quite much these $\mathcal{N}=2\to \mathcal{N}=1$ models (it is a little more subtle 
since we have a twisting).
In the analysis done in \cite{Herzog:2001fq} for the Klebanov-Strassler
(KS) model, it was 
found that confining strings there do not present a sine law behavior. In this aspect, the family of deformed 
solutions we are studying, behaves like  the KS solution.
\subsection{Rotating Strings} 
We will study rotating strings in the same spirit as advocated by Gubser, 
Klebanov and Polyakov in the paper
\cite{Gubser:2002tv}. These rotating strings should be dual to 
large-operators in 
the gauge theory (composed of a large number of adjoint 
fields). We will see that these strings display 
characteristic energy-angular momentum
relations and sample (for short strings) the 
different dynamics of the KK modes, that is, they depend on the particular 
value of $\mu$.
There are a number of solutions already found in \cite{Pons:2003ci, Bobev:2005ng} that show this behavior. So, consider for example a 
string 
that rotates on $S^2$ and is stretched in the radial direction  
parametrized by
\beq
t= \kappa \tau, \; \varphi=\omega\tau,\; \theta=\frac{\pi}{2}, \; 
\rho=\rho(\sigma).
\label{configkk}
\eeq
After writing the Nambu-Goto action, the energy and angular momentum can 
be computed (for a string that stretches from the origin to a position 
$\rho_0$) one gets \cite{Pons:2003ci, Bobev:2005ng}
\bea
& & E\sim \int_{0}^{\rho_0} \frac{e^{2\phi + 2 k}}{\sqrt{e^{2\phi + 2 
k}(\kappa^2 -  \omega^2 (e^{2h} +\frac{e^{2g}}{4} a^2)  )}}\nonumber\\
& & J\sim \int_{0}^{\rho_0} \frac{e^{2\phi + 2 k}  (e^{2h} 
+\frac{e^{2g}}{4} a^2)}{\sqrt{e^{2\phi + 2
k}(\kappa^2 -  \omega^2 (e^{2h} +\frac{e^{2g}}{4} a^2)  )}}
\label{ejkkstring}
\eea
By doing an expansion near $\rho_0\sim 0$, that is by considering short 
strings, it is easy to 
see that the dispersion relations for this string depends on the 
parameter $\mu$.
This means that the operator dual to this string configuration contains a 
gauge invariant combination of fields, including the KK fields discussed 
above. In other words, these operators will have different 
energy-angular 
momentum relations depending on the member of the family of solutions we 
consider.

Now, we would like to study string configurations whose energy and 
angular momentum expressions {\it do not} depend on the $\mu$ parameter. 
The latter are more interesting than the solutions discussed above, since 
they should be dual to pure 
gauge theory operators made out of $A_\mu, \lambda$ and derivatives; so 
we will be more detailed. For this, we consider a 
configuration for a 
bosonic string that rotates on ``the gauge theory space" and on an 
internal direction 
(chosen so that it will give the string some $R$-charge) and 
also stretches along the radial direction.
We consider the ansatz (see appendix \ref{appd} for details)\footnote{See also \cite{Bigazzi:2004yt} for the discussion of a similar configuration in the backgrounds of  \cite{Maldacena:2000yy, Gubser:2001eg}.}:
\begin{gather}
t=\kappa \t \,,\qquad x=R(\s)\cos(\omega_1 \tau)
\,,\qquad y=R(\s)\sin(\omega_1 \tau)\,,\rc
\r=\r (\s)\,,\qquad \psi = 2 \omega_2 \tau
\end{gather}
Let us take a look at some particularly interesting solutions.

\paragraph{Rotating folded closed string}
\

Let us consider constant $\rho$. Then, one has to take $\rho=\o_2=0$
(notice that, at $\rho=0$, we have $\partial_\r k=\partial_\r \phi=0$)
and:
\beq
R= \frac{1}{\o_1} \cos{\o_1 \s}
\eeq
This is dual to a high spin glueball. Notice that the slope of the
Regge trajectory will depend on the tension of the string at the 
bottom of the geometry and therefore on $e^{\phi_0}$.
\paragraph{Pointlike strings rotating in $\psi$}
\

Consider:
\beq
R=const\,\qquad \r=const\,\qquad 
\eeq
Then, we need $\o_1 =0$ and:
\bear
\partial_\rho k =0  \Rightarrow \rho =0 \rc
e^{2k} \o_2^2 =1
\eear
These strings are dual to (long) R-charged operators .

It is interesting to notice that when computing observables for these 
strings, like energy and angular momentum ($R$-charges), they do not depend 
on the parameter $\mu$ hence, they are the same for different members of 
the family. This can be interpreted as an indication\footnote{Notice that most probably $\mu$-independence is a necessary but not sufficient condition to ensure that the operator does not contain KK modes.} that these strings are a purely gauge 
theory effect and are dual to (large) operators that are gauge 
invariant 
combinations of $A_\mu,\lambda$. On the contrary, moving strings like the 
ones 
considered in (\ref{configkk})-(\ref{ejkkstring}) 
(originally found in \cite{Pons:2003ci, Bobev:2005ng}), show dependence on 
the $\mu$ parameter when computed in these new backgrounds. This is because 
they are dual to large operators containing KK modes, as established in  
\cite{Bobev:2005ng} (following the proposal of \cite{Gursoy:2005cn}).
It would be interesting to recompute our solutions above in the 
deformed background proposed in \cite{Gursoy:2005cn}.

\subsection{Penrose Limits and PP-waves}
This subsection is constructed on the work done in the very good paper
\cite{Gimon:2002nr}. Let us summarize their findings. They studied a 
Penrose limit of the background (\ref{metricaa})-(\ref{RR})
on a geodesic that is 
mostly localized near 
$\rho=0$. The limit gives a PP-wave geometry on which the 
string 
theory can be quantized. The authors of \cite{Gimon:2002nr} associated 
this geodesic on the string side with an operator on the gravity side that 
looks like a large chain composed  of KK modes, that they called 
annulon. The spectrum of excitations of the annulon was put in 
correspondence with the spectrum of the string on the PP-wave geometry 
\cite{Gimon:2002nr}.
We will follow their approach and take the same Penrose limit on our 
one-parameter family of solutions. 
The details of the computation are spelled out in appendix \ref{appd}.
After following the approach in \cite{Gimon:2002nr}
adapted to the present case,
we get a plane wave background that looks
\bea
& & e^{-2k(0) -\phi(0)/2}L^2 ds^2= -2 dx^+ dx^- +  d\vec{x}_3^2 
+d\vec{y}_2^2 
+ dz^2 
 +\frac{1}{4}d\vec{v}_2^2  - (\frac{\vec{v}_2^2}{4} + \frac{\mu^2}{4} 
\vec{y}_2^2 )d\phi_+^2, \nonumber\\
& & F_3= (2 dv_1 \wedge dv_2 + \frac{\mu}{2} dz_1 \wedge dz_2)\wedge dx^+ .
\eea
So, we see that we have four massless directions ($\vec{x}_3, z$) and four 
massive ones; $\vec{v}_2$ with mass\footnote{Notice that the 
mass of the two modes parametrized by 
$\vec{v}_2$ is not affected by the $\mu$-deformation  and is common a feature to all members of 
the family.} $m^2=1$ and $\vec{y}_2$ with 
$m^2=\mu^2/4$. The excitations of this PP-wave are dual to  operators 
composed 
out of KK modes and one can clearly see that the parameter $\mu$ is 
related to the mass of the KK modes. As we mentioned before, it may be the 
case that $\mu$ is also influencing the dynamics of the KK's or 
condensates, but this is 
erased in the Penrose limit and we only see the effect in the masses.
So, following the treatment in \cite{Gimon:2002nr}, we can compute the 
spectrum for strings on this family of pp-wave backgrounds and put the 
dynamics of this string in correspondence with the dynamics of the family 
of annulons. 

\subsection{Wilson Loop}
\label{unflavwl}
We would like to analyze here a subtle point related to Wilson loops.
We will follow  the treatment developed in \cite{Brandhuber:1998er}
and carefully revised in 
\cite{Sonnenschein:1999if}. We will consider a probe brane that is 
wrapping the same 2-cycle as the branes that originate the geometry, 
but this probe 
brane is very far away from the origin (this probe was proven to be SUSY 
preserving in~\cite{Nunez:2003cf}). This leads to very massive 
fields that we 
associate with quarks, and we compute the expectation value of the Wilson 
loop as 
\beq
<W>\sim e^{S_{NG}}
\eeq
where $S_{NG}$ is the Nambu-Goto action for a string in the background 
(\ref{nonabmetric}).
After choosing a configuration for a string that extends in the 
coordinates $x, \rho, t$ and is parametrized by the worldsheet 
coordinates $\sigma, \tau$ according to $x=\sigma, \rho=\rho(\sigma), 
t=\tau$, the NG action will read \beq
S_{NG}= \frac{1}{2\pi\a'}\int d\sigma \sqrt{|g_{xx}g_{tt}| + |g_{tt} 
g_{\rho\rho}|\rho'^2}= \frac{1}{2\pi\a'}\int d\sigma \sqrt{F^2(\sigma) + 
G^2(\sigma)\rho'^2}
\eeq
Notice that this action implies the existence of a conserved 
quantity, that we associate with $F(\rho_0)$, where $\rho_0$ is the point 
of maximal proximity of the probe string to the origin of space $\rho=0$.
The separation of the two very heavy quarks is given by
$L= 2 F(\rho_0) \int_{\rho_0}^{\rho_1}\frac{G}
{F\sqrt{F^2 - F(\rho_0)^2}} d\rho,$ 
where $\rho_1$ is the distance where we put the probe brane (in principle 
one should take $\rho_1 \to \infty$ to make the quarks very massive).

So, following the treatment reviewed in \cite{Sonnenschein:1999if}, one 
gets that the energy of the quark anti-quark pair is given by (normalized 
after we subtract the infinite mass of the non-dynamical quarks)
\beq
E_{q\bar{q}}= F(\rho_0) L + 2 \int_{\rho_0}^{\rho_1}\frac{G}{F}
\sqrt{F^2 - F(\rho_0)^2} d\rho - 2 \int_{0}^{\rho_1} G d\rho
\label{eqq}
\eeq
The authors of \cite{Brandhuber:1998er}, proved a very nice theorem
stating under which conditions the energy of the quark anti-quark leads to 
confinement. Some hypotheses are done on the behavior of the functions 
$F(\rho)= e^{\phi}$ and $G(\rho)= e^{\phi+ k}$

1) $F(\rho) $ is analytic and near $\rho=0$, $F= F(0) + \sum_k a_k \rho^k$

2) $G(\rho) $ is analytic and near $\rho=0$, $G=  \sum_j b_j \rho^j$

3) $F(\rho), G(\rho)$ are positive

4) $F'(\rho)$ is positive

5) The integral $\int^{\infty}\frac{G}{F^2}d\rho$ converges.

We can see that the functions for our family of backgrounds behave, for 
small values of the radial coordinate, as
\beq
F\sim e^{\phi_0} \Big(1 + \frac{(2+\mu)^2}{4}\rho^2 +..\Big ), \;\; \
G\sim \frac{e^{\phi_0}g_s N_c \a'}{\sqrt{6 + 3\mu}} \Big(2 + 
\frac{(20 + 4 \mu + \mu^2)}{20} \rho^2 +..\Big)
\eeq
So, one can check that hypothesis (1)-(4) above are satisfied but, unless 
$\mu=-2/3$, hypothesis (5) fails to be satisfied by the members of this 
family. This is quite remarkable, since it separates the exact solution 
occurring when $\mu=-2/3$ from the other members of the family (just 
like what happens when computing string tensions in subsection \ref{sub: k-string}).

We want to argue nevertheless that all this family is dual to confining 
theories, since the SYM string tension will behave like $T= F(0)$, but the correction to the leading behavior might be different from the 
one predicted in \cite{Brandhuber:1998er, Sonnenschein:1999if}. It 
would be interesting to evaluate the integrals in (\ref{eqq}), to obtain 
the correction to the main confining behavior.
\subsection{Beta Function}
\label{betaunflavoredsection}
There is a 
computation of the $\mathcal{N}=1$ SYM  beta function done in 
\cite{DiVecchia:2002ks}. In spite of being quite controversial (because 
some of the assumptions are not clear to be correct 
and one might criticize the regime of validity in which the computation 
is done), 
it works in a quite remarkable way, not only because the final result is 
correct, but also because it keeps working  even when deforming the 
background in quite dramatic ways (like a non-commutative  or dipole 
deformation 
\cite{toni,Gursoy:2005cn}). 
One way in which we might view this is that the authors of ~\cite{DiVecchia:2002ks} have found a two-cycle (needed to define the 
coupling) that has very interesting and  robust geometrical properties. 
What we will 
attempt in this subsection is to do again this calculation of the beta 
function. We will see that the result seems to change drastically from the 
one in \cite{DiVecchia:2002ks}, but we will also point out  an interesting subtlety 
regarding this point.

Let us start by taking the expansion for large values of $\rho$ given in 
 eq. (\ref{expandinga}), (\ref{conifoldpar}). We see that the coupling 
constant, 
that is defined as the inverse volume of the cycle 
$\tilde{\theta}=\theta$, $\varphi=2\pi-\tilde{\varphi}$, $\psi=\pi$, 
is given by
\beq
\frac{4\pi^2}{g^2 N}= e^{2h}+ e^{2g}\frac{(a-1)^2}{4}
\label{coupling}
\eeq
Using the fact that the gaugino condensate is related to the function 
$b(\rho)$ \cite{Apreda:2001qb} allows us to write $\log(E/\Lambda)= -1/3 
\log (b(\rho))$. 
Then computing the beta function gives
\beq
 \frac{d g}{d(\log(\mu/\Lambda))}= 
-\Big(\frac{g^3 N}{2}\Big) \Big(\frac{d (g^2N)^{-1}}{d\rho}\Big) 
\Big(\frac{d\rho}{d\log(\mu/\Lambda)}\Big)\sim - g
\eeq
We see that this is not a running that we recognize as the one of a SYM 
theory. So, since this computation can be done with a member of the family 
very close to $\mu=-2/3$ (for which the computation above reproduces the 
NSVZ result), we see that separating slightly from the exact solution 
(\ref{exacta1}) changes drastically the result. This may cast doubts on 
this computation being relevant for the SYM beta function. 

\begin{figure}[htb] 
   \includegraphics[width=0.45\textwidth]{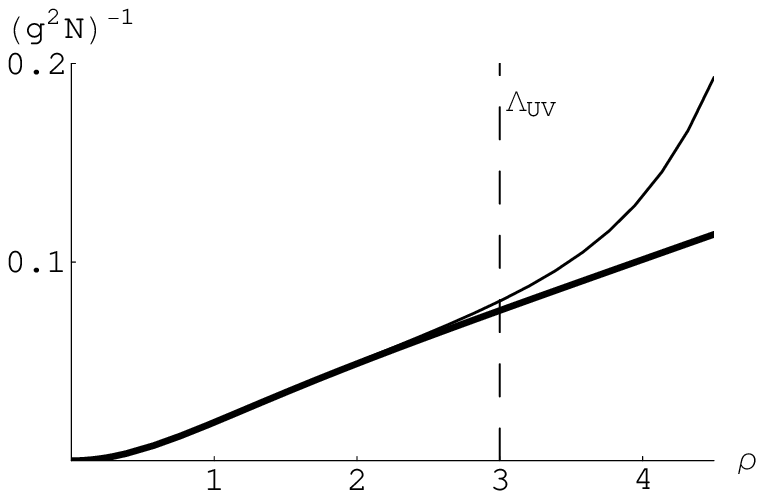} 
   \includegraphics[width=0.45\textwidth]{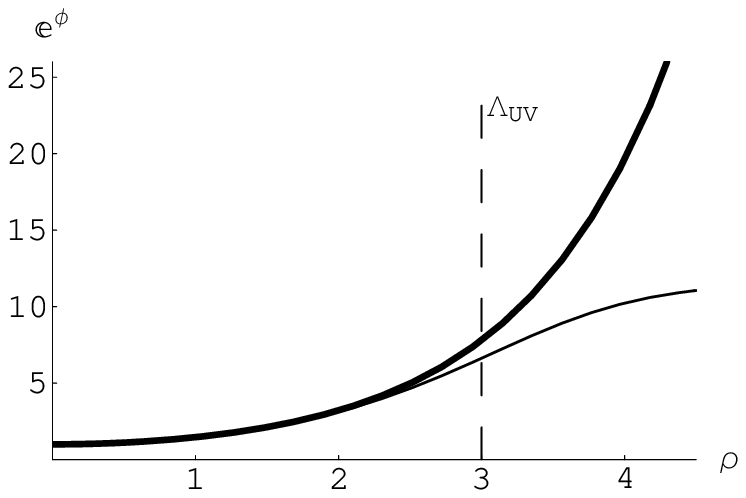} 
   \caption{We plot the inverse 't Hooft coupling
    and $e^{\phi}$ for $\mu=-.67$ (thin line)
    compared with the usual $\mu=-\frac23$ case (thick line). 
    Starting from some
    value of $\rho$ ($\Lambda_{UV}$), the UV completion sets in and
    the geometry asymptotes to a (Ricci flat) conifold.
    $\Lambda_{UV}$ decreases as $\mu$ decreases.
    }
   \label{fige2hlambda}
\end{figure}

Nevertheless, there is an interesting fact that should not be overlooked.
In Figure \ref{fige2hlambda} one can see that a coupling defined as in 
eq (\ref{coupling}) will behave, near the UV of the field theory, like 
\beq
\frac{1}{g^2 N}\sim \rho,
\eeq
if we take  the UV of the SYM field theory to be around the region 
bounded from above by
the cut-off $\Lambda_{UV}$, after which  the stringy completion to the field 
theory should set in. With this caveat, following the procedure 
described in \cite{DiVecchia:2002ks}, one gets the typical NSVZ result.
This addresses one of the criticism that were raised to the papers
\cite{DiVecchia:2002ks}. We believe that this puts the whole result 
in a better context.

\section{Conclusions and Future Directions} 
\label{concl}
In this paper we have studied two very interesting problems. The first was 
to find supergravity (string) backgrounds dual to minimally SUSY field 
theories in four dimensions containing fundamental matter
with $N_f \sim N_c$. After writing 
the BPS eqs that characterize the backgrounds, we solved them 
asymptotically and showed numerically the smooth behavior of these 
solutions.
Then, we studied numerous gauge theory aspects of these solutions, 
checking known results and making interesting predictions.
We believe the material presented in this part of the paper
may be quite useful in future approaches to non-perturbative SQCD-like 
theories.

In a second part of the paper, we discussed a family of solutions dual to SYM 
plus some UV completion, that 
generalize the background of \cite{Maldacena:2000yy}. We studied the gauge theory aspects 
of this family of UV completions by analyzing the supergravity solutions.
It is  quite interesting to uncover the differences between gauge theory 
observables of different members of the family of solutions.

We would like to comment now on possible
future directions that this paper opens. Indeed, the first thing that 
comes 
to mind is that the formalism developed 
in the first part of the paper could be immediately 
generalized to find string duals to field theories with flavor in 
different number of dimensions and with different number of SUSY's, using 
backgrounds like the ones in 
\cite{Maldacena:2001pb}. 

It would be interesting, following for example the technology developed 
in \cite{Gubser:2001eg, Aharony:2002vp}, to find non-supersymmetric deformations of the 
solutions we have 
presented here. This might prove a way to get information on QCD-like 
theories with an arbitrary number of colors and flavors.

We left some open points, specially in section \ref{gaugethnonsing}. 
Indeed, it would be very nice to get a more complete understanding of
Seiberg duality, the tension of the domain walls and the 
field theory living on them. Improving our understanding of the 
gauge theory, following for example the ideas in  \cite{Andrews:2006aw},
would be desirable. It would be very nice to explore some 
hydrodynamical aspects of the solution dual to the field theory with 
$N_f=2N_c$ at finite 
temperature. Also, finding a black hole in the solutions with arbitrary 
$N_f$ and $N_c$ is complicated, but might prove very useful. It would be 
interesting 
to follow the formalism in \cite{Berg:2005pd} for our background, in 
order to construct an effective 5-d action with which the computation of 
any correlation function might be feasible (a lot of work should be done 
in this direction). Also, completing the approach we sketched in section \ref{sec: 
fluxes} may have interesting consequences.

Finally, it would be of great interest  to find a sigma model describing 
strings in our backgrounds. In this line, the background with 
$N_f=2 N_c$, due to its simplicity, is the most suitable to start studying 
these issues.

We believe that the solution to the problems presented above, among others, 
could significantly improve the understanding of gauge-strings duality for 
QCD-like theories.

\section*{Acknowledgments}
It is a pleasure to thank Adi Armoni, 
Gaetano Bertoldi, Francesco Bigazzi, Agostino Butti, Aldo Cotrone, 
Johanna Erdmenger,
Nick Evans, Umut Gursoy, Sean Hartnoll, Tim Hollowood, Emiliano Imeroni, 
Harald Ita, Elias Kiritsis, Prem Kumar, Karl Landsteiner, Biagio Lucini, Javier Mas, 
 Asad Naqvi, Michela Petrini, Alfonso Ramallo and Alberto Zaffaroni 
for comments and discussions that improved our 
understanding of the topics and the way they were presented in this paper. RC and AP are supported by European Commission Marie Curie Postdoctoral Fellowships, under contracts MEIF-CT-2005-024710 and MEIF-CT-2005-023373, respectively. The work of  RC and AP  is also partially supported by
INTAS grant 03-51-6346, CNRS PICS \# 2530,
RTN contracts MRTN-CT-2004-005104 and
MRTN-CT-2004-503369 and by a European Union Excellence Grant 
MEXT-CT-2003-509661.

\appendix

\section{Appendix: The (abelian) BPS Equations from a Superpotential}
\label{appendixsup}
\renewcommand{\theequation}{A.\arabic{equation}}
\setcounter{equation}{0}

It is instructive  to obtain the BPS systems of section
\ref{singularsolution} using an alternative approach, computing the
so-called superpotential. Obtaining the non-abelian system of section
\ref{flavnonsing} with this technique would be much harder since it involves
algebraic constraints.

Generically, for a lagrangian of
the form ($\eta$ is the ``time" variable, $A$, $\varphi^a$
$a=1,2,\dots $ are the fields):
\beq
L=e^{c_1 A}\left[ c_2 (\partial_\eta A)^2 - \frac12 G_{ab}(\varphi)
\partial_\eta \varphi^a \partial_\eta \varphi^b - V(\varphi)\right]
\label{genlagr}
\eeq
if one can find $W$ such that:
\beq
V(\varphi)=\frac{c_3^2}{2}G^{ab}\frac{\partial W}{\partial \varphi^a}
\frac{\partial W}{\partial \varphi^b}-\frac{c_1^2 c_3^2}{4c_2}W^2
\label{genVeq}
\eeq
then the equations of motion are automatically satisfied by the first order
system:
\beq
\frac{dA}{d\eta}=\mp \frac{c_1 c_3}{2c_2} W \,\,,\qquad
\frac{d\varphi^a}{d\eta}=\pm c_3 G^{ab} \frac{\partial W}{\partial
\varphi^b}
\label{genBPS}
\eeq
Now, let us  recover the system
(\ref{newbps1})-(\ref{newbps4}). The relevant
action in Einstein frame is (\ref{stotal}),
{\it i.e.} (\ref{gravaction})+(\ref{BIterm}), since
(\ref{WZterm}) does not depend on the metric nor on the dilaton
and therefore does not enter the Einstein equations (its effect was
taken into account when writing (\ref{new3form})).

In order to obtain an effective lagrangian, we substitute
the ansatz (\ref{metric}), (\ref{new3form}). We also impose by
hand the condition
$\phi = 4f$. After integrating by parts the second derivatives 
arising from the
Ricci scalar and defining
$4Y=8f+2g+2h+k$,
we find the effective lagrangian:
\beq
L= \frac{1}{2\kappa_{(10)}^2}\,\frac18 \sin\theta
\sin \tilde \theta
e^{4Y}\Big[16Y'^2 - 2g'^2 -2h'^2-k'^2 - V\Big]
\label{Leff}
\eeq
with:
\beq
V=-2e^{-2h}-8e^{-2g}
+2e^{-4g+2k}
+\frac18 e^{-4h+2k} +2 N_c^2 e^{-4g-2k}+
\frac{(N_c - N_f)^2}{8}e^{-4h-2k} + N_f\, e^{-2g-2h}\,\,.
\label{newV}
\eeq
The last term of this
potential comes from
the Dirac brane action (\ref{BIterm}) after using (\ref{T5value}). 
The lagrangian (\ref{Leff})
 is of the form (\ref{genlagr}) and, choosing $c_3=2$, eq.
(\ref{genVeq}) reads,
\beq
V=\frac12(\partial_g W)^2+\frac12(\partial_h W)^2+
(\partial_k W)^2-W^2
\label{Veq}
\eeq
Remarkably, this equation has a very simple solution:
\beq
W=-\frac14 (N_c - N_f) e^{-2h-k} + 
N_c e^{-2g-k} - \frac14 e^{-2h+k}
- e^{-2g+k} - 2  e^{-k}\,\,.
\label{newsup}
\eeq
In particular, it is amusing to see how simple is the modification
of this $W$ when the flavor branes are added.
Inserting (\ref{newsup}) in (\ref{genBPS})
(with the upper sign), 
we recover (\ref{newbps1})-(\ref{newbps4}).

\section{Appendix: The BPS Eqs. A Detailed Derivation }
\label{appendixa}
\renewcommand{\theequation}{B.\arabic{equation}}
\setcounter{equation}{0}
In this  appendix we analyze in detail the computations that led us to 
the system of eqs (\ref{bghexpression})-(\ref{feqflav}) when 
flavoring the non-singular solution (\ref{metricaa})-(\ref{RR}).
The analysis of the generalized unflavored ansatz of section 
\ref{sect: unflavnonab} is incorporated in the same
formalism just by taking $N_f=0$. 

It is important to notice that the ansatz for the metric
we will use in the following
is a subcase of that considered in \cite{Butti:2004pk} so we can 
profit from their analysis (although $N_f=0$ in their case).
 For the sake of clarity, we rewrite here
the ansatz of the main text (\ref{nonabmetric42}), (\ref{F3beinflav}).
We take the Einstein frame 
metric\footnote{Comparing to \cite{Butti:2004pk} (we write a
subindex $B$ for quantities defined in that paper):
${h_1}_B={h_2}_B=\chi_B=K_B=Q_B=0$. Noticing that in  \cite{Butti:2004pk}
the metric is written in string frame:
$A_B=2f$, $dt_B=2e^{-k}dr$, $e^{x_B}=\frac12 e^{4f+h+g}$,
$e^{g_B}=2e^{h-g}$, $e^{6p_B}=8e^{-8f-2k-h-g}$. Afterwards,
we will define $\tilde g=g_B$ and $\rho=t_B/2$. Notice also the definitions
$du_B=e^{-4p_B}dt_B$, $v_B=e^{6p_B+2x_B}$ and that
in \cite{Butti:2004pk} $N_c$ has been set to 2. In order
to use the most usual convention, we take $a=-a_B$,
$b=-b_B$.}:
\ba
ds^2 &=& e^{2 f(r)} \Big[dx_{1,3}^2 + dr^2 + e^{2 h(r)} 
(d\theta^2 + sin^2\theta d\varphi^2) +\rc
&+&\frac{e^{2 g(r)}}{4} 
\left((\omega_1+a(r)d\theta)^2 
+ (\omega_2-a(r)\sin\theta d\varphi)^2\right)
 + \frac{e^{2 k(r)}}{4} 
(\omega_3 + cos\theta d\varphi)^2\Big]  \,\,.
\label{nonabmetrica}
\ea
The vielbein we consider for this metric is the straightforward
generalization of (\ref{vielbein}) by the inclusion of the $a(r)$
dependence in $e^1$ and $e^2$.
Apart from the dilaton, there is also the RR 3-form field strength:
\ba
F_{(3)}&=&-2 N_c e^{-3f-2g-k} e^1\wedge e^2 \wedge e^3 +
\frac{N_c}{2} b' e^{-3f-g-h} e^r\wedge (-e^\theta \wedge e^1
+e^\varphi \wedge e^2)+\rc
&+&\frac{1}{2}e^{-3f-2h-k}\left(N_c(a^2 -2ab +1) 
-N_f \right)e^\theta \wedge e^\varphi
\wedge e^3 + \rc
&+& e^{-3f-h-g-k} N_c (b-a)
\left(-e^\theta \wedge e^2 +
e^1 \wedge e^\varphi \right) \wedge e^3
\label{F3flava}
\ea
Let us analyze the dilatino and gravitino transformations. For the
functions we have in the ansatz, they read:
\ba
\delta \lambda &=& \frac12 i\Gamma^\mu \epsilon^* \partial_\mu \phi+
\frac{1}{24}e^{\frac{\phi}{2}}\Gamma^{\mu_1\mu_2\mu_3}\,
\epsilon \,F_{\mu_1\mu_2\mu_3}\,\,,\rc
\delta \psi_\mu &=& \partial_\mu \epsilon +\frac14 \omega_\mu^{ab}
\Gamma_{ab}\epsilon + \frac{e^{\frac{\phi}{2}}}{96}
(\Gamma_\mu^{\ \mu_1\mu_2\mu_3} - 9 \delta_\mu^{\mu_1}
\Gamma^{\mu_2\mu_3}) i \epsilon^* F_{\mu_1\mu_2\mu_3}
\label{IIBproja}
\ea
The projections one has to impose on the Killing spinor are:
\beq
\epsilon=i \epsilon^*\,\,,\qquad
\Gamma_{\theta\varphi}\epsilon=\Gamma_{12}\epsilon\,\,,\qquad
\Gamma_{r123}\epsilon=({\cal A}+ {\cal B}\,\Gamma_{\varphi 2})\epsilon\,\,.
\label{killingproj}
\eeq
Notice that consistency implies:
\beq
{\cal A}^2 + {\cal B}^2 =1
\label{alg1a}
\eeq
so one can parameterize:
\beq
{\cal A}=\cos\alpha\,\,,\qquad {\cal B}=\sin\alpha
\eeq
and then the spinor reads:
\beq
\epsilon = e^{-\frac{\alpha}{2}\Gamma_{\varphi 2}}\epsilon_0
\eeq
where $\epsilon_0$ satisfies the unrotated projections:
\beq
\epsilon_0=i \epsilon_0^*\,\,,\qquad
\Gamma_{\theta\varphi}\epsilon_0=\Gamma_{12}\epsilon_0\,\,,\qquad
\Gamma_{r123}\epsilon_0=\epsilon_0\,\,.
\eeq
With this, it is straightforward to obtain the set of differential equations 
and constrains coming from (\ref{IIBproja}). 
The spin connection ($de^a + \omega^a_{\ b}\wedge e^b =0$) 
is needed. We write it 
here
for completeness:
\ba
\omega^{x_i r}= e^{-f} f' e^{x_i}&,&\quad\omega^{3r}=e^{-f}(f'+k')e^3\rc
\omega^{\theta 1}= -
\omega^{\varphi 2} =  \frac14 a' e^{-f-h+g} e^r\,\,&,&\quad
\omega^{1 \varphi}= \omega^{2 \theta} =
a \sinh (k-g) e^{-f-h} e^3 \,,\rc
\omega^{\theta r}=e^{-f} (f'+h') e^\theta +
\frac14 a' e^{-f-h+g} e^1&,&\quad
\omega^{\varphi r}=e^{-f} (f'+h') e^\varphi -
\frac14 a' e^{-f-h+g} e^2\,,\rc
\omega^{1 r}=e^{-f} (f'+g') e^1 +
\frac14 a' e^{-f-h+g} e^\theta&,&\quad
\omega^{2 r}=e^{-f} (f'+g') e^2 -
\frac14 a' e^{-f-h+g} e^\varphi\,,\rc
\omega^{23}=-e^{-f+k-2g} e^1 + a e^{-f-h} \cosh (k-g) e^\theta&,&\quad
\omega^{21}=(-e^{-f+k-2g}+2e^{-f-k})e^3-e^{-h-f} \cot\theta e^\varphi\,,
\rc
\omega^{13}=e^{-f+k-2g} e^2 + a e^{-f-h} \cosh (k-g) e^\varphi&,&\quad
\omega^{\theta\varphi} = -\frac14 e^{-2h-f+k} (a^2-1) e^3
-e^{-h-f} \cot \theta e^\varphi\,,\nonumber
\ea
\ba
\omega^{3\varphi}&=&-\frac14 e^{-2h-f+k} (a^2-1) e^\theta+
ae^{-f-h} \sinh (k-g) e^1 \,\,,\rc
\omega^{3\theta}&=&\frac14 e^{-2h-f+k} (a^2-1) e^\varphi+
ae^{-f-h} \sinh (k-g) e^2 \,\,.
\ea
Comparing the equation for the variation of the dilatino and that
of $\delta\psi_x$, we immediately find the condition
\beq
\phi = 4f
\eeq
Moreover, from $\delta\psi_x=0$ we get:
\ba
f'&=&\frac{N_c}{4} {\cal A} e^{-2g-k}-
\frac{1}{16}{\cal A} e^{-2h-k}
\left(N_c (a^2 -2ab +1) - N_f\right)+\rc
&&+
\frac{1}{4} {\cal B} e^{-h-g-k} N_c (b-a) \,\,,
\label{fequationflava}\\ \rc
b'&=&\frac{1}{N_c}\Big[
2 N_c e^{h-g-k} {\cal B}-\frac12 e^{g-h-k} {\cal B} 
\left(N_c (a^2 -2ab +1) -N_f\right)-\rc
&&-2 {\cal A} e^{-k} N_c (b-a)\Big]
\label{bequationflava}
\ea
From $\delta\psi_\theta =0$ (or equivalently, from $\delta\psi_\varphi =0$)
and using the previous expressions:
\ba
h'&=&-{\cal B}e^{-h}\left(a \cosh (k-g) + \frac{1}{2}
e^{-k-g} N_c (b-a)
\right)-\frac14 {\cal A} e^{-2h + k} (a^2-1)+\rc
&&+\frac{1}{4}{\cal A} e^{-2h-k} 
\left(N_c (a^2 -2ab +1) - N_f \right)\,\,,
\label{hequationflava}\\ \rc
a'&=&-4e^{-g} {\cal A} a \cosh (k-g) + {\cal B} e^{-g-h+k} (a^2-1)
-2 N_c {\cal B} e^{h-3g-k} - \rc
&&- \frac12  {\cal B}
e^{-h-g-k}\left(N_c (a^2 -2ab +1) - N_f \right)
\label{aequationflava}
\ea
From $\delta \psi_1=0$ (or  $\delta \psi_2=0$):
\ba
g'&=&{\cal A} e^{k-2g} - {\cal B} a e^{-h} \sinh (k-g)
-\frac{1}{2} {\cal B} e^{-h-g-k} N_c(b-a)-
 N_c {\cal A}
e^{-2g-k}\,\,, \rc
0&=&-4 a {\cal A} e^{k-g-h} - 4 {\cal B} e^{k-2g} +
{\cal B} e^{-2h+k} (a^2 -1)
\label{alg2a}
\ea
From $\delta \psi_3=0$ one finds:
\ba
k'&=&-{\cal A} e^{k-2g}+2 {\cal A} e^{-k} +
\frac14 {\cal A} e^{-2h+k} (a^2 -1) + 2 a {\cal B} e^{-h} \sinh (k-g)
-N_c {\cal A} e^{-2g-k}+\rc
&&+\frac{1}{4} {\cal A} e^{-2h-k} 
\left(N_c (a^2 -2ab +1) - N_f \right)
 - 
 {\cal B} e^{-h-k-g} N_c (b-a)\,\,,
\label{kequationflava}
\ea
and the algebraic equation:
\ba
0&=&{\cal B} e^{k - 2g} -2{\cal B}e^{-k} -\frac14 {\cal B}
e^{-2h +k} (a^2-1) + 2a {\cal A} e^{-h} \sinh (k-g)
+N_c {\cal B} e^{-2g-k}- \rc
&& - \frac{1}{4} {\cal B} e^{-2h-k}
\left( N_c (a^2 -2ab +1) - N_f \right)
 -  {\cal A} e^{-h-g-k} N_c (b-a)
\label{alg3flava}
\ea
Finally, $\delta \psi_r=0$ yields:
\beq
\epsilon_0 = e^\frac{\phi}{8} \eta \,\,,
\eeq 
where $\eta$ is some constant spinor satisfying the same projections as
$\epsilon_0$ and:
\beq
\alpha' = - \frac12 e^{-h+g} (a' + N_c b' e^{-2g})
\label{alphaconda}
\eeq
In order to simplify these equations, let us make some definitions:
\beq
e^{2\tilde g} = 4 e^{2h-2g}
\label{gtildedefa}
\eeq
and, following \cite{Butti:2004pk}:
\ba
S&=&\frac{1}{2a}\sqrt{a^4 + 2a^2 (-1 + e^{2\tilde g})
+(1+ e^{2 \tilde g})^2}\,\,,\rc
C&=&\frac{1}{2a}(1+a^2 + e^{2 \tilde g})\,\,.
\label{SCdefa}
\ea
Then, the set of algebraic constraints (\ref{alg1a}), 
(\ref{alg2a}), (\ref{alg3flava}) can be written as:
\ba
{\cal A}=\frac{C-a}{S}\,\,,\qquad
{\cal B}= - \frac {e^{\tilde g}}{S}
\label{ABconda}\\
e^{2g}=\frac{N_c (b C -1) +\frac12 N_f}{aC-1}
\label{constraintflava}
\ea
Although the system seems overdetermined, it is not. By
deriving in (\ref{ABconda}) and using the rest of expressions, one
arrives at (\ref{alphaconda}), which is therefore redundant.
By deriving (\ref{constraintflava}) and using all the differential equations
and constraints, one finds an identity. Therefore,
apart from (\ref{alphaconda}),
 one of the other
differential equations  (say
the $g'$ one) is also redundant.
Let us count the number of functions. We have $f,g, \tilde g, k,a,b$.
They are not independent because of (\ref{constraintflava}). But since
the equation of $g'$ plays no role, one can think of having
five independent functions with a first order equation for each.

Remarkably, the system of first order equations and
algebraic constraints (\ref{fequationflava})-(\ref{alg3flava})
solves the set of Einstein
second order equations. The equations are (\ref{dileq}) and
(\ref{Einseq}) along with the equation of motion for the RR form
$\partial_\mu \left(e^\phi \sqrt{-g_{10}} F^{\mu\nu\rho} \right)=0$.
We have checked this using 
{\it Mathematica}.

We then would like to solve, after
substituting the expressions (\ref{ABconda}),(\ref{constraintflava}),
the equations (\ref{fequationflava}),
(\ref{bequationflava}), (\ref{aequationflava}), (\ref{kequationflava}) and the
equation for $\tilde g$ that can be obtained from the above:
\beq
\tilde g'=-{\cal B} a e^{-h+g-k} -\frac{1}{4} {\cal A}
e^{-2h+k}(a^2-1)-{\cal A}  e^{k-2g}
+ \frac{1}{4} {\cal A} e^{-2h-k}\left(N_c(a^2 -2ab +1)-N_f\right) + 
N_c {\cal A} e^{-2g-k}.
\label{gtildeeqa}
\eeq
Let us 
define:
\beq
d\rho = e^{-k} dr,
\label{rhodef}
\eeq
then, one can prove that:
\beq
\partial_\rho C = 2 S\,\,,\qquad \partial_\rho S = 2 C
\label{SCderivsa}
\eeq
and by also noticing the identity $C^2 - S^2=1$ and by fixing an
integration constant (the origin of $\rho$) we can write:
\beq
C=\cosh (2\rho) \,\,,\qquad S=\sinh (2\rho)
\label{SCexpra}
\eeq
From this one can obtain for instance $\tilde g$ in terms of $a$:
\beq
e^{2\tilde g} = 2a \cosh 2\rho -1 -a^2
\label{gtilexpra}
\eeq
The equation for $b$ is reduced to:
\beq
\partial_\rho b = \frac{1}{\sinh(2\rho)}
\left(2 (1-b\cosh(2\rho))-\frac{N_f}{N_c}\right)
\label{bexplicita}
\eeq
which is explicitly solved by:
\beq
b=\frac{\left(2-\frac{N_f}{N_c}\right) \rho + const}
{\sinh (2\rho)}
\label{bexpressiona}
\eeq
We will fix this constant to zero. In the usual unflavored case,
setting the analogous constant to zero is required in order to have
regularity at the origin. As we have seen in section
\ref{flavnonsing}, the flavored case presents a singularity anyway.
However, setting the constant to zero seems to allow a 
milder singular behavior at the origin.
Since $f$ never appears in the right hand
sides of the differential equation, we are left with
having to solve two coupled equations for $a$ and $k$. 
These are the equations (\ref{eq1}) and (\ref{eq2}) of the
main text. It does not seem possible to find an explicit general
solution of these equations.
The equation for $f$ is written in (\ref{feqflav}).

Finally, let us comment on three particular explicit solutions of
this system:
When $N_f =0$, one has, of course, the usual unflavored 
solution of \cite{Chamseddine:1997nm, Maldacena:2000yy}:
\beq
e^{2k}=1\,,\quad a=\frac{2\rho}{\sinh 2\rho}\,,\quad
e^{4f}=e^{4f_0}\frac{\sinh^{\frac12} 2\rho}{2^\frac12\, 
(\rho \coth 2\rho -\frac{\rho^2}{\sinh^2 2\rho} -\frac14)^\frac14}\,,
\qquad
 (N_f=0)\,\,.
\eeq
When $N_c=N_f=0$ one finds a Ricci flat geometry which is nothing
but the deformed conifold (see appendix \ref{appendixb} for more
details). And finally, when $N_f = 2N_c$, one has the solution presented
in section \ref{nf2nc}. Notice that for these two last cases the manipulations
leading to (\ref{SCderivsa})-(\ref{bexpressiona}) and 
(\ref{eq1})-(\ref{eq2}) are ill-defined, but the solutions can be
obtained from the system (\ref{fequationflava})-(\ref{alg3flava}).

\section{Appendix: Solutions Asymptoting to the Conifold}
\label{appendixb}
\renewcommand{\theequation}{C.\arabic{equation}}
\setcounter{equation}{0}

When $N_f=N_c=0$, the Ricci-flat solution compatible with the ansatz
(\ref{nonabmetric}) is the deformed conifold. Actually, 
(\ref{nonabmetric}) is the natural
way of writing the deformed conifold metric when it is obtained from
gauged supergravity \cite{Edelstein:2002vw}. The different functions of the
ansatz take the following values:
\ba
&&a(\rho)=\frac{1}{\cosh(2\rho)}, \;\;\; \quad \quad \quad \quad \
e^{2k}= \frac23 \epsilon^{\frac43}
{\cal K(\rho)}^{-2},\;
\rc
&&e^{2h}=\frac14 \epsilon^{\frac43} \frac{\sinh^2(2\rho)}{\cosh(2\rho)}
{\cal K(\rho)}\;,\;\;\quad 
e^{2g}= \epsilon^{\frac43} \cosh(2\rho) \cal K(\rho)\,,\;\;\quad 
\label{conifoldpar}
\ea
where $\epsilon$ is the deformation parameter,
the dilaton is constant and we have defined:
\beq
{\cal K(\rho)}=\frac{(\sinh(4\rho) - 4\rho)^{1/3}}{2^{1/3}\sinh(2\rho)}\,\,.
\eeq
We have argued, and it is apparent from figure \ref{fig1},
that the unflavored equations (\ref{akeqs})-(\ref{solnonab}) as well as the
flavored equations (\ref{bghexpression})-(\ref{feqflav}) have large $\rho$ solutions
in which the effect of the branes becomes asymptotically negligible
and they must therefore approach the Ricci flat
geometry (\ref{conifoldpar}). In this limit, the only difference between
flavored and unflavored is a rescaling of $e^{2k}$, $e^{2h}$, $e^{2g}$
by $(1-\frac{N_f}{2N_c})$.

In fact, this can be proved explicitly by discarding exponentially suppressed
terms and considering an expansion starting with:
\beq
a \cosh 2\rho -1 = \beta e^{-\frac{4\rho}{3}}(2\rho -1) + \dots
\eeq
 We can write:
\beq
a=2 e^{-2\rho} (1 + \beta e^{-\frac{4\rho}{3}}(2\rho -1))+ \dots\,\,,
\qquad
e^{2\tilde k}= \frac{2}{3\beta} e^{\frac{4\rho}{3}}+ \dots
\label{expandinga}
\eeq
and similar expressions for the rest of functions
which agree at leading order with (\ref{conifoldpar}). By comparing
(\ref{expandinga}) to (\ref{conifoldpar}),
 we see that $\beta$ is a constant related to
the deformation parameter of the conifold.

\section{Appendix: An Interesting Solution with $N_f= 2 N_c$}
\label{appendixnf2nc}
\renewcommand{\theequation}{\thesection.\arabic{equation}}
\setcounter{equation}{0}

The differential equations  (\ref{eq1}) and (\ref{eq2}) are 
not valid for the special value $x=2$. To study the solutions in 
this case start then from the unsimplified equations 
(\ref{fequationflava})-(\ref{alg3flava}). In this appendix,
we will show that there is another solution apart from
the one presented in section \ref{nf2nc}.
The manipulation of the 
differential equations and algebraic constraints goes along similar 
lines to the $x<2$ case, with the important exception that now we use 
the constraint (\ref{alg3flava}) to solve for $a$ and $b$ rather than 
for $e^{2g}$. We impose that the two independent terms in the equation 
vanish simultaneously, which gives
\be\label{anew and bnew}
a=\frac{1}{\cosh 2\rho}\qquad\quad\mathrm{and}\qquad\quad b=0
\ee
The choice of this solution to (\ref{alg3flava}) is suggested by the 
behavior of the $x<2$ solutions for $x$ that goes towards 2, see
the plots at the bottom of figure \ref{flavoredgraphs}.

At this point, it is easy to show that some of the remaining equations 
can be used to evaluate
\be\label{new2alg}
\mathcal{A}=\tanh 2\rho\qquad\quad \mathcal{B}=-\frac{1}{\cosh 2\rho} 
\qquad\quad e^{h-g}=\frac{1}{2}\tanh 2\rho
\ee
and now, as before,  we can 
reduce our system of BPS equations to a system of two 
coupled first order  differential equations
\be\label{newpartial}
\begin{split}
&\partial_\rho e^{h+g}=e^{2k}-N_c\\
&\partial_\rho k=-(e^{2k}+N_c)e^{-h-g}+2\coth 2\rho
\end{split}
\ee
with the only remaining unknown function determined 
once we find the solution to (\ref{newpartial}):
\be\label{new2f}
\partial_\rho f=\frac{N_c}{4}e^{-g-h}
\ee
To solve equations (\ref{newpartial}), we 
can obtain  $e^{2k}$ from the first one, and substitute it 
into the second one to obtain a second order differential 
equation containing only $e^{g+h}$
\be
\frac{1}{2}e^{g+h}\partial_\rho^2 e^{g+h} +
( \partial_\rho e^{g+h})^2+\partial_\rho e^{g+h} 
(3N_c-2 e^{g+h}\coth 2\rho)+2 (N_c-e^{g+h}\coth 2\rho)=0
\ee
In the large $\rho$ region, we can approximate $\coth 2\rho \simeq 1$ 
up to exponentially suppressed terms, which gives us that  
$e^{g+h}=N_c$ is an exact solution at large $\rho$. 
This also fixes the value of $e^{2k}$ in the large $\rho$ region, 
$e^{2k}=N_c$, and again the only corrections to this large $\rho$ 
solution are exponentially suppressed. From (\ref{new2alg}) and 
(\ref{new2f}) we find also
\be
\begin{split}
&e^{2h}\simeq \frac{N_c}{2} +O(e^{-2\rho})\\
&e^{2g}\simeq 2N_c +O(e^{-2\rho})\\
&f\simeq \frac{\rho}{4}+\ldots=\frac{r}{4\sqrt{N_c}}
\end{split}
\ee
Notice that this behavior is very similar to the one of the other 
flavored $x\neq 2$ cases we described in subsection \ref{sub:infty}, 
even though the constants at infinity do not match. 

\begin{figure}[htb] 
   \includegraphics[width=0.45\textwidth]{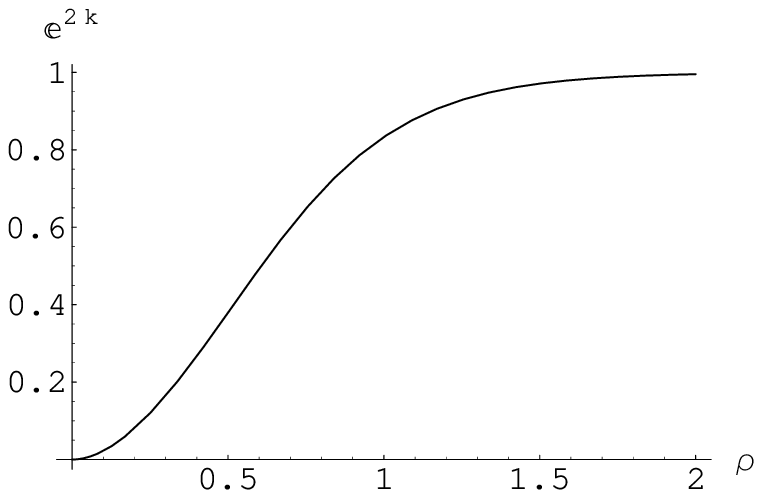} 
   \includegraphics[width=0.45\textwidth]{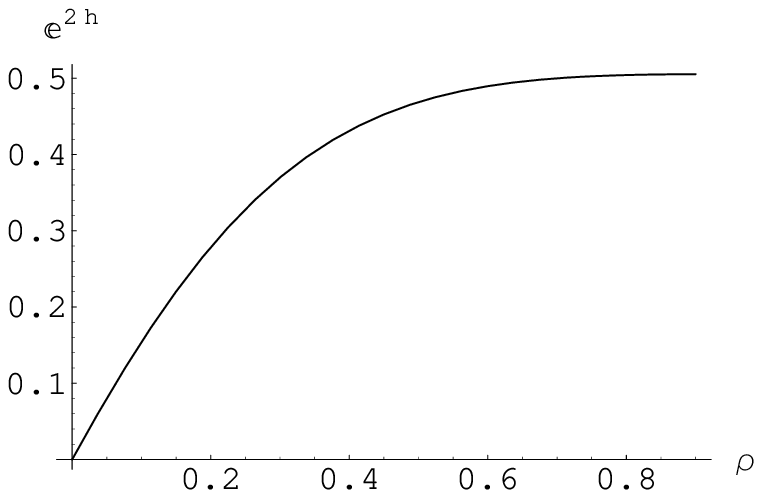} \\
    \includegraphics[width=0.45\textwidth]{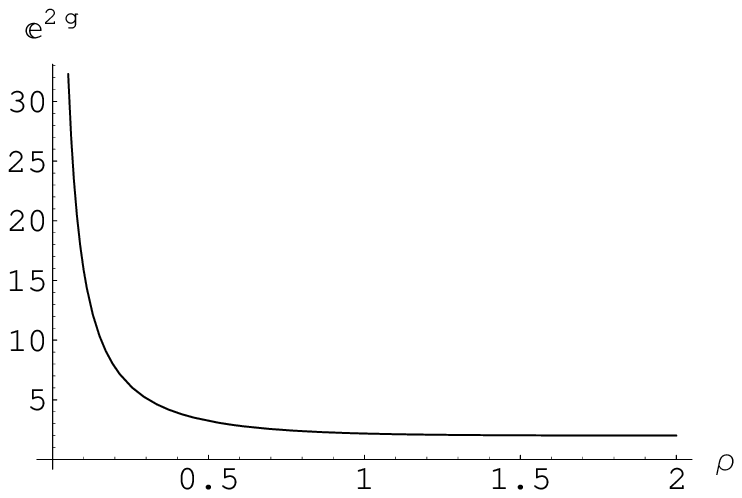} 
   \includegraphics[width=0.45\textwidth]{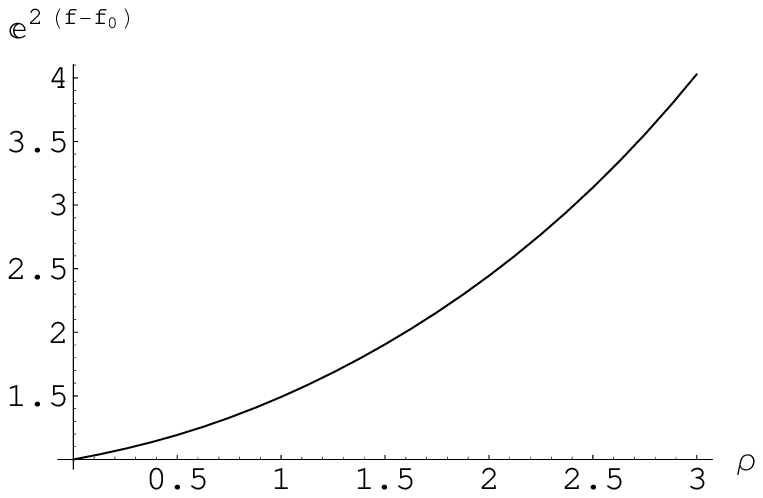} \\
   \caption{Behavior of the background functions for the $x=2$ solution with $a=\frac{1}{\cosh 2\rho}$. From the top left we have $e^{2k(\rho)}$, $e^{2h(\rho)}$, $e^{2g(\rho)}$,  $e^{2(f(\rho)-f_0)}$. Here we have taken $N_c=1$.}
   \label{figx=2new}
\end{figure}

We can now plug these asymptotic conditions in {\it Mathematica} and obtain 
the numerical solution to the BPS equations. The functions we 
obtain are shown in figure \ref{figx=2new}.
Notice that this solution has a very nice property. 
In section \ref{seibergsubsection} we had introduced the gravity dual 
description of Seiberg duality for 
the $\mathcal{N}=1$ SQCD-like theories we 
are considering. The solution we have presented 
in this subsection is exactly 
Seiberg duality invariant.  In fact, it 
is easy to check that this condition imposes a single equation
\be
e^{2h}+\frac{e^{2g}}{4}(a^2-1)=0
\ee
which is satisfied exactly by our 
$a=1/\cosh 2\rho$ and $e^{h-g}=\frac{1}{2}\tanh 2\rho$.

\section{Appendix: Some Technical Details in Section \ref{unflavoredcase}}
\label{appd}
\renewcommand{\theequation}{\thesection.\arabic{equation}}
\setcounter{equation}{0}

In this appendix, we will just cover the technical details that were left 
aside in the analysis of gauge theory aspects of the unflavored solutions.

\paragraph{Massless Glueballs}\

Let us start with the fluctuated eqs of motion, that lead us to the fact 
that a normalizable massless glueball does not exist.

The eqs for the fluctuation were analyzed in 
\cite{Caceres:2005yx} and  read, for the fluctuated Ricci tensor 
equation 
\bea
& &  \frac{1}{2}[\nabla_\alpha\nabla_\mu h_{\nu}^{\alpha} +
\nabla_\alpha\nabla_\nu h_{\mu}^{\alpha} - \nabla_\nu\nabla_\mu 
h_{\alpha}^{\alpha}
- \nabla^2 h_{\mu\nu}     ] = \frac{1}{2}[\partial_\mu \phi 
\partial_\nu \delta\phi +
\partial_\mu \delta\phi \partial_\nu \phi ]\nonumber\\
& & + \frac{g_s e^{\phi}}{4}
[\delta\phi F_{\mu **}F_{\nu}^{**} +\delta F_{\mu **} F_\nu^{**} +
F_{\mu **} \delta F_\nu^{**} - 2 h^{ab}F_{\mu a *}F_{\nu b }^* ] 
\nonumber\\
& & -\frac{g_s e^{\phi}}{48}[ g_{\mu\nu}( 2 F_3 \delta F_3 - 3 
h^{ab} 
F_{a**}F_b^{**} + \delta\phi
F_3^2) + h_{\mu\nu} F_3^2] 
\label{deltarmunu3} 
\eea
For the fluctuation of the Ricci scalar,
\bea
& &\nabla_\mu\nabla_\nu h^{\mu\nu} -\nabla^2 h_\mu^\mu -
R_{\mu\nu}h^{\mu\nu}= g^{\mu\nu}\partial_\mu\phi 
\partial_\nu\delta\phi -\frac{h^{\mu\nu}}{2}\partial_\mu\phi
\partial_\nu\phi+
\nonumber\\ & & \frac{g_s e^{\phi}}{24}[2 F_3 \delta F_3 -
3 h^{kl} F_{k **} F_l^{**} + \delta\phi F_3^2],
\label{riccivaried2b} 
\eea
For the fluctuated dilaton equation,
\beq
\nabla^2\delta\phi - h^{\mu\nu}\nabla_\mu\partial_\nu\phi 
-\frac{1}{2} g^{rr}\partial_r\phi(2\nabla^\mu h_{r\mu} - 
\partial_rh_\mu^\mu) -\frac{g_s e^\phi}{12}(\delta\phi F^2 + 2 F 
\delta F - 3 h^{\mu\nu}F_{\mu **}F_\nu^{**})=0
\label{dileq2}
\eeq
For the fluctuated Maxwell and Bianchi eqs.
\eqn{delatmaxwell}{\partial_\mu[\sqrt{g}e^\phi ( (\frac{1}{2} 
h_\rho^\rho
+\delta\phi)F^{\mu\nu\alpha}+\delta F^{\mu\nu\alpha} - h^{\alpha 
c}F^{\mu\nu}_c + h^{\nu c}F^{\mu\alpha}_c -
h^{\mu c} F_c^{\nu\alpha})]=0, \;\;d\delta F=0.}
One can easily see that a solution to these eqs is the one found by the 
authors of \cite{Gubser:2004qj},
\beq
\delta F_3= *_4 d a, \;\;\; h_{\mu\nu}= \delta\phi=0; \;\; F_3 .\delta 
F_3=0
\eeq
the analysis of this solution shows that is non-normalizable, hence is 
not a  mode in the dual gauge theory. The rest is covered in the main part 
of the paper.

\paragraph{Rotating strings}\

Now, let us turn to the analysis of new solutions for strings rotating in 
our backgrounds.
So, the Polyakov action for the configuration we will propose reads ($ 
\tau, \sigma$ are the 
worldsheet coordinates with the usual metric diag(-1,1)) reads,
\beq
{\cal S}=\frac{1}{2\pi\alpha'}\int d\sigma d\tau G_{\mu\nu}\partial_\a 
X^\mu \partial_b X^\nu \eta^{\a\b}
\eeq
and the Virasoro constraints:
\bear
\partial_\t X^\mu \partial_\s X_\mu =0 \rc
G_{\mu\nu} \partial_\t X^\mu \partial_\t X^\nu
+ G_{\mu\nu} \partial_\s X^\mu \partial_\s X^\nu = 0
\eear
Consider the ansatz, 
\bear
t=\kappa \t \,,\qquad x=x(\t,\s)\,,\qquad y=y(\t,\s)\,,\rc
\r=\r (\s)\,,\qquad \psi =\psi (\t)
\eear
This is consistent if $\th =\tilde \th =\frac{\pi}{2}$. 
We consider the induced string frame metric
\beq
ds^2 = e^\phi \left( -dt^2 + dx^2 +dy^2 + e^{2k} d\r^2 +
\frac{e^{2k} }{4}d\psi^2\right)
\eeq
where $\phi(\r)$ and $k(\r)$. The equations of motion are consistent with:
\beq
x=R(\s) \cos (\o_1 \t)\,,\qquad
y=R(\s) \sin (\o_1 \t)\,,\qquad
\psi = 2 \o_2 \t
\eeq
One obtains (prime is derivative with respect to $\s$):
\bear
R'' + (\partial_\r \phi) \r' R' + \o_1^2 R =0 \rc
\r'' + (\partial_\r k) \rho'^2 - (\partial_\r \phi) e^{-2k} R'^2
+ (\partial_\r k) \o_2^2 =0 \rc
-\kappa^2 + \o_1^2 R^2 +R'^2 + e^{2k} \o_2^2 + e^{2k} \r'^2 =0
\eear
These equations are not independent since by deriving the last one
and using the other two, one reaches an identity. The rest follows as in 
section \ref{unflavoredcase}.

\paragraph{Penrose Limit and PP-waves}\

Now, let us present the technical details of the Penrose limit and plane 
wave geometry.
The 
computation is very similar to the one in \cite{Gimon:2002nr}.

After doing  a gauge transformation (that can be better understood as a 
coordinate transformation or reparametrization of the left invariant 
forms of $SU(2)$, see \cite{Gursoy:2005cn}), we leave the geometry in the 
form
(\ref{nonabmetric}), but with the $SU(2)$-valued one form $A^a \sigma^a$ 
being
\beq
A= \mu \rho^2 \Big[(\cos\varphi d\theta -\cos\theta\sin\theta\sin\varphi 
d\varphi)\sigma^1 +(\sin\varphi d\theta + \cos\theta 
\sin\theta\cos\varphi d\varphi)\sigma^2 + \sin^2\theta d\varphi \sigma^3 
\Big].
\label{atransformed}
\eeq
Now, we have to rescale and redefine the coordinates according to,
\beq
\rho= \frac{r}{L}, \;\; \tilde{\theta}= \frac{v}{L},\;\; 
\vec{x}=\frac{\vec{x}}{L},\;\; 2\phi_+ = \psi + \tilde{\varphi}
\label{redefrescal}
\eeq
Then we take the limit of $L\to \infty$ and keep only terms suppressed in 
$L^{-2}$. We get a PP-wave that contains the so called magnetic terms 
(that is crossed terms $d\phi_+ d\varphi$ and  $d\phi_+ d\tilde{\varphi}$ 
), that reads,
\bea
& & e^{-2k(0) -\phi(0)/2}L^2 ds^2= L^2(-dt^2e^{-2k(0)} + d\phi_+^2) + dr^2 
+ 
r^2 (d\theta^2 + \sin^2\theta d\varphi^2) + \nonumber\\
& & \frac{1}{4}(dv^2 + v^2 d\tilde{\varphi}^2) 
 - (\frac{v^2}{2} d\tilde{\varphi} + \mu r^2 
\sin^2\theta d{\varphi})d\phi_+.
\eea
We remind the reader that $\mu$ is the parameter labelling the different 
solutions near 
$\rho=0$. Even when this can be quantized, is better in order to have a 
better intuition on the spectrum, to make a last change in variables
\beq
d\varphi \to d\varphi +\frac{\mu}{2}d\phi_+, \;\;\; d\tilde{\varphi}\to 
d\tilde{\varphi} + d\phi_+ , 
\eeq
then, defining as usual,
$x^+= t e^{-k(0)}, \;\;\; x^-= L^2 (t - \phi_+)$, we have that the 
geometry looks,
\beq
e^{-2k(0) -\phi(0)/2}L^2 ds^2= -2 dx^+ dx^- +  d\vec{x}_3^2 +d\vec{y}_2^2 
+ dz^2 
 +\frac{1}{4}d\vec{v}_2^2  - (\frac{\vec{v}_2^2}{4} + \frac{\mu^2}{4} 
\vec{y}_2^2 )d\phi_+^2.
\eeq
For the rest of the analysis, please go back to the relevant part in 
section \ref{unflavoredcase}.

\renewcommand{\theequation}{D.\arabic{equation}}
\setcounter{equation}{0}

\end{document}